%% file: main.tex
\documentclass[12pt]{article}
\pdfoutput=1
\usepackage{amsmath,amssymb,amsthm}
\usepackage{bm}
\usepackage{pdflscape}
\usepackage[usenames,dvipsnames]{xcolor}
\usepackage{graphicx}
\usepackage[para]{threeparttable}
\usepackage{longtable}
\usepackage{booktabs}
\usepackage{float}
\usepackage{array}
\usepackage{arydshln}
\usepackage{multirow}
\usepackage{rotfloat}
\usepackage{caption}
\usepackage{subcaption}
\usepackage[normalem]{ulem}
\usepackage{cite}
\usepackage{hyperref}
\usepackage{cleveref}
\usepackage{tikz}
\usepackage{tikz-cd}

\newcommand{\fsu}{\mathfrak{su}}
\newcommand{\fu}{\mathfrak{u}}
\newcommand{\fso}{\mathfrak{so}}
\newcommand{\fsp}{\mathfrak{sp}}
\newcommand{\fg}{\mathfrak{g}}
\newcommand{\ff}{\mathfrak{f}}
\newcommand{\fe}{\mathfrak{e}}

\numberwithin{equation}{section}
\numberwithin{figure}{section}

\usepackage{parskip}
\usepackage{setspace}
\theoremstyle{plain}

\theoremstyle{plain}
\newtheorem{definition}{Definition}
\theoremstyle{plain}
\newtheorem{classification}{Classification}
\theoremstyle{plain}
\newtheorem{algorithm}{Algorithm}

\def\node#1#2{\overset{#1}{\underset{#2}{\circ}}}

\usetikzlibrary{arrows}
\usetikzlibrary{calc}
\tikzstyle{every picture}+=[remember picture]
\tikzstyle{na} = [baseline=-.5ex]
\tikzstyle{mine}= [arrows={angle 90}-{angle 90},thick]

\def\Llleftarrow{%
\lower2pt\hbox{\begingroup
\tikz
\draw[shorten >=0pt,shorten <=0pt] (0,3pt) -- ++(-1em,0) (0,1pt) -- ++(-1em-1pt,0) (0,-1pt) -- ++(-1em-1pt,0) (0,-3pt) -- ++(-1em,0) (-1em+1pt,5pt) to[out=-105,in=45] (-1em-2pt,0) to[out=-45,in=105] (-1em+1pt,-5pt);
\endgroup}
}

\tikzset{
  doublearrow/.style={
    double distance=2pt,
    shorten <=0.2pt,
    shorten >=0.2pt,
    postaction={decorate},
    decoration={markings,
      mark=at position 0.1 with {
			  \draw[line cap=round] (0.06,0.14) -- (0.16,0) -- (0.06,-0.14);
      }
    }
  }
}

\tikzset{
  doublearrowright/.style={
    preaction={draw,transform canvas={yshift=1pt}},
    draw,transform canvas={yshift=-1pt},
    postaction={decorate},
    decoration={markings,
      mark=at position 0.1 with {
			  \draw[line cap=round] (0.08,0.14) -- (0.19,0.035) -- (0.08,-0.075);
      }
    }
  }
}

\tikzset{
  doublearrowleft/.style={
    preaction={draw,transform canvas={yshift=1pt}},
    draw,transform canvas={yshift=-1pt},
    postaction={decorate},
    decoration={markings,
      mark=at position 0.16 with {
			  \draw[line cap=round] (0.19,0.15) -- (0.08,0.035) -- (0.19,-0.075);
      }
    }
  }
}

\usetikzlibrary{arrows,shapes.misc,positioning,decorations.pathmorphing,decorations.markings,decorations.pathreplacing,matrix,patterns,backgrounds,calligraphy}

\tikzset{
scale cd/.style={every label/.append style={scale=#1}, cells={nodes={scale=#1}}}}

\tikzset{gauge/.style={rounded rectangle, draw=black!100, thick, minimum size=5mm},  gaugeD/.style={rounded rectangle, draw=black!100,double,thick,minimum size=5mm},  empty/.style={rounded rectangle, draw=white!100, thick, minimum size=5mm}, flavor/.style={rectangle, draw=black!100, thick, minimum size=5mm},flavorD/.style={rectangle, draw=black!100, double,thick, minimum size=5mm}}

\tikzset{
node/.style={circle, thick, draw=black!100,fill=white!100,  minimum size=2mm, inner sep=0pt},
sqnode/.style={rectangle
, thick, draw=black!100,fill=white!100,  minimum size=2mm, inner sep=0pt
},
sonode/.style={circle, thick, draw=black!100,fill=red!100,  minimum size=3mm, inner sep=0pt},
spnode/.style={circle, thick, draw=black!100,fill=blue!100,  minimum size=3mm, inner sep=0pt},
fnode/.style={rectangle, thick, draw=black!100,fill=white!100,  minimum size=3mm, inner sep=0pt},
tnode/.style={rounded rectangle, outer sep=0pt, thick, minimum size=5mm},
Rightarrow/.style={double equal sign distance,>={Implies},->},
triplearrow/.style={-,preaction={draw,Rightarrow}},
 triple/.style={
        double distance=2pt,
        line width=0.4pt,
    }
}

\begin{document}
 
~\vspace{2cm}
\begin{center}
\vspace*{-3cm} 
\begin{flushright}
{\tt ZMP-HH/25-19}

\end{flushright}

{\LARGE\bfseries The Weight of Spin(32)$/\mathbb{Z}_2$ Little Strings: T-duality and Hasse Diagrams
}
\vspace{1.2cm}


Hamza Ahmed$^{\,a,}$\footnote{\href{mailto:ahmed.ha@northeastern.edu}{ahmed.ha@northeastern.edu}}, 
Florent Baume$^{\,b,}$\footnote{\href{mailto:florent.baume@desy.de}{florent.baume@desy.de}}, 
Paul-Konstantin Oehlmann$^{\,c,}$\footnote{\href{mailto:oehlmann@ucsb.edu}{oehlmann@ucsb.edu}}
\\[10mm]
\bigskip
{  
	{\it ${}^{\text{a}}$Department of Physics, Northeastern University, Boston, MA 02115, USA}\\[.5em]
    {\it ${}^{\text{b}}$II. Institut f\"ur Theoretische Physik, Universit\"at Hamburg, 22607 Hamburg, Germany}\\[.5em]
	{\it ${}^{\text{c}}$Department of Physics, University of California, Santa Barbara, CA 93106, USA}\\[.5em]
}
\end{center}
\setcounter{footnote}{0} 
\bigskip\bigskip

\begin{abstract}
		\noindent
		We study the worldvolume theories of stacks of $\text{Spin}(32)/\mathbb{Z}_2$
		heterotic NS5-branes probing a transverse singularity $\mathfrak{g}$. We revisit
		and extend the original classification by Blum and Intriligator, and
		show that the resulting $6d$ Little String Theories (LSTs) are
		naturally labeled by affine dominant coweights of the singularity
		$\mathfrak{g}$. This in turn enables us to efficiently arrange these
		theories into groupings satisfying all known necessary conditions to be
		T-dual. Using this formulation, we then study the partial order of
		those coweights, and extend a recently proposed slice-subtraction
		algorithm to construct Hasse diagrams for LSTs directly from their
		six-dimensional generalized quivers, allowing us to probe certain
		properties of their Higgs branch. Along the way, we exploit these
		techniques to show that the number of duality orbits at maximal flavor
		rank is determined by the center of the transverse singularity $\mathfrak{g}$,
		and provide a simplified proof of a monotonicity theorem for this class
		of theories. Finally, we show how some of our techniques can be
		extended to other classes of 6d theories, such as Type-II LSTs and
		SCFTs.
\end{abstract}

\clearpage
\tableofcontents
\clearpage

\newpage

\section{Introduction}

Six-dimensional supersymmetric Quantum Field Theories (SQFTs) occupy a special
place in the landscape of field theories due to a combination of a high degree
of supersymmetry and a large number of spacetime dimensions, which severely
restricts the consistent matter spectra and allowed symmetries. This leads to a
very constrained set of theories that provides an attractive playground to study
a host of field-theoretic features, ranging from non-perturbative effects to
generalized symmetries \cite{Gaiotto:2014kfa}.

Moreover, the presence of anti-symmetric tensor fields and the BPS non-critical
strings to which they couple is intimately related to the possible ultraviolet
(UV) completions of six-dimensional SQFTs in the absence of gravity: when BPS
strings become tensionless, new massless, non-perturbative degrees of freedom
arise \cite{Witten:1995zh, Seiberg:1996qx}. If all BPS strings can be made
tensionless at the same time, all other scales---in particular inverse gauge
coupling constants---are also eliminated by supersymmetry. One obtains a
local, strongly coupled, superconformal field theory (SCFT). On the other hand,
there are cases where not all BPS strings can be made tensionless
simultaneously. The remnant scale acts as an ultraviolet cutoff above which the
theory develops a Hagedorn density of states \cite{Aharony:1998ub}. The
resulting non-local theories, known as Little String Theories (LSTs), share
several hallmark features of ten-dimensional critical strings---most notably
T-duality---despite being non-gravitational. Below this scale, LSTs reduce to
effective Quantum Field Theories \cite{Seiberg:1996qx, Aharony:1999ks} whose
dynamics can be analyzed using conventional field-theoretic techniques.

A tool of choice for studying SQFTs in six dimensions has long been geometric
engineering through string theory, in particular F-theory. Indeed, arguably one of the
greatest successes of F-theory has been an enumeration of all possible
geometries leading to SCFTs \cite{Heckman:2013pva, Heckman:2015bfa} and LSTs
\cite{Bhardwaj:2015xxa, Bhardwaj:2015oru, Bhardwaj:2019hhd} with eight supercharges in spacetime
dimensions $d=6$ via compactification, see reference \cite{Heckman:2018jxk} for
a review.

Little String Theories have recently experienced a renewed interest, as they
exhibit generalized symmetries \cite{Gaiotto:2014kfa}---see recent reviews
\cite{Schafer-Nameki:2023jdn, Shao:2023gho, Luo:2023ive, Bhardwaj:2023kri,
Iqbal:2024pee, Brennan:2023mmt, Costa:2024wks} and references therein---which
can be understood with the geometric techniques of
F/M-theory \cite{Morrison:2020ool, DelZotto:2015isa, Albertini:2020mdx}. They are
in particular all endowed with a unique continuous two-group
symmetry \cite{Cordova:2018cvg, Cordova:2020tij, DelZotto:2020sop,
Apruzzi:2021mlh}, which is absent for SCFTs or in supergravity. The presence of
a non-dynamical tensor multiplet $B_{\mu\nu}^\text{LST}$ associated with the
little string induces a $\mathfrak{u}(1)^{(1)}$ one-form symmetry that mixes
non-trivially with ordinary zero-form symmetries, namely the
$\mathfrak{su}(2)_R$ R-symmetry, spacetime Poincar\'e invariance
$\mathfrak{p}$, and a possible flavor symmetry $\mathfrak{f}$, to form a
two-group:
\begin{equation}
		\left(\mathfrak{su}(2)_R\oplus \mathfrak{p}\oplus \mathfrak{f}\right)^{(0)}\,\times_{\kappa_R,\kappa_P,\kappa_F}\, \mathfrak{u}(1)^{(1)}\,.
\end{equation}
The coefficients $\kappa_R,\kappa_P,\kappa_F$ are the structure constants of
the two-group, and encode the non-trivial mixing between
$B_{\mu\nu}^\text{LST}$ and the other background fields under a symmetry
transformation.

In addition, like their critical ten-dimensional cousins, LSTs manifest a form
of T-duality: upon reduction on a circle, two different LSTs may lead to the
same five-dimensional theory. This feature was already recognized three decades
ago by Aspinwall and Morrison, who explicitly realized the duality between
$\mathfrak{so}_{32}$ and $\mathfrak{e}_8\oplus\mathfrak{e}_8$ heterotic strings
by identifying inequivalent elliptic fibrations of the same birational
Calabi--Yau threefold \cite{Aspinwall:1997ye}. Geometric engineering
has since been the preferred approach to verify T-duality, and has been
recently applied to a plethora of models \cite{DelZotto:2022ohj,
		DelZotto:2022xrh,DelZotto:2023ahf, Bhardwaj:2022ekc, Ahmed:2023lhj,
Mansi:2023faa, Baume:2024oqn, Ahmed:2024wve}. An important question is
therefore what structure T-dual models may exhibit, and how different T-dual
pairs are related. Indeed, there can also be more than two LSTs that are dual
to one another, and there are many explicit examples of so-called
``T-$n$-ality'' \cite{DelZotto:2020sop, DelZotto:2022ohj, DelZotto:2022xrh,
		Ahmed:2023lhj, Mansi:2023faa, Lawrie:2023uiu, Baume:2024oqn,
Ahmed:2024wve}.

A list of quantities that are invariant under T-duality has been put forward,
greatly simplifying the search for dual pairs \cite{DelZotto:2022ohj,
DelZotto:2022xrh, Ahmed:2023lhj, Lawrie:2023uiu}:
\begin{equation}\label{t-dual-invariants}
		\kappa_P\,,\quad \kappa_R\,,\quad \text{dim}(\text{CB})\,,\quad \text{rk}(\mathfrak{f})\,,
\end{equation} 
with $\text{dim}(\text{CB})$ the dimension of the resulting five-dimensional
theory, given by the sum of dynamical tensor multiplets and the rank of the
total gauge symmetry of the six-dimensional theory, and
$\text{rk}(\mathfrak{f})$ the rank of the flavor symmetry. Notably, the
two-group constants above are part of the T-duality invariants and are
therefore very desirable to understand.\footnote{The third two-group constant
		$\kappa_F$ is also expected to match across duality after embedding the
flavor symmetry in an appropriate larger symmetry \cite{Ahmed:2023lhj,
Lawrie:2023uiu}.} It has furthermore been observed that the two-group
structure constant associated with the Poincar\'e symmetry only takes two
values: $\kappa_P=0,2$. This has been identified in the M-theory picture as
counting the number of M9-branes \cite{DelZotto:2020sop}. In string
constructions, $\kappa_P=2$ then labels those LSTs that can be engineered from
$\mathfrak{so}_{32}$ or $\mathfrak{e}_8\oplus\mathfrak{e}_8$ heterotic strings,
while theories with $\kappa_P=0$ are those with an origin in Type-IIB string
theory.\footnote{Possible spectra of five-dimensional theories which could
potentially descend from the twisted compactification of LSTs resulting
in $\kappa_P=1$ have been identified in reference \cite{Ahmed:2024wve},
but whether they admit a geometric realization is as yet unknown.} The
list given in Equation \eqref{t-dual-invariants} is by no means
exhaustive, and it is not known what the complete list of invariants is.
For instance, the defect group and one-form symmetries lead to one-form
symmetries in five dimensions, and the former is exchanged under
T-duality \cite{Baume:2024oqn}. It therefore remains an open question
whether the matching of the invariants given in Equation
\eqref{t-dual-invariants} is enough to differentiate T-dual theories,
or form only a necessary condition. However, we are not aware of any
case where two theories share the same invariants, but can be shown to
not be T-dual.

In this work, we focus on charting the network of Little String Theories with
an $\mathfrak{so}_{32}$ heterotic origin. At a generic point of their tensor
branch---the space spanned by expectation values of the scalar fields that are
part of tensor multiplets---these theories admit a very simple field-theoretic
realization: their matter spectrum, in addition to tensor and vector
multiplets, involves only hypermultiplets transforming in the bifundamental
representation of classical algebras $\mathfrak{su}_k\,,\mathfrak{so}_k\,,
\mathfrak{sp}_k$. It was moreover shown by Blum and Intriligator
\cite{Blum:1997mm, Blum:1997fw, Intriligator:1997dh} that they follow an ADE
classification, and that given a choice of a simple algebra,
\begin{equation}
		\mathfrak{g}\in\{\mathfrak{su}_n\,,\mathfrak{so}_{2n}\,,\mathfrak{e}_6\,,\mathfrak{e}_7\,,\mathfrak{e}_8\}\,,
\end{equation}
all possible consistent $Spin(32)/\mathbb{Z}_2$ LSTs can be labeled by an
integer $Q$ and a choice of embedding of the finite group $\Gamma\subset SU(2)$
McKay dual to $\mathfrak{g}$ into $Spin(32)/\mathbb{Z}_2$. We will therefore
denote these LSTs as:
\begin{equation}
		\widetilde{\mathcal{K}}_Q(\mu^\vee;\mathfrak{g})\,,\qquad \mu^\vee\in \text{Hom}(\Gamma, Spin(32)/\mathbb{Z}_2)\,.
\end{equation}
In the heterotic string constructions, $Q$ is related to the number of
instantons probing a $\mathbb{C}^2/\Gamma$ singularity, and $\mu^\vee$ corresponds to a
choice of holonomy at infinity that breaks the original $\mathfrak{so}_{32}$
flavor symmetry to one of its subalgebras.

Our goal is to use this classification to find structures in the network of
T-duality that can occur in heterotic LSTs. To do so, we translate the
embedding of the singularity into the more modern language of dominant
coweights. The embedding $\mu^\vee$ is then interpreted as a particular
dominant coweight of $\mathfrak{g}$. We will therefore exploit the versatility
of Lie theory to easily separate putative T-dual theories into ``orbits''
leading to the same duality invariants. It turns out that this procedure only depends
on the coweight labeling the LST and does not rely explicitly on its geometric
engineering. While the invariants only form a set of necessary rather than
sufficient conditions, it converts an \emph{a priori} complicated geometric
question into a simpler linear-optimization problem easier to solve.

We will in particular show that theories with the same T-duality invariants
partition the set of dominant coweights of $\mathfrak{g}$ into convex integer
cones, and will give a simple prescription to find their generators.
This comes with certain advantages, as some properties of the T-duality
invariants follow from properties of finite and affine root systems. For
instance, $\kappa_R$ has been conjectured to follow a monotonicity
``$\kappa$-theorem'' \cite{DelZotto:2022ohj}, and decreases monotonically along
Renormalization Group (RG) flows. In our setting, this property follows directly from
the partial order of dominant coweights.

In the context of 3d $\mathcal{N}=4$ quiver gauge theories, the language of
coweights and related techniques using the technology of magnetic quivers have
proven to be a powerful tool to explore the Higgs branch of six-dimensional
theories \cite{Mekareeya:2017jgc, Fazzi:2022yca, Fazzi:2022hal,
DelZotto:2023myd, Fazzi:2023ulb, Lawrie:2023uiu, Bourget:2024mgn,
Giacomelli:2024ycb, Giacomelli:2025zqn, Mansi:2023faa, Lawrie:2024zon,
Cabrera:2019dob, Cabrera:2019izd, Bao:2024wls, Bao:2025pxe,
Bennett:2025edk, Bennett:2024llh}. These techniques have recently led to an
algorithm called ``slice subtraction'' \cite{Lawrie:2024zon}, used to study a
large class of 6d SCFTs, and giving a remarkably simple and elegant way to
recursively obtain all possible SCFTs that can be reached through an RG flow,
and which operator participates in the Higgs mechanism. For these SCFTs, the
algorithm uses the partial order of dominant coweights. As LSTs are similarly
labeled, we propose a generalization of slice subtraction that applies to LSTs,
working also for all types of $\mathfrak{so}_{32}$ LSTs.

This work is organized as follows. In Section~\ref{sec:review} we give a review
of $Spin(32)/\mathbb{Z}_2$ LSTs. Although inspired by the string theory
construction, we mostly take a field-theoretic point of view of certain known
results. We in particular compute the full anomaly polynomial of theories with
a perturbative description in Section~\ref{sec:anomalies}, generalizing certain
known results to the complete set of continuous flavor symmetries. In
Section~\ref{sec:classification-spin32-lst} we review the classification of
$Spin(32)/\mathbb{Z}_2$ LSTs, and describe it in the language of coweights.

In Section~\ref{sec:t-duality} we show how coweights can be used to find
putative T-dual orbits easily. In Section~\ref{sec:higgs} we explore the
partial order of coweights and its relation to Higgs-branch RG flows, leading
us to a generalized slice-subtraction algorithm. In Section~\ref{sec:extension}
we explore possible extensions of our results to more general classes of SCFTs
and LSTs. We give our conclusions in Section~\ref{sec:conclusion}, where we summarize our results.

In addition, we collate all heterotic LSTs corresponding to a trivial embedding
in various tables in Appendix \ref{app:unbroken-LST}, give additional details of the
derivation of the anomaly polynomial in Appendix \ref{app:anomaly-polynomial},
and a short review of finite and affine root systems in Appendix
\ref{app:roots}, setting our conventions.

\section{\texorpdfstring{$Spin(32)/\mathbb{Z}_2$}{Spin(32)/Z2} Heterotic Little String Theories}\label{sec:review}

The simplest realization of Little String Theories corresponds to a small
instanton in $\mathfrak{so}_{32}$ heterotic string theory that shrinks to zero
size, first shown by Witten to be associated with an $\mathfrak{su}_2$ gauge
theory describing the worldvolume theory of an NS5-brane
\cite{Witten:1995gx}. A stack of $Q$ such branes then leads to an enhanced
$\mathfrak{sp}_Q$ gauge theory. To cancel gauge anomalies, 32 half
hypermultiplets must be present and are rotated by an $\mathfrak{so}_{32}$
symmetry, thereby giving rise to the familiar $Spin(32)/\mathbb{Z}_2$ group of the
heterotic string. This construction admits several dual descriptions, most
notably Type-I string theory---which offers a perturbative regime---but also
through Type-IIA and IIB (orientifold) vacua.

A further generalization is obtained by considering orbifold singularities of
the form $\mathbb{C}^2/\Gamma$, with $\Gamma$ a finite subgroup of $SU(2)$, in
the directions transverse to the branes. This gives rise to new physical
degrees of freedom corresponding to additional hyper-, tensor, and vector
multiplets \cite{Douglas:1996sw,Intriligator:1997kq}. The $\mathfrak{so}_{32}$
flavor symmetry can then be broken to obtain new LSTs. In the heterotic
picture, this is done by a choice of a flat connection at infinity, encoded in
the fundamental group $\pi_1(S^3/\Gamma)\simeq \Gamma$, corresponding to a choice
of embedding of the finite group $\Gamma$ into $Spin(32)/\mathbb{Z}_2$.

Although \emph{a priori} not perturbative, there exists an elegant Type-IIB
description of heterotic LSTs through F-theory, where many of the relevant
field-theoretic features can be obtained from the data of an
elliptically-fibered Calabi--Yau threefold:
\begin{align}
    \begin{array}{ccc}
        T^2  & \hookrightarrow & X_3 \\
        &&\downarrow \\
         & & B_2
    \end{array}
\end{align}
The base $B_2$ must be non-compact in order to decouple gravity. Furthermore,
to obtain a Little String Theory in six dimensions, we must engineer a BPS
string that can never be tensionless at any point of the tensor-branch moduli
space: the little string itself. In F-theory, BPS strings arise from D3-branes
wrapping two-real-dimensional curves in the base, their volume setting the
tension of the strings. The LST criterion is then that $B_2$ must have a curve
that cannot be shrunk to zero volume anywhere in the K\"ahler moduli space of
$B_2$, and in addition that there is a point of that moduli space where all
other curves go to zero volume simultaneously. This forces the base to be in
one of two classes, with the heterotic choice required to be birational to
$\mathbb{P}^1 \times \mathbb{C}^1$.\footnote{On the other hand, for Type-II
LSTs the base $B_2$ is required to be birational to $T^2 \times \mathbb{C}$
\cite{Bhardwaj:2015oru, Baume:2024oqn}.} Geometrically, this means that the
base $B_2$ admits a compact genus-zero curve $\Sigma_{\text{LST}}$ of
self-intersection $\Sigma_{\text{LST}}^2 = 0$, further implying that $X_3$
comes with a nested elliptic K3 fibration over $\mathbb{C}^1$, as expected for
theories with a heterotic dual. We will not delve into a detailed explanation
of the geometric construction here, and defer to the review
\cite{DelZotto:2023ahf} and references therein.

Rather, the key point is that the tensor-branch geometry of the theory is
encoded in the curve structure of the base $B_2$. Let us choose a basis of
curves $\Sigma^I \in H^{2}(B_2)$ with $I=0, \ldots, h^{1,1}(B_2)$ of genus zero.
By Grauert's criterion, to be able to send the volume of all other curves than
$\Sigma_\text{LST}$ to zero simultaneously, they must have negative
self-intersection. Consistency of the elliptic fibration then requires that the
self-intersection of the curves $\Sigma^I\cdot\Sigma^I=-n_I$ lie within the
range $ 0 \leq n_I \leq 12$, and have a negative semi-definite intersection
form:\footnote{We use a convention where $\eta^{IJ}$ is a \emph{positive}
		semi-definite matrix, while the intersection pairing itself is
\emph{negative} semi-definite. We do so because $\eta$ is then interpreted as
the Dirac pairing of the associated BPS strings.}
\begin{align}\label{def-intersection}
    \Sigma^I \cdot \Sigma^J = -\eta^{IJ} \preceq 0\,.
\end{align}
Without loss of generality, we do not consider product theories in this work,
so that $\eta^{IJ}$ is irreducible, and has a unique null vector up to
rescaling. In the absence of gravity, product theories are decoupled and can
therefore be treated independently. As there is one more curve than the
independent generators of $H^2(B_2)$, there is one linear relation among the
curves:
\begin{align}\label{LST-charge-def}
		\Sigma_\text{LST} = \ell_I \Sigma^I \,, \qquad \eta^{IJ}\,\ell_J = 0\,,
\end{align}
with $\ell_I\in\mathbb{N}$ and $\text{gcd}(\ell_I)=1$. It is a vector in the
string charge lattice, interpreted as the charges of the little string under the
Dirac pairing.

The base therefore defines the spectrum of tensor multiplets, the associated
BPS strings, and their intersection pattern. On the other hand, the gauge
sector is encoded in possibly-singular elliptic fibers over each curve. These
singularities follow the Kodaira--Tate classification, and can be associated
with an algebra that defines the gauge symmetry on that curve. Furthermore,
global aspects of gauge and flavor groups are encoded in the finite part of the
Mordell--Weil group of $X_3$ \cite{Aspinwall:1998xj, Dierigl:2020myk,
Hubner:2022kxr, Apruzzi:2020zot}. Indeed, the dual F-theory model of the
$Spin(32)/\mathbb{Z}_2$ heterotic string admits an order-two Mordell--Weil
group. The matter spectrum is then associated with enhanced singularities at
the intersection of two curves, and can also be easily obtained from the
classification, see e.g. reference \cite{Katz:2011qp}.

Geometric engineering of LSTs through F-theory is therefore very practical, and
reduces to considering a collection of curves whose elliptic fibers are singular,
and from which the spectrum can easily be obtained. Furthermore, as we have
reviewed above, $\mathfrak{so}_{32}$ heterotic LSTs are very attractive due to
their perturbative nature, and it follows that their F-theory realization is
particularly simple: the base only involves curves of self-intersection
$n_I\in\{0,1,2,4\}$ that have at most a simple intersection $\eta_{IJ}=-1$, and
the algebras associated with the singular fibers are only of classical types
$\mathfrak{sp}_k\,,\mathfrak{so}_{k}\,,\mathfrak{su}_k$. This is in stark
contrast with Type-II LSTs, where more complicated intersection patterns occur,
or those of $\mathfrak{e}_8\oplus\mathfrak{e}_8$-heterotic type where any
simple algebra can appear, leading to the presence of non-perturbative states.

As the transverse singularities satisfy an ADE classification, given a choice
of algebra $\mathfrak{g}=\mathfrak{su}_k\,,
\mathfrak{so}_{2k}\,,\mathfrak{e}_k$, the base is related to the topology of
its affine Dynkin diagram. There is however a caveat: for
reasons that will become clear when we review the classification of the allowed
flavor and gauge symmetries, when the algebra $\fg$ associated with the
transverse singularity of the LST admits complex representations, the topology
of the base is associated with the \emph{folded} affine version of
$\mathfrak{g}$, which we denote by $\widetilde{\mathfrak{b}}$. For instance,
with $\mathfrak{g} = \mathfrak{e}_6$ the base is
$\widetilde{\mathfrak{b}}=F^{(1)}_4$. The correspondence is given in Table
\ref{tab:folding}. Note that we have not included the case
$\mathfrak{g}=\mathfrak{su}_{2k+1}$ by design, as it has additional subtleties
which will be treated separately in Section~\ref{sec:antisymmetric}.

\begin{table}
		\centering
		\begin{tabular}{ccccccc}
				\toprule
				$\mathfrak{g}$ & $\mathfrak{su}_{2k}$ & $\mathfrak{so}_{4k}$ & $\mathfrak{so}_{4k+2}$ & $\mathfrak{e}_{6}$ & $\mathfrak{e}_{7}$ & $\mathfrak{e}_{8}$\\\midrule
				$\widetilde{\mathfrak{g}}$ & $A^{(1)}_{2k-1}$ & $D^{(1)}_{2k}$ & $D^{(1)}_{2k+1}$ & $E^{(1)}_{6}$ & $E^{(1)}_{7}$ & $E^{(1)}_{8}$\\
				$\widetilde{\mathfrak{b}}$ & $\bm{C^{(1)}_k}$ & $D^{(1)}_{2k}$ & $\bm{B^{(1)}_{2k}}$ & $\bm{F^{(1)}_{4}}$ & $E^{(1)}_{7}$ & $E^{(1)}_{8}$  \\
				\bottomrule
		\end{tabular}
\caption{Algebras arising in $Spin(32)/\mathbb{Z}_2$ heterotic LSTs and the
		associated bases $\widetilde{\mathfrak{b}}$, possibly folded from the
		affine algebra $\widetilde{\mathfrak{g}}$. The bases that are different
		from $\widetilde{\mathfrak{g}}$ are in bold. The special case of
$\mathfrak{g}=\mathfrak{su}_{2k+1}$ is treated separately in
Section~\ref{sec:antisymmetric}.}
		\label{tab:folding}
\end{table}

\paragraph{Notation and conventions:} we will often have to jump
between quantities related to a simple Lie algebra $\mathfrak{g}$, its affine
version $\widetilde{\mathfrak{g}}$, as well as the base algebra
$\widetilde{\mathfrak{b}}$ and its non-affine counterpart $\mathfrak{b}$. 

For a simple algebra $\mathfrak{g}$ we use the usual fraktur font, e.g.
$\mathfrak{e}_6$, while we use Kac's notation \cite{Kac:1990gs} $X_r^{(n)}$ for
its affine version, e.g. $E^{(1)}_6$. This makes the distinction easier, and
will also enable us to treat the case $\mathfrak{g}=\mathfrak{su}_{2k+1}$, as
well as other types of $\mathfrak{so}_{32}$ LSTs whose bases correspond to
twisted affine algebras.

Given an affine algebra $\widetilde{\mathfrak{b}}$ representing the base, we
label the nodes of its Dynkin diagram---or equivalently the curves in the
base---with uppercase letters from the middle of the Latin alphabet $I, J,
\dots\in\{0, 1, \dots r_{\mathfrak{b}}\}$. If we need to exclude the affine
node $I=0$, we use lowercase letters $i, j, \dots\in\{1, \dots
r_{\mathfrak{b}}\}$.

For quantities related to arbitrary algebras $\mathfrak{g}$, including the ADE
classification of LSTs---but not their base---we instead use letters from the
beginning of the Latin alphabet, with the same distinction between affine and
simple nodes: $A, B, \dots\in\{0, 1, \dots r_{\mathfrak{g}}\}$ and
$a,b,\dots\in\{1, \dots r_{\mathfrak{g}}\}$ for 
$\widetilde{\mathfrak{g}}$ and $\mathfrak{g}$, respectively. The affine node of
a Dynkin diagram is always labeled by $I=0$ or $A=0$.

In addition, if an index-free quantity of a finite algebra has a natural
extension to its affine version we denote it with a tilde. For instance the
highest root $\theta = \theta_a\,\alpha^a$ of $\mathfrak{g}$ is naturally
extended to the null root $\widetilde{\theta} = \theta_A\alpha^A$ of
$\widetilde{\mathfrak{g}}$. When indices are present we do not make the
distinction, e.g. $C^{AB}$ is an affine Cartan matrix while $C^{ab}$ is that of
the associated simple algebra.

Finally, we will also extensively use the properties of finite and affine root
systems. A short review of the relevant concepts can be found in Appendix
\ref{app:roots}.

\paragraph{Pictorial representation of LSTs:} we will use the now-standard
notation encoding the geometry of the elliptic fibration in terms of
``generalized quivers'' \cite{Heckman:2015bfa}. Consider a curve $\Sigma\subset
B_2$ with self intersection $\Sigma^2 = -n$, hosting a singular fiber
corresponding to an algebra $\mathfrak{g}$. We also assume that there is a
flavor symmetry $\mathfrak{f}$ rotating the possible matter fields associated
with that curve.  We can pictorially represent this configuration as follows:
\begin{equation}
		\underset{[\mathfrak{f}]}{\overset{\mathfrak{g}}{n}}\,.
\end{equation}
If the fiber is not associated with an algebra, $\mathfrak{g}= \varnothing$, or
there is no flavor symmetry associated with the configuration, we omit them.
If there is more than one curve, they are arranged in a pattern reproducing
the adjacency of the pairing matrix $\eta^{IJ}$: if two neighboring curves
$\Sigma^I$ and $\Sigma^J$ intersect with $\eta^{IJ}=-1$, they are depicted side
by side. 

The possible combinations of $n$ and $\mathfrak{g}$ are furthermore restricted
by anomaly cancellation, or in geometric terms, the consistency of the elliptic
fibration $X_3$. For the heterotic LSTs of type $\mathfrak{so}_{32}$ we will
consider, we only have algebras of classical types, and there are only four
possibilities:\footnote{We use the usual convention $\mathfrak{sp}_1 =
\mathfrak{usp}_2 \simeq \mathfrak{su}_2$ for symplectic algebras. Therefore the
fundamental representation of $\mathfrak{sp}_k$ has dimension $2k$.}
\begin{equation}
	\underset{[\mathfrak{so}_{m}]}{\overset{\mathfrak{sp}_k}{1}}\,,\qquad
	\underset{[\mathfrak{sp}_{m}]}{\overset{\mathfrak{so}_k}{4}}\,,\qquad
	\underset{[\mathfrak{su}_{m}]}{\overset{\mathfrak{su}_k}{2}}\,,\qquad
	\underset{[\mathfrak{su}_{m}]}{\overset{\mathfrak{su}_k}{1}}\,,
\end{equation}
which can be traced back to the fact that those LSTs admit perturbative
Type-I or Type-IIA dual descriptions which only have mutually-local 5-branes
and O-planes. 

In each case, there is a number of hypermultiplets transforming in the
bifundamental representations of $\mathfrak{g}\oplus\mathfrak{f}$, where the
precise rank of $\mathfrak{f}$ is set by anomaly-cancellation conditions, as we
will review below. The last case is special, and hosts an additional
hypermultiplet transforming in the anti-symmetric representation of
$\mathfrak{su}_k$, and arises for LSTs with $\mathfrak{g}=\mathfrak{su}_{2k+1}$.
The presence of this additional representation explains why we will defer
discussing those theories to Section~\ref{sec:antisymmetric}, and the majority of this
work will focus on the first three cases.

Furthermore, if two compact curves $\overset{\mathfrak{g}}{m}$ and
$\overset{\mathfrak{h}}{n}$ intersect, part of the hypermultiplets will
transform in the bifundamental representation of
$\mathfrak{g}\oplus\mathfrak{h}$, and the flavor symmetries on each curve will
be reduced as there are fewer matter fields to rotate. In the F-theory picture, the
precise matter content can be obtained by studying the enhanced singularity at
the intersection, but the remnant flavor symmetries associated with each gauge
algebra can equivalently be obtained from anomaly-cancellation conditions, as
reviewed below.

As an example, let us consider $\mathfrak{so}_{32}$ heterotic LSTs with a base
labeled by $\mathfrak{g}=\mathfrak{e}_6$ and instanton number $Q$. We have
argued above---and will give additional details in the next section---that the
base is associated with $\widetilde{\mathfrak{b}} = F_4^{(1)}$. With generic
flavor symmetries, i.e. a generic embedding $\mu^\vee$, the quiver has the
following shape:
\begin{equation}\label{generic-e6-quiver}
	\widetilde{\mathcal{K}}_Q(\mu^\vee; {\mathfrak{e}_6}):\qquad 
	\underset{[\mathfrak{so}_{m^0}]}{\overset{\mathfrak{sp}_{Q+p_0}}{1}}
	\underset{[\mathfrak{sp}_{m^1}]}{\overset{\mathfrak{so}_{2Q+p_1}}{4}}
	\underset{[\mathfrak{so}_{m^2}]}{\overset{\mathfrak{sp}_{3Q+p_2}}{1}}
	\underset{[\mathfrak{su}_{m^3}]}{\overset{\mathfrak{su}_{4Q+p_3}}{2}}
	\underset{[\mathfrak{su}_{m^4}]}{\overset{\mathfrak{su}_{2Q+p_4}}{2}}\,,
	\qquad\qquad 
	\eta^{IJ} = \left(\begin{smallmatrix}
			1& -1 & 0 & 0 & 0\\
			-1& 4 & -1 & 0 & 0\\
			0& -1& 1 & -1 & 0\\
			0& 0 & -1 & 2 & -1\\
			0& 0& 0 & -1 & 2
	\end{smallmatrix}\right)\,.
\end{equation}
We therefore have four hypermultiplets transforming in the bifundamental
representation of $\mathfrak{g}_I\oplus \mathfrak{g}_{I+1}$, and five
transforming in the bifundamental of $\mathfrak{g}_I\oplus \mathfrak{f}_I$. We
stress that the Cartan matrix of $F_4^{(1)}$ and $\eta^{IJ}$ are different.
While $\eta^{IJ}=0,-1$ off diagonal, this is not the case for the Cartan
matrix, see Equation \eqref{cartan-F4}. The relation between the two, as well
as the relation between the coefficients $m^I$ and $p_I$ and the choice of
embedding $\mu^\vee$, will be clearer when we review anomaly-cancellation
conditions.

We can also have trivalent intersections, for instance in the case of LSTs with
$\mathfrak{g}=\mathfrak{e}_7$.  With a trivial choice of embedding of the
associated finite group $\Gamma$, the LST with an $\mathfrak{so}_{32}$ flavor
symmetry arising from $Q$ NS5-branes probing the corresponding ALE space is
given by:
\begin{equation}
		\widetilde{\mathcal{K}}_Q(\varnothing; {\mathfrak{e}_7}):\qquad \underset{[\mathfrak{so}_{32}]}{\overset{\mathfrak{sp}_{Q}}{1}} \overset{\mathfrak{so}_{4Q - 16}}{4} \overset{\mathfrak{sp}_{3Q - 24}}{1}\overset{\displaystyle\overset{\mathfrak{sp}_{2Q - 20}}{1}}{\overset{\mathfrak{so}_{8Q - 64}}{4}}\overset{\mathfrak{sp}_{3Q - 28}}{1} \overset{\mathfrak{so}_{4Q - 32}}{4} \overset{\mathfrak{sp}_{Q - 12}}{1}\,.
\end{equation}
The other quivers for LSTs with any ADE algebra $\mathfrak{g}$ and a trivial
embedding can be found in Appendix \ref{app:unbroken-LST}, along with their
T-dual invariants.

\subsection{Anomalies of \texorpdfstring{$Spin(32)/\mathbb{Z}_2$}{Spin(32)/Z2} LSTs}\label{sec:anomalies}

Heterotic LSTs of type $\mathfrak{so}_{32}$ have the attractive property that
all algebras involved are of classical type. As mentioned several times above,
their ranks are constrained by requiring the absence of gauge anomalies. A
careful analysis of these constraints was performed by Blum and Intriligator
\cite{Intriligator:1997dh, Blum:1997mm, Blum:1997fw}, see also references
\cite{Douglas:1996sw,Brunner:1997gf}, who achieved a classification of all
consistent heterotic $Spin(32)/\mathbb{Z}_2$ LSTs in terms of embeddings of the
discrete group $\Gamma\subset SU(2)$ into $Spin(32)/\mathbb{Z}_2$, giving a
prescription to enumerate them in each case.

We now analyze the anomaly polynomial of theories with general classical
algebras in detail. Beyond pedagogy, our purpose is twofold: first, this
enables us to review the original classification, and generalize it to the
complete set of continuous (flavor) anomalies, the underlying two-group
structure constants, and highlight the difference with respect to other LSTs
and their SCFT cousins. Moreover, this will make the relationship between
$\mathfrak{so}_{32}$ heterotic LSTs and Lie theory manifest. We will
furthermore be able to find several closed-form formulas for various quantities
of interest, paving the way to reinterpret the classification in the modern
language of coweights in the next section.

To compute the anomaly polynomial of an LST, we can use the fact that scalar
fields falling in tensor multiplets are singlets under all global symmetries.
These symmetries are therefore conserved under tensor-branch deformations. This
enables us to compute the anomaly polynomial at a point of the tensor branch
where the theory is weakly coupled, and by 't Hooft anomaly matching the
obtained expression is then valid at the strongly-coupled point corresponding
to the LST. This procedure is straightforward and algorithmic. Here, we only
briefly recall the salient points, and defer to \cite{Ohmori:2014kda,
Heckman:2018jxk, Baume:2021qho, Baume:2023onr} for detailed explanations in the
case of six-dimensional SQFTs. We follow the same conventions as in references
\cite{Ohmori:2014kda, Baume:2021qho, Baume:2023onr}.

In six dimensions, the anomaly polynomial is made out of two different
contributions:
\begin{equation}\label{I_8-def}
		I_8 = I_8^\text{1-loop} + I_8^\text{GS}\,.
\end{equation}
The former is a ``one-loop'' contribution obtained by summing the contribution
of each multiplet in the weakly-coupled regime, while the latter corresponds to
a Green--Schwarz--West--Sagnotti mechanism \cite{Green:1984bx, Sagnotti:1992qw}---which
we will refer to as a GS term for brevity---necessary to completely
cancel gauge anomalies.

We will compute the anomaly polynomial for a six-dimensional quiver theory with
the following spectrum:
\begin{enumerate}
		\item a collection of $r_\mathfrak{b}$ dynamical tensor multiplets;
		\item $r_\mathfrak{b} + 1$ vector multiplets transforming in the adjoint representation of a
			\emph{classical} gauge algebra
			$\mathfrak{g}_I\in\{\mathfrak{sp}_{k_I}, \mathfrak{so}_{k_I},
			\mathfrak{su}_{k_I}\}$, $I= 0,1,\dots,r_\mathfrak{b}$;
		\item hypermultiplets transforming in the bifundamental $(\bm{d}_I,
			\overline{\bm{d}}_J)$ of $\mathfrak{g}_I\oplus\mathfrak{g}_J$.
			Their symmetric adjacency matrix is given by $\frac{1}{2}A^{IJ}$,
			with $A^{II}=0$, and differentiates between half- and full
			hypermultiplets (see below);
		\item hypermultiplets transforming in the bifundamental
			$(\bm{d}_I,\overline{\bm{f}}^J)$ of
			$\mathfrak{g}_I\oplus\mathfrak{f}_J$ with a \emph{diagonal} adjacency matrix
			$\frac{1}{2}D^{IJ}$, again taking into account the difference
			between half- and full hypermultiplets.
\end{enumerate}

The algebras $\mathfrak{f}_I$ correspond to flavor symmetries of the theory.
Moreover, $r_{\mathfrak{b}}$ will correspond to the rank of the folded base
$\widetilde{\mathfrak{b}}$, see Table \ref{tab:folding}. We have also
anticipated that the tensor corresponding to the LST can be thought of as being
a non-dynamical background field \cite{Bhardwaj:2015oru}. If one considers an
SCFT instead, there is an additional dynamical multiplet; however this will not
influence the constraints between the gauge and flavor symmetries. 

We stress that we must differentiate between full and half hypermultiplets.
Generically, a hypermultiplet transforms in the representation
$\mathcal{R}\oplus\overline{\mathcal{R}}$ of the associated symmetry. However,
if the representation is (pseudo)-real, $\mathcal{R}=\overline{\mathcal{R}}$,
we are overcounting and only have a half hypermultiplet. In practice, it means
that $\frac{1}{2}D^{II} = 1$ (respectively $\frac{1}{2}A^{IJ} = 1$) if
${\mathfrak{g}_I\oplus\mathfrak{f}_I}$ (respectively
${\mathfrak{g}_I\oplus\mathfrak{g}_J}$) contains an $\mathfrak{su}_k$ factor,
or $\frac{1}{2}$ otherwise.

The matter interactions taking this property into account are then encoded in
the matrix:
\begin{equation}\label{def-symm}
		G^{IJ} = 2D^{IJ} - A^{IJ}\,,\qquad G^{IJ} = D^I_K\, C^{KJ}\,.
\end{equation}
Here and throughout this work, indices are raised and lowered with the flat metric $\delta^I_J$. Furthermore, for the cases we consider, $G^{IJ}$ must be a symmetric matrix with
$G^{II}\in\{2,4\}$ and have off-diagonal entries $-2,-1,0$. This is precisely
equivalent to the classification of the \emph{symmetrized} Cartan matrices of
both finite and affine Lie algebras \cite{Kac:1990gs} excluding those with
$-3$ off diagonal, such as $\mathfrak{g}_2$. The matrix $C$ defined above is
the associated Cartan matrix.

Indeed, recall that given the (generalized) Cartan matrix $C$ of a Kac--Moody
algebra, there exists a positive-integer-valued diagonal matrix $D$ such that $G
= DC$ is symmetric with integer values. The determinant of $G$ differentiates
between an SCFT and an LST, as for the latter it implies that there is a combination
of the tensor multiplets that is not dynamical.\\   
\begin{classification}[base topology of $6d$ SQFTs]\label{class:base-topology}
	If the irreducible quiver of a
	six-dimensional SQFT with a finite number of nodes involves only hypermultiplets in the fundamental
	representation of classical gauge symmetries, the hypermultiplet
	adjacency matrix $G$ must be the symmetrized Cartan matrix of:
	\begin{itemize}
		\item a finite algebra $\mathfrak{b}$ for those describing the tensor branch of an SCFT;
		\item an affine (possibly-twisted) algebra $\widetilde{\mathfrak{b}}$ for those describing the tensor branch of an LST.
	\end{itemize}
	Cartan matrices with off-diagonal elements $C^{IJ} < -2$ are not realized
	as $6d$ SQFTs with only bifundamental matter. The precise type of algebras
	$\mathfrak{b}$, $\widetilde{\mathfrak{b}}$ is then read off the diagonal
	matrix $D$. If it includes both half and full hypermultiplets, it is not
	simply-laced.
\end{classification}

Although found through combinatorial arguments, this is of course in agreement
with the known classifications \cite{Heckman:2013pva, Heckman:2015bfa,
Bhardwaj:2015oru, Bhardwaj:2015xxa, Bhardwaj:2019hhd} when restricted to
classical gauge algebras $\mathfrak{g}_I$ and $\mathfrak{f}_I$. Note that not
all finite and affine algebras are realized. Since the base must be
supplemented with a choice of algebras associated with the fiber data---or
equivalently gauge algebras---it is not always possible to find a consistent
choice. Furthermore, we cannot have bases for which $\text{det}(G)<0$ since
they lead to negative definite Dirac pairing. The only exceptions are
supergravity theories which have $\text{det}(\eta)=-1$, see e.g. reference
\cite{Bhardwaj:2015xxa}.

For the LST of type $\mathfrak{g}=\mathfrak{e}_6$
discussed around Equation \eqref{generic-e6-quiver}, we have a base
$\widetilde{\mathfrak{b}}= F^{(1)}_4$ and it is straightforward to see that we indeed have 
\begin{equation}\label{cartan-F4}
		G^{IJ} = \left(\begin{smallmatrix}
				 2& -1&  0&  0&  0\\
				-1&  2& -1&  0&  0\\
				 0& -1&  2& -2&  0\\
				 0&  0& -2&  4& -2\\
				 0&  0&  0& -2&  4
 \end{smallmatrix}\right)\,,\qquad
		C^{IJ}  = \left(\begin{smallmatrix}
				 2& -1&  0&  0&  0\\
				-1&  2& -1&  0&  0\\
				 0& -1&  2& -2&  0\\
				 0&  0& -1&  2& -1\\
				 0&  0&  0& -1&  2
	 \end{smallmatrix}\right)\,,\qquad
	 D^{IJ} = \left(\begin{smallmatrix}
				 1& 0&  0&  0&  0\\
				 0&  1& 0&  0&  0\\
				 0& 0&  1& 0&  0\\
				 0&  0& 0&  2& 0\\
				 0&  0&  0& 0&  2
	 \end{smallmatrix}\right)\,.
\end{equation}
This explains why the algebra $\widetilde{\mathfrak{b}}=F_4^{(1)}$ arises
naturally in this picture, rather than e.g. $\widetilde{\mathfrak{b}}=A^{(1)}_4$ or any
other linear Dynkin diagram. A complete list of the bases that can occur can be
found in Table \ref{tab:dynkin-bases} of Section~\ref{sec:conclusion}.

Furthermore, in the definition of the spectrum we have denoted the dimension of
the fundamental representation of a gauge algebra $\mathfrak{g}_I$ by $d_I$,
while that of the flavor symmetry $\mathfrak{f}_I$ is denoted by $f^I$. For the
rest of this work, $d_I$ and $f^I$ will always be associated with the
fundamental representation, and all closed-form expressions we will derive will
depend on them explicitly. However, for the sake of clarity, it will sometimes
be more convenient to use $\mathfrak{sp}_{k_I}$ rather than
$\mathfrak{sp}_{d_I/2}$. We can conveniently pass from one to the other through
\begin{equation}
		d_I = V_I^J k_J\,,\qquad V_I^J=
				\begin{cases}
						V_I^I = 1\,,\quad \text{if}\quad \mathfrak{g}_I\in\{\mathfrak{su}_k, \mathfrak{so}_k\}\,;\\
						V_I^I = 2\,,\quad \text{if}\quad \mathfrak{g}_I = \mathfrak{sp}_{k_I}\,;\\
						V_I^J = 0\,, \quad \text{if}\quad I \neq J\,.
				\end{cases}
\end{equation}
Of course, using either $d_I$ or $k_I$ is equivalent, but the dimension of the
fundamental representation is the natural object to consider in the language of
dominant coweights, as we will see below. 

The matrix $V$ is also useful to relate the dual Coxeter numbers
$h^I=h^\vee_{\mathfrak{g}_I}$ and the quartic Casimir $x_I$ of the gauge
algebra $\mathfrak{g}_I$ in terms of the dimension of the fundamental
representation of the associated flavor symmetry:
\begin{equation}\label{classical-relations}
		d_I = V^{I}_Jh^J - 2 S^I\,,\qquad x^I = D^{IJ} d_J +8 S^I\,,\qquad
	S^I=
	\begin{cases}
			-1\,,\quad \text{if}\quad \mathfrak{f}_I= \mathfrak{sp}_k\,;\\
			\phantom{-}1\,,\quad \text{if}\quad \mathfrak{f}_I= \mathfrak{so}_k\,;\\
		\phantom{-}0\,, \quad \text{if}\quad \mathfrak{f}_I= \mathfrak{su}_k\,.
	\end{cases}
\end{equation}
These are easily shown by inspection. For $Spin(32)/\mathbb{Z}_2$ LSTs, the
quantity $S^I$ can be interpreted as the Frobenius--Schur indicator of the
associated finite representation of $\Gamma$, as we will review in detail in
Section~\ref{sec:classification-spin32-lst}. These relations allow us to
greatly simplify the computation of the anomaly polynomial, and will lead to
simple closed-form expressions for many quantities relevant to this work. We
will furthermore assume that $d_I$ is always large enough so that the gauge
algebras have a non-vanishing quartic Casimir.

\paragraph{``One-loop'' contribution:} with the above spectrum, we can easily
compute the first term in Equation \eqref{I_8-def}, obtained by summing the
contribution of each supermultiplet:\footnote{For brevity, we will ignore
Abelian factors, and only reintroduce them in relevant examples.}
\begin{equation}\label{I8-ABJ}
		I_8^\text{1-loop} = r_\mathfrak{b} I_8^\text{tensor}
		+ \sum_I I_8^\text{vec}(\mathfrak{g}_I) 
		+ \frac{1}{4}A^{IJ}I_8^\text{hyp}(\mathfrak{g}_I,\mathfrak{g}_J)
		+ \frac{1}{2}D^{IJ}I_8^\text{hyp}(\mathfrak{g}_I,\mathfrak{f}_J)\,.
\end{equation}
where the extra factor $\frac{1}{2}$ for gauge hypermultiplets takes into
account double counting, as $A^{IJ}$ is symmetric. The contribution of a given
supermultiplet is universal, and can be written in terms of characteristic
classes associated with the 0-form symmetries, such as the Chern character
$\text{ch}_{\mathcal{R}}$ in a representation $\mathcal{R}$ for the background
flavor, gauge, and R-symmetry, denoted $F_I$, $\widetilde{F}_I$, and $R$,
respectively \cite{AlvarezGaume:1983ig}. Terms encoding gravitational
background terms depend on the A-roof genus $\widehat{A}(T)$, and the
Hirzebruch genus $L(T)$ of the spacetime tangent bundle $T$. As the spectra
solely involve classical algebras and their adjoint and fundamental
representations, we can use various group-theoretical relations to simplify
certain expressions, see e.g. Equation \eqref{classical-relations}. The steps
necessary to obtain the one-loop contribution are given in Appendix
\ref{app:anomaly-polynomial} along with further details on characteristic
classes.

We find that the one-loop contribution to the anomaly polynomial can be
rewritten in a relatively compact form:\footnote{As noted above, in the case of
an SCFT with classical algebras, this result holds with the exception
that the number of tensors is different. Furthermore, there is no
relation between $\mathfrak{g}$ and $\mathfrak{b}$ in that case, and
the first two terms have slightly different coefficients.}
\begin{equation}\label{1-loop-so32}
	\begin{aligned}
			I_8^\text{1-loop} &= \left.\frac{r_\mathfrak{g}}{2}\widehat{A}(T)\text{ch}_{\bm{2}}(R) - \frac{r_\mathfrak{b}}{8}L(T) + \frac{d_I}{4}\big(2D^{IJ}\text{ch}_{\bm{F}}(F_J) +A^{IJ}d_J - (S^I+D^{IJ}d_J)\text{ch}_{\bm{2}}(R)\big)\widehat{A}(T)\right|_\text{8}\\
					& - \frac{1}{2}c_2(\widetilde{F}_I)(\eta^{IJ} c_2(\widetilde{F}_J) - 2J^I) 
				 + \frac{1}{48}\left(f^I - C^{IJ}d_J + 16S^I\right)\left( D_I^K\text{Tr}\widetilde{F}^4_K + V^K_I c_2(F_K)p_1(T) \right)
	\end{aligned}
\end{equation}
The first line of Equation \eqref{1-loop-so32} is given in terms of the
characteristic classes and the dimension of the fundamental representation
$d_I$ of each gauge symmetry, and is understood as taking the eight-form
component of the resulting expression. The second line depends on the second
Chern class of the background flavor symmetry, denoted $c_2(F_I)$, while
$p_2(T)$ denotes the second Pontryagin class of the spacetime background.
Furthermore, we have written the first terms of the second line with the Dirac
pairing $\eta^{IJ}$, see Equation \eqref{def-intersection}, using the
group-theoretic relations of classical groups. We have also defined the
following quantity, which will be related to the GS term:
\begin{equation}
		J^I = \delta^{Ia}c_2(F_a) + \frac{1}{4}a^Ip_1(T) - h^I c_2(R)\,,\qquad a^I = 2 - \eta^{II}\,,\qquad h^I = h^\vee_{\mathfrak{g}_I}\,.
\end{equation}
As is, the one-loop contribution to the anomaly polynomial contains gauge
anomalies, which can be separated into two classes. The first corresponds to
quadratic gauge anomalies proportional to $c_2(\widetilde{F}_I) =
\frac{1}{4}\text{Tr}(\widetilde{F}^2_I)$. Due to the presence of dynamical
tensor multiplets and their dual BPS strings, such anomalies can be canceled
through a GS term \cite{Green:1984bx, Sagnotti:1992qw}. The others are quartic
terms of the form $\text{Tr}(\widetilde{F}^4_I)$ which cannot be canceled
through such a mechanism, and therefore impose constraints on the matter
spectrum.

\paragraph{Cancellation of quartic gauge anomalies:} as the anomaly polynomial
cannot contain terms proportional to $\text{Tr}\widetilde{F}^4_I$, the last term in
Equation \eqref{1-loop-so32} must vanish identically, constraining the allowed
values of the dimensions of the fundamental representation of the flavor
symmetry $f^I$:
\begin{equation}\label{quartic-cancellation}
		C^{IJ}d_J = f^I - 16S^I\,.
\end{equation}
The matrix $C^{IJ}$ has been defined in Equation \eqref{def-symm} and---through
the argument given there---must correspond to the Cartan matrix of a simple
(for SCFTs) or affine (for LSTs) Lie algebra. The vector $S^I$ is defined in
Equation \eqref{classical-relations}, and will also be interpreted in terms
of group theory in Section~\ref{sec:classification-spin32-lst}.

Equation \eqref{quartic-cancellation} is well known, and reproduces the correct
curve self-intersection in the geometric approach. In fact, it is one of the
central results of the classification of $Spin(32)/\mathbb{Z}_2$ heterotic
Little String Theories \cite{Blum:1997mm}, and most of the results we will
obtain can have their origin traced back to this constraint. Note that in the
language of quiver gauge theory, Equation \eqref{quartic-cancellation} is
equivalent to demanding that the quiver is ``balanced''. In six dimensions,
this is therefore a necessary condition to ensure cancellation of gauge
anomalies.

For LSTs, the Cartan matrix is of affine type and cannot be inverted. However,
following reference \cite{Blum:1997mm}, we can express $d_I$ in terms of a
one-parameter family of solutions:
\begin{equation}\label{sol-dI}
		d_I= 2Q\,K_I + X_{IJ}(f^J - 16S^J)\,.
\end{equation}
The integer $Q$ is then interpreted as the instanton number in the heterotic
construction. Furthermore, the matrix $X_{IJ}$ is expressed in terms of the
inverse Cartan matrix of the associated \emph{finite} Lie algebra:
\begin{equation}
		X_{0I}=0=X_{I0}\,, \qquad X_{ij} = C_{ij}\,,\qquad C_{ij}\, C^{jk} = \delta_i^k\,,
\end{equation}
while the coefficients $K_I$ correspond to the comarks of
$\widetilde{\mathfrak{b}}$. For future reference, we recall that they can be
obtained from the Cartan matrix, along with the marks $\theta_I$:\footnote{The
		comarks are often (confusingly) given different names, such as Kac
		labels or Dynkin multiplicities. We follow Kac's nomenclature
		\cite{Kac:1990gs} and use the terms marks and comarks to dispel any
		doubts, as both will arise in computations; we denote them by
		$\theta_I$ and $K_I$, respectively.
}
\begin{equation}\label{comarks-definition}
		C^{IJ}K_J = 0 = \theta_I C^{IJ}\,,\qquad \theta_I = D^J_I K_J\,. 
\end{equation}
They are normalized such that the greatest common divisor of their elements is
one. As a cross check, we can see from Table \ref{tab:het-so32-unbroken} that
on the affine node, $Spin(32)/\mathbb{Z}_2$ LSTs have an $\mathfrak{sp}_Q$ gauge
algebra, with a fundamental representation of dimension $d_0=2Q$. 

Depending on the base algebra, this equation can \emph{a priori}
lead to non-integer fundamental representations $d_I$ or odd-valued dimensions
for $\mathfrak{sp}_k$ algebras, in which case these solutions to the quartic
constraint must be discarded.

We also note that for LSTs, acting with the marks $\theta_A$ on the left in
Equation \eqref{sol-dI}, we obtain a weaker constraint, albeit giving a
relation between the flavor symmetries:
\begin{equation}\label{level}
		\theta_I f^I = 16\, \theta_I S^I\,,
\end{equation}
We can therefore always write the flavor symmetry on the affine node in
terms of those associated with nodes of the simple algebra.

\paragraph{Green--Schwarz term and two-group invariants:} while cancellation of
quartic anomalies imposes strong constraints on the spectrum, the remaining
gauge anomalies do not further constrain the quiver. They can be canceled
through a Green--Schwarz--West--Sagnotti mechanism \cite{Green:1984bx,
Sagnotti:1992qw}, which generically takes the form:
\begin{equation}
		I_\text{GS} = \frac{1}{2}I^i\eta_{ij}I^j\,,\qquad 
		I^i = \eta^{iI}c_2(F_I) - J^i\,,
\end{equation}
with $\eta_{ij}\,\eta^{jk} = \delta_i^k$. If we consider LSTs, we again run
into a zero eigenvalue making the Dirac pairing $\eta^{IJ}$ non-invertible.
In physical terms, it means that one of the tensor multiplets is non-dynamical
and therefore cannot participate in a GS mechanism. Instead, we can omit the
tensor associated with the affine node $I=0$. We denote the remaining ones with Latin
indices $i,j = 1, \dots, \text{rk}(\mathfrak{b})$ running over nodes of the
finite algebra $\mathfrak{b}$, see the paragraph on conventions above. The
second line of Equation \eqref{1-loop-so32} can then be written as:
\begin{equation}\label{GS-cancellation}
		- \frac{1}{2}c_2(F_I)(\eta^{IJ}c_2(F_J) - 2J^I) = -I_\text{GS} + \frac{1}{2}J^i\eta_{ij}J^j - c_2(F_0)(J^0 - \eta^{0i}\eta_{ij}J^j)
\end{equation}
The absence of a term of the form $c_2(F_0)^2$ which cannot be canceled by the
GS term constrains the consistent adjacency matrix, and is straightforwardly
shown to depend on the LST charge vector $\ell_I$, see Equation
\eqref{LST-charge-def}.

We stress that from the field-theoretic perspective, the Dirac pairing
$\eta^{IJ}$---and by extension the curve intersections
$\Sigma^I\cdot\Sigma^J$---and the little string charge arise naturally from
group-theoretical relations and imposing anomaly cancellation. While this
reproduces the geometric expectation, it does so without any string theoretic
input. The only assumption beyond the usual field-theoretic ones is
completeness of the spectrum, so that each tensor couples to a BPS string. In
particular, it is easy to show that both the Dirac pairing and the LST charge
satisfy:
\begin{equation}\label{eta-from-G}
		\ell_I = \frac{1}{2}V_I^J\,\theta_J\,,\qquad
		\eta^{IJ} = 2((VD)^{-1}\, G\, (VD)^{-1})^{IJ}\,.
\end{equation}
Since the determinant of $\eta$ is proportional to that of $G$ and $VD$ is
diagonal with non-negative integer coefficients, this justifies Classification
\ref{class:base-topology}.

Summing the two contributions in Equation \eqref{I_8-def}, we arrive at the
final formula for the anomaly polynomial of any LST containing only classical
algebras:
\begin{equation}\label{I8-final}
	\begin{aligned}
			I_8 =& \left.\left(\frac{1}{4}d_I\big(2D^{IJ}\text{ch}_\text{F}(F_J) +A^{IJ}d_J - (S^I+D^{IJ}d_J)\text{ch}_{\bm{2}}(R)\big)\widehat{A}(T)\right)\right|_\text{8-form}\\
					& + \left.\left(\frac{r_\mathfrak{g}}{2}\widehat{A}(T)\text{ch}_{\bm{2}}(R) - \frac{r_\mathfrak{b}}{8}L(T) \right)\right|_{\text{8-form}}+ \frac{1}{2}J^i\eta_{ij} J^j -  c_2(F_0)\, \ell_IJ^I\,.
	\end{aligned}
\end{equation}
The last term has been shown to correspond to the two-group structure constants up to a possible prefactor
\cite{Cordova:2020tij, DelZotto:2020sop}. Indeed, we find
\begin{equation}
	\kappa_R = \ell_I\, h^I\,,\qquad 
	\kappa_P = \ell_I\, a^I\,,\qquad 
	\kappa_F(\mathfrak{f}_I) = \ell_I\,,
\end{equation}
reproducing the usual formulas. Using the relations above, it is then easy to
show that for $\mathfrak{so}_{32}$ heterotic LSTs:
\begin{equation}\label{closed-form-kappa}
	\kappa_P = 2\,,\qquad 		
	\kappa_R = \ell_I h^I = \frac{1}{2}\theta_I\, (\delta^{IJ}d_J + 2S^I)\,.
\end{equation}
The former is expected, as we have already mentioned in the introduction that
in the M-theory picture, $\kappa_P$ counts the number of M9-branes---in this
case two, consistent with an anomaly inflow argument on the little string
\cite{Baume:2024oqn}. On the other hand, $\kappa_R$ changes from one LST to the
other. Using the constraint between the gauge and flavor symmetries, it can be
rewritten solely in terms of the flavor symmetry. As we will show below, it
admits a very simple Lie-theoretic interpretation as a pairing in the root
system of the associated affine algebra.

\paragraph{Higgs-branch dimension:} another important property of a
six-dimensional SQFT is the dimension of its Higgs branch. It was shown in
reference \cite{Mekareeya:2017sqh} that it is also encoded in the anomaly
polynomial through the coefficient $\delta$ of the second Pontryagin class
$p_2(T)$. For $\mathfrak{so}_{32}$ heterotic LSTs,
\begin{equation}\label{I8-delta} 
	I_8\supset \frac{\delta}{24}p_2(T)\,,\qquad
	\delta = -\frac{1}{60}\left(30Q + r_{\mathfrak{g}} +
	\frac{1}{2}f^iC_{ij}(f^j - 2S^j)  \right)\,,
\end{equation}
where we have
used the various relations above to simplify the expression; in
particular $C_{ij}$ is the inverse Cartan matrix of the simple algebra
$\mathfrak{b}$ associated with the base, and the sum does not run over
the affine node.\footnote{For SCFTs, the constant term must be modified
slightly to take into account the additional dynamical tensor
multiplet, but the term depending on the flavor symmetry is the same.}

Given a Higgs-branch RG flow between two theories $\mathcal{T}_\text{UV}$ and
$\mathcal{T}_\text{IR}$, it can be shown that the change in the dimension of
the Higgs branch is then found to be proportional to the difference between the
two coefficients $\delta$:
\begin{equation}\label{Delta-HB}
		\Delta\text{HB} = \text{dim}(\text{HB}_\text{UV}) - \text{dim}(\text{HB}_\text{IR}) = -60(\delta_\text{UV} - \delta_\text{IR})\,.
\end{equation}
This formula will be useful when discussing the Higgs branch of LSTs in Section~\ref{sec:higgs}.

\paragraph{Extra abelian flavor symmetries:} we
have so far neglected possible $\mathfrak{u}_1$ factors in our analysis of the
anomaly polynomial. There is formally an additional $\mathfrak{u}_1$ flavor
symmetry arising for each hypermultiplet transforming in a complex
representation, with unit charge under that symmetry. However, not all
those would-be Abelian symmetries are genuine flavor symmetries of the theory. Indeed, the
anomaly polynomial generically contains a term
\begin{equation}\label{ABJ-I8}
		I_8\supset -\frac{1}{6}\text{Tr}\widetilde{F}^3_I\, \mathcal{Q}^{IJ}\,P_J\,,
\end{equation}
where $P_I$ is the background field strength for the flavor
$\mathfrak{u}_{1}$, and the so-called charge matrix $\mathcal{Q}^{IJ}$ is made out
of combinations of $d_I,f^I$. The presence of terms $\text{Tr}\widetilde{F}^3_J$
is related to non-trivial cubic Casimir operators for $\mathfrak{su}_{k}$ gauge algebras with $k > 2$.
Such a gauge term in the anomaly polynomial does not make the theory
inconsistent, but rather leads to an ABJ-type anomaly and the
$\mathfrak{u}_1$ flavor symmetries are broken at the quantum level
\cite{Lee:2018ihr,Apruzzi:2020eqi}. Note that this distinguishes them from flavor
symmetries of type $\mathfrak{so}_2\simeq\mathfrak{u}_1$ symmetries, which
arise naturally from the quiver and cannot be broken by quantum effects since
$\mathfrak{sp}_{k}$ does not have a cubic Casimir.

If $\mathcal{Q}^{IJ}$ has a non-trivial kernel however, there might be combinations
that survive. A careful analysis of those remnant $\mathfrak{u}_1$ factors was
performed for heterotic LSTs in reference \cite{Ahmed:2023lhj} to show that
$\text{rk}(\mathfrak{f})$ was a T-duality invariant after these subtle
contributions were taken into account. Hence, we will only focus on the
dimension of the kernel of $\mathcal{Q}^{IJ}$, as the number of surviving
$\mathfrak{u}_{1}$ factors is what is relevant for T-duality, rather than the
minutiae of the specific charges. We however note that those charges might be
``delocalized'' over the whole quiver rather than only a single hypermultiplet
being charged under the remnant symmetry. 

In practice however, for $\mathfrak{so}_{32}$ LSTs with a generic instanton
number $Q$, one can show that there is a remnant $\mathfrak{u}_1$ for each
non-trivial $\mathfrak{su}_{f^I}$ flavor symmetry.

\subsection{Classification of \texorpdfstring{$Spin(32)/\mathbb{Z}_2$}{Spin(32)/Z2} LSTs and Coweights}\label{sec:classification-spin32-lst}

We have seen that the allowed flavor symmetries are constrained by cancellation
of quartic anomalies, see Equation \eqref{quartic-cancellation}. We now review
how this can be used to classify all possible consistent spectra based on the
McKay correspondence. We mostly follow the work of references \cite{
Blum:1997fw, Blum:1997mm, Intriligator:1997dh}, albeit reformulated in a more
modern language that will make certain closed-form formulas simpler, and make a
systematic study of the T-duality structure of these theories possible.

\paragraph{The McKay correspondence:} consider a simply-laced simple Lie
algebra $\mathfrak{g}$ of rank $r_{\mathfrak{g}}$. Each finite group
$\Gamma\subset SU(2)$ can be described by the Dynkin diagram of the affine
algebra $\widetilde{\mathfrak{g}}$ of $\mathfrak{g}$. Let us briefly summarize
how the correspondence arises.

In the
standard basis of simple roots, the (fundamental) null root
$\widetilde{\theta}$ is defined as:\footnote{The fundamental null root $\widetilde{\theta}$ is
		traditionally written as $\delta$ \cite{Kac:1990gs}. We however reserve
		the symbol $\delta$ to denote one of the coefficients of the anomaly
		polynomial, see Equation \eqref{I8-delta}, as it is the standard
		convention in the literature of six-dimensional SQFTs.}
\begin{equation}
		\widetilde{\theta} = \sum_{A=0}^{r_{\mathfrak{g}}}\,\theta_A \,\alpha^A = \alpha^0 + \theta\,,
\end{equation}
where the marks $\theta_A$ have already been encountered in Equation
\eqref{comarks-definition} and form the null vector of the transpose of the
Cartan matrix (in the proper normalization). Note that for simply-laced
algebras, the Cartan matrix is symmetric and marks and comarks are the same:
$K_A=\theta_A$.

The highest root of a simple algebra $\mathfrak{g}$ is given by $\theta =
\theta^a\alpha_a$ with $a=1,\dots,r_{\mathfrak{g}}$. The null root
$\widetilde{\theta}$ therefore defines the affine simple root $\alpha^0$ from
the data of the finite algebra $\mathfrak{g}$. A short review of root systems
can be found in Appendix \ref{app:roots}.

McKay realized that the finite subgroups $\Gamma\subset SU(2)$ can be
associated with a simply-laced affine algebra $\widetilde{\mathfrak{g}}$
\cite{mckay1980graphs}: to each node of its affine Dynkin diagram, labeled by
$A = 0, 1, \dots, \text{rk}_{\mathfrak{g}}$, one can associate one of its
irreducible representations $\mathcal{R}_A$ of $\Gamma$, and vice versa. Their
dimension is then given by the value of the corresponding (co)mark of
$\widetilde{\mathfrak{g}}$: 
\begin{equation}
		\text{dim}(\mathcal{R}_A) = \theta_A\,.
\end{equation}
The McKay correspondence therefore allows us to relate the dimensions of the
irreducible representations $\mathcal{R}_A$ to quantities related to
$\mathfrak{g}$. For instance, it is well known---and can easily be checked by
inspection---that ADE algebras satisfy the following properties:
\begin{equation}\label{ADE-mark-sums}
		\text{ADE algebras:}\qquad\qquad
		\sum_{A=0}^{r_\mathfrak{g}}\, \theta_A = h^\vee\,,\qquad
		\sum_{A=0}^{r_\mathfrak{g}}\, (\theta_A)^2 = \Gamma\,,
\end{equation}
where by abuse of notation $\Gamma$ denotes both the finite subgroup of $SU(2)$
and its order.

In the context of $Spin(32)/\mathbb{Z}_2$ Little String Theories, the
irreducible representations of $\Gamma$ have an impact on the relevant
branching rules of $Spin(32)/\mathbb{Z}_2$ classifying them \cite{Blum:1997mm}. We will
therefore sometimes decompose the indices $A=0,1,\dots, r_{\mathfrak{g}}$
depending on the type of the representation. For all real representations
$\mathcal{R}_A$, we collectively denote the set of indices labeling the
corresponding nodes by $\mathcal{R}$; for pseudo-real representations, they are
denoted by $\mathcal{P}$. We also separate complex representations into sets
$\mathcal{C}$ and $\overline{\mathcal{C}}$ so that we may distinguish conjugate
representations: if $A\in \mathcal{C}$ and $\overline{\mathcal{R}}_A =
\mathcal{R}_B$, then $B\in \overline{\mathcal{C}}$. This is summarized in
Table~\ref{tbl:node-color}.

\begin{table}[t]
    \centering
    \begin{threeparttable}
        \begin{tabular}{ccccc}
            \toprule
			node & $S^A$ & type & flavor \\\midrule
			$\begin{tikzpicture}\node[node, fill=red] (A0) at (0,0) {};\end{tikzpicture}$    & $+1$ & $\mathcal{R}$ & $\mathfrak{so}_{f^A}$   \\
			$\begin{tikzpicture}\node[node, fill=blue] (A0) at (0,0) {};\end{tikzpicture}$   & $-1$ & $\mathcal{P}$ & $\mathfrak{sp}_{f^A/2}$\\
		    $\begin{tikzpicture}\node[node, fill=white] (A0) at (0,0) {};\end{tikzpicture}$  & $0$  & $\mathcal{C}$ & $\mathfrak{su}_{f^A}$  \\
			$\begin{tikzpicture}\node[node, fill=black] (A0) at (0,0) {};\end{tikzpicture}$  & $0$  & $\overline{\mathcal{C}}$ & $\mathfrak{su}_{f^A}$  \\
			\bottomrule
        \end{tabular}
    \end{threeparttable}
	\caption{Relations between the representations of ADE finite subgroups
	$\Gamma\subset SU(2)$ and quiver data. $S^A$ denotes the Frobenius--Schur
	indicator of the representation $\mathcal{R}_A$.}
    \label{tbl:node-color}
\end{table}

We can then arrange the group-theoretical data of $\Gamma$ that will be
relevant for the study of LSTs into a colored Dynkin diagram, with the color
differentiating between the type of representations. They are summarized,
along with other useful group-theoretical quantities, in Table
\ref{tab:algebra-values}.

\paragraph{The Frobenius--Schur indicator:} when discussing anomalies, we have
defined the quantity $S^I$, which is a constant depending on the associated
flavor symmetry of the quiver. In the context of the McKay correspondence, it
simply corresponds to the Frobenius--Schur indicator of the finite
representation $\mathcal{R}_A$, whose values depend on whether it is
(pseudo)-real or complex:
\begin{equation}
	S^A=
	\begin{cases}
			\phantom{-}1\,,\quad \text{if}\quad A\in \mathcal{R}\,;\\
			-1\,,\quad \text{if}\quad A\in \mathcal{P}\,;\\
		\phantom{-}0\,, \quad \text{if}\quad A\in \mathcal{C}\cup\overline{\mathcal{C}}\,.
	\end{cases}
\end{equation}
This reproduces the definition given in Equation \eqref{classical-relations}.
It is further possible to show that for any ADE algebra excluding
$\mathfrak{su}_{2k+1}$, it satisfies the relation
\begin{equation}\label{thetaS2}
		\theta_A\, S^A = 2\,, \qquad \mathfrak{g}\in \{\mathfrak{su}_{2k}, \mathfrak{so}_{k}, \mathfrak{e}_6, \mathfrak{e}_7,\mathfrak{e}_8\}\,.
\end{equation}
These algebras are precisely those appearing in the $Spin(32)/\mathbb{Z}_2$
LSTs we are focusing on, and will give us a universal way of rewriting Equation
\eqref{level}. For $\mathfrak{g}=\mathfrak{su}_{2k+1}$ we instead have
$\theta_A\, S^A = 1$, once again illustrating the special case of LSTs with
that type of singularity.

\begin{table}
		\centering
		\begin{tabular}{ccccccc}
				$\mathfrak{g}$ & $\text{rk}_{\mathfrak{g}}$ & $\text{dim}(\mathfrak{g})$ & $h^\vee_\mathfrak{g}$ & $\Gamma$ & $Z(\mathfrak{g})$ & Affine Dynkin diagram \\
				\toprule
				$\mathfrak{su}_{K=2k+1}$ & $K-1$ & $K^2-1$ & $K$ & $K$ & $\mathbb{Z}_K$ & $\begin{matrix}\input{figures/Aeven.tex}\end{matrix}$\\
				$\mathfrak{su}_{K=2k}$ & $K-1$ & $K^2-1$ & $K$ & $K$ & $\mathbb{Z}_K$ & $\begin{matrix}\input{figures/Aodd.tex}\end{matrix}$\\
				$\mathfrak{so}_{2K=4k}$ & $2k$ & $K(2K-1)$ & $2K-2$ & $4K-8$ & $\mathbb{Z}_2\times\mathbb{Z}_2$ & $\begin{matrix}\input{figures/Deven.tex}\end{matrix}$\\
				$\mathfrak{so}_{2K=4k+2}$ & $2k+1$ & $K(2K-1)$ & $2K-2$ & $4K-8$ & $\mathbb{Z}_4$ & $\begin{matrix}\input{figures/Dodd.tex}\end{matrix}$\\
				$\mathfrak{e}_{6}$ & $6$ & $78$ & $12$ & $24$ & $\mathbb{Z}_3$ & $\begin{matrix}\input{figures/E6.tex}\end{matrix}$\\
				$\mathfrak{e}_{7}$ & $7$ & $133$ & $18$ & $48$ & $\mathbb{Z}_2$ & $\begin{matrix}\input{figures/E7.tex}\end{matrix}$\\
				$\mathfrak{e}_{8}$ & $8$ & $248$ & $30$ & $120$ & $\varnothing$ & $\begin{matrix}\input{figures/E8.tex}\end{matrix}$\\
				\bottomrule
		\end{tabular}
		\caption{
				Relevant quantities of simply-laced Lie algebras. We abuse the
				notation and use $\Gamma$ to denote the order of the finite
				subgroup of $SU(2)$ McKay dual to $\mathfrak{g}$. The labels of
				the affine Dynkin diagram refer to marks $\theta_A$. The
				color of a node denotes the type of the representation of the
				discrete group: 
				$\begin{tikzpicture}\node[node, fill=red] (A0) at (0,0) {};\end{tikzpicture} \leftrightarrow\mathcal{R}$; 
				$\begin{tikzpicture}\node[node, fill=blue] (A0) at (0,0) {};\end{tikzpicture} \leftrightarrow \mathcal{P}$;
				$\begin{tikzpicture}\node[node, fill=white] (A0) at (0,0) {};\end{tikzpicture} \leftrightarrow \mathcal{C}$; 
				$\begin{tikzpicture}\node[node, fill=black] (A0) at (0,0) {};\end{tikzpicture} \leftrightarrow \overline{\mathcal{C}}$.
		}
		\label{tab:algebra-values}
\end{table}

\paragraph{Embeddings of the finite group:} with the conventions we have
introduced above, the classification of the possible embeddings of $\Gamma$
into $Spin(32)/\mathbb{Z}_2$ then corresponds to decompositions of the
fundamental representation of $Spin(32)/\mathbb{Z}_2$ into representations of
$\Gamma$ \cite{Blum:1997mm}:
\begin{equation}
		\mathcal{R} = \bigoplus_A \mu^A\,\mathcal{R}_A\,,\qquad 
		\mu^A \, \theta_A = 32\,.
\end{equation}
Taking into account the reality condition of $Spin(32)/\mathbb{Z}_2$, we must have
$\mu^I=\mu^J$ if $\overline{\mathcal{R}}_I = \mathcal{R}_J$. In addition,
pseudo-real representations must be even, and are rotated by a symplectic
transformation, forcing the decomposition of $Spin(32)/\mathbb{Z}_2$ to be given by:
\begin{equation}\label{BRso32}
\begin{aligned}
		Spin(32)/\mathbb{Z}_2 & \longrightarrow\quad \Gamma \times \prod_{A\in\mathcal{R}}SO(\mu^A)\times \prod_{A\in\mathcal{P}}SP(\frac{\mu^A}{2})\times \prod_{A\in\mathcal{C}}U(\mu^A)\,.\\
		\bm{32} & \longrightarrow\quad \bigoplus_A\, \theta_A\, \mu^A\,.
\end{aligned}
\end{equation}
The McKay correspondence therefore gives an elegant way to find the
decomposition of $Spin(32)/\mathbb{Z}_2$ into subgroups that are relevant to the study of
$Spin(32)/\mathbb{Z}_2$ LSTs.

\paragraph{Recovering the quiver:} the integers $\mu^I$ can therefore be
understood as forming a weighted Dynkin diagram of $\widetilde{\mathfrak{g}}$.
The mapping between the coefficients $\mu^A$ and a $Spin(32)/\mathbb{Z}_2$ LST
is then straightforward: through Equation \eqref{BRso32}, we see that they
define the dimension of the fundamental representation of a flavor symmetry
$\mathfrak{f}_I$ on each node. This flavor symmetry rotates hypermultiplets
transforming in the fundamental representation of a gauge symmetry
$\mathfrak{g}_I$. The type of gauge symmetry can be read off the Dynkin
diagram as well: if a flavor symmetry is real, i.e. of type $\mathfrak{so}_f$,
it rotates gauge hypermultiplets in the fundamental representation of
$\mathfrak{sp}_k$, and vice versa.

A subtlety however arises when the discrete group $\Gamma$ has complex
representations: since $\mathfrak{so}_{32}$ only has real representations, we
must identify complex representations with their conjugate \cite{Blum:1997fw}.
We then obtain a full hypermultiplet rotated by a $\mathfrak{su}_k$ symmetry.
At the level of the affine Dynkin diagram $\widetilde{\mathfrak{g}}$, this
corresponds to a \emph{folding} of the diagram to obtain that of the base
$\widetilde{\mathfrak{b}}$.

This explains why, while in the string theory construction we have five-branes
probing a $\mathbb{C}^2/\Gamma$ singularity, the quiver does not always have
the topology of the Dynkin diagram of $\widetilde{\mathfrak{g}}$. For instance,
in the case of $\widetilde{\mathfrak{g}} = E^{(1)}_6$, we obtain the affine
algebra $\widetilde{\mathfrak{b}} = F^{(1)}_4$:
\begin{equation}\label{foldingE6}
    \begin{matrix}
    \begin{tikzpicture}
      \node[node, label=below:{\footnotesize $\mu^0$}, fill=red] (A1)  {};
      \node[node, label=below:{\footnotesize $\mu^6$}, fill=blue] (A2) [right=6mm of A1] {};
      \node[node, label=below:{\footnotesize $\mu^3$}, fill=red] (A3) [right=6mm of A2] {};
      \node[node, label=below:{\footnotesize $\mu^2$}, fill=white] (A4) [right=6mm of A3] {};
      \node[node, label=below:{\footnotesize $\mu^1$}, fill=white] (A5) [right=6mm of A4] {};
      \node[node, label=left:{\footnotesize $\mu^4$}, fill=black] (A6) [above=6mm of A3] {};
      \node[node, label=left:{\footnotesize $\mu^5$}, fill=black] (A0) [above=6mm of A6] {};
      \draw (A1.east) -- (A2.west);
      \draw (A2.east) -- (A3.west);
      \draw (A3.east) -- (A4.west);
      \draw (A4.east) -- (A5.west);
      \draw (A3.north) -- (A6.south);
      \draw (A6.north) -- (A0.south);
	  \draw[<->, dashed, bend left=30, shorten >=6pt, shorten <=6pt]  (A0) to node [auto] {} (A5);
	  \draw[<->, dashed, bend left=30, shorten >=6pt, shorten <=6pt]  (A6) to node [auto] {} (A4);
    \end{tikzpicture}
    \end{matrix}
	\qquad\Rightarrow\qquad
    \begin{matrix}
	\begin{tikzpicture}
	    \node[node, label=below:{\footnotesize $f^0$}, fill=red] (A0)  {};
	    \node[node, label=below:{\footnotesize $f^1$}, fill=blue] (A1) [right=6mm of A0] {};
	    \node[node, label=below:{\footnotesize $f^2$}, fill=red] (A2) [right=6mm of A1] {};
	    \node[node, label=below:{\footnotesize $f^3$}] (A3) [right=6mm of A2] {};
	    \node[node, label=below:{\footnotesize $f^4$}] (A4) [right=6mm of A3] {};
		\node[yscale=1.4] (C) [right=.2mm of A2] {$>$};
	    \node (D) [right=6mm of C] {};
	    \draw (A0.east) -- (A1.west);
	    \draw (A1.east) -- (A2.west);
		\draw ([yshift=1.5pt]A2.east) -- ([yshift=1.5pt]A3.west);
	    \draw ([yshift=-1.5pt]A2.east) -- ([yshift=-1.5pt]A3.west);
	    \draw (A3.east) -- (A4.west);
	\end{tikzpicture}
    \end{matrix}
\end{equation}
Note that we have arranged the nodes of the $E^{(1)}_6$ Dynkin diagram
differently than in Table \ref{tab:algebra-values} to highlight their relation
to $F^{(1)}_4$. This reproduces correctly the adjacency of hypermultiplets for
the LST discussed around Equation \eqref{cartan-F4}. Indeed, taking this folding into
account the coefficients $\mu^A$ in Equation \eqref{BRso32}
precisely give the flavor fundamental $f^I$ of the $Spin(32)/\mathbb{Z}_2$ LST,
with $I=0,1\dots,r_{\mathfrak{b}}$. We see that we recover the correct spectrum
from the coefficients $\mu^I$, and the symmetrized Cartan matrix $G^{IJ}$ of
$F^{(1)}_4$ correctly reproduces the complete adjacency of the hypermultiplet
spectrum: four $\mathfrak{g}_I\oplus\mathfrak{g}_{J}$ arranged linearly, and
five $\mathfrak{g}_I\oplus\mathfrak{f}_{J}$. This means that given $\mu^A$, we
can directly write down the generalized quiver in the curve notation described
at the beginning of this section:
\begin{equation}\label{general-quiver-f4}
		\underset{[\mathfrak{so}_{f^1}]}{\overset{\mathfrak{sp}_{k_1}}{1}} \, 
	\underset{[\mathfrak{sp}_{\frac{f^2}{2}}]}{\overset{\mathfrak{so}_{k_2}}{4}} \, 
	\underset{[\mathfrak{so}_{f^3}]}{\overset{\mathfrak{sp}_{k_3}}{1}} \, 
    \underset{[\mathfrak{su}_{f^4}]}{\overset{\mathfrak{su}_{k_4}}{2}} \, 
    \underset{[\mathfrak{su}_{f^5}]}{\overset{\mathfrak{su}_{k_5}}{2}}\,,\qquad
\end{equation}
the gauge algebras are then fixed up to an integer $Q$ via anomaly
cancellation, see Equation \eqref{sol-dI}. Furthermore, we see that the
bifundamental $\mathfrak{sp}_{k_3}\oplus\mathfrak{su}_{k_4}$ hypermultiplet
corresponds to the double link of $F_4^{(1)}$, with the arrow pointing in the
direction of the node hosting the complex gauge algebra $\mathfrak{su}_{k_4}$.

More generally, if we have a folding $\mathcal{F}
$ mapping the indices associated with
$\widetilde{\mathfrak{g}}$ to that of $\widetilde{\mathfrak{b}}$, we can define
the flavor and a folded version of the Frobenius--Schur indicator as:
\begin{equation}\label{f-S-folding}
		f^I = \mu^{\mathcal{F}(A)}\,,\qquad
		S^I = S^{\mathcal{F}(A)}\,.
\end{equation}
Under that identification the relation involving $\theta_A$ and the
group-theoretical quantities in Equation \eqref{ADE-mark-sums} holds, as the
comarks of nodes associated with complex representations are doubled after
folding: If $A\in\mathcal{C}$, then $\theta_{\mathcal{F}(A)} = 2\,\theta_A$.
Moreover, as $S_A=0$, the relation $\theta_I S^I=2$ remains correct for the
folded algebra. Similarly, if $\widetilde{\mathfrak{b}}$ is the algebra
coming from folding $\widetilde{\mathfrak{g}}$, one can show that the relations
in Equation \eqref{ADE-mark-sums} hold
\begin{equation}\label{folded-mark-sums}
	\widetilde{\mathfrak{b}} = \mathcal{F}(\widetilde{\mathfrak{g}}):\qquad 
	\sum_{I=0}^{r_\mathfrak{b}}\, \theta_I^{\mathfrak{b}} = h_{\mathfrak{g}}^\vee\,,\qquad
	\sum_{I=0}^{r_\mathfrak{b}}\, (\theta_I^{\mathfrak{b}})^2 = \Gamma_{\mathfrak{g}}\,,
\end{equation}
which is easily shown by inspection.

\paragraph{$\mathbf{Spin(32)/\mathbb{Z}_2}$ LSTs and weight systems:} the various
formulas above enable us to interpret the flavor symmetries in a more
Lie-algebraic context. While this is certainly interesting at an aesthetic
level, it will more importantly also allow us to use results that are standard
in that context in order to easily prove certain properties of LSTs.

Let us first recall that the root system of a \emph{finite} Lie algebra
$\mathfrak{g}$ has a root lattice $\mathcal{Q}$ generated by the simple roots
$\alpha^a$, and the coweight lattice $\mathcal{P}^\vee$ by the fundamental
coweights $\omega_a^\vee$. These two lattices are dual, and we therefore have:
\begin{equation}
	\langle \alpha^a,\omega_b^\vee\rangle = \delta^a_b = \langle \widetilde{\omega}_a,\alpha^{b\,\vee}\rangle\,,
\end{equation}
where $\langle\cdot,\cdot\rangle$ is the natural pairing of the root system. A
coweight $\mu^\vee = \mu^a \omega_a^\vee$ is called dominant, collectively
denoted $\mathcal{P}^\vee_+$, if all its coefficients $\mu^a$ are non-negative
integers. 

Given a finite algebra $\mathfrak{g}$, one can extend its root system to that
of the affine algebra $\widetilde{\mathfrak{g}}$. We then define a set of
affine roots $\alpha^A$ and coweights $\widetilde{\omega}_A$ satisfying
\begin{equation}
		\langle \alpha^A,\widetilde{\omega}_B^\vee\rangle = \delta^A_B = \langle \widetilde{\omega}_A,\alpha^{B\,\vee}\rangle\,,
\end{equation}
where the pairing is extended from that of the finite algebra. We have also
defined the coroot $\alpha^{A\,\vee}$ and weights $\widetilde{\omega}_A$ for
future reference. We review the construction of simple and affine root systems
in Appendix \ref{app:roots}. For our present purpose, it will be sufficient to
simply recall that a dominant affine coweight $\widetilde{\mu}^A\in
\widetilde{\mathcal{P}}^\vee_+$ is then defined as being of the form
\begin{equation}\label{def-coweight}
		\widetilde{\mu}^\vee = \mu^A\, \widetilde{\omega}_A^\vee + n\, K \,\in\, \widetilde{\mathcal{P}}^\vee_+\,,\qquad \mu^A,n\geq 0\,,
\end{equation}
where $K=K_A\alpha^{A\,\vee}$ is the so-called central element, whose
coefficients $K_A$ are the comarks of $\widetilde{\mathfrak{g}}$. The level of
an affine coweight is defined as $\text{lv}(\widetilde{\mu}^\vee)=\theta_A
\mu^A$. 

It can often be convenient to think of an \emph{affine} coweight
$\widetilde{\mu}^\vee$ of $\widetilde{\mathfrak{g}}$ as a \emph{finite}
coweight $\mu^\vee=\mu^a\omega_a^\vee$ of $\mathfrak{g}$ supplemented by a
level and an integer $n$, encoded in the triplet:
\begin{equation}\label{triplet-coweight}
		\widetilde{\mu}^\vee = (\mu^\vee, \text{lv}(\widetilde{\mu}^\vee), n)\,,\qquad \text{lv}(\widetilde{\mu}^\vee)=\theta_A \mu^A\,.
\end{equation}

In this language, all the constraints we have reviewed take a Lie-algebraic
meaning. Indeed, we have already seen that the type of flavor and gauge
structure depends on the choice of $\mathfrak{g}$ through the Frobenius--Schur
indicator, and that the dimension of their fundamental representation is
constrained by gauge-anomaly cancellation, see Equation
\eqref{quartic-cancellation}, from which we have seen that $\theta_I f^I = 16\,
\theta_I S^I =32$. This is precisely the relation that defines the embedding of
$\Gamma$ into $Spin(32)/\mathbb{Z}_2$, see Equations \eqref{level} and
\eqref{BRso32}, and the flavor symmetry is therefore encoded in an affine
coweight of level 32. 

Other properties can be rephrased in those terms. For instance, the matrix used
in Equation \eqref{sol-dI} to solve the gauge-anomaly cancellation condition is
given as the pairing $\langle\widetilde{\omega}_I,
\widetilde{\omega}^\vee_J\rangle = X_{IJ}$ in the weight system of the base
algebra, or equivalently for $\widetilde{\mathfrak{g}}$. In turn, we see that
the dimension of the fundamental representation on the affine node $d_0=2Q$ is
nothing but the coefficient $n$ in the equation above. The position of the indices therefore reflects their relations with coroots and coweights, and are therefore raised and lowered with the flat metric $\delta^A_B$, or similarly for finite or unfolded algebras.

In summary, a choice of an ADE algebra and a dominant coweight completely fixes
the data defining a $Spin(32)/\mathbb{Z}_2$ LST.\footnote{While this data
completely fixes the data at a generic point of the tensor branch of the theory, in a few cases the spectrum at the
singular point might require a choice of $\theta$-angle for the
$\mathfrak{sp}_k$ gauge algebras in order to distinguish between two different
types of spectra. We will not consider this subtlety here, see reference
\cite{Distler:2022yse}.} This is precisely the classification by Blum and
Intriligator.\\ 
\begin{classification}[$Spin(32)/\mathbb{Z}_2$ Little String Theories
		\cite{Blum:1997mm}]\label{class:spin32}

		Given a simply-laced algebra
		$\mathfrak{g}\neq\mathfrak{su}_{2n+1}$, every LST
		$\mathcal{K}_Q(\mu^\vee; {\mathfrak{g}})$ is realized by a generalized
		quiver where the adjacency of the hypermultiplets is given by
		the symmetrized Cartan matrix $G^{IJ}$ of the associated folded
		affine algebra $\widetilde{\mathfrak{b}}$, see Table \ref{tab:folding}.
		The types of gauge and flavor symmetries are obtained from the
		Frobenius--Schur indicator $S_A$, see Tables \ref{tbl:node-color} and
		\ref{tab:algebra-values}.

	The rest of the spectrum is then determined by a dominant coweight
	$\widetilde{\mu}^\vee~\in~ \widetilde{\mathcal{P}}^\vee_+$ of $\widetilde{\mathfrak{g}}$ at level 32:
	\begin{equation}\label{LST-coweight}
			\widetilde{\mu}^\vee = (\mu^\vee, 32, 2Q) = \mu^A\, \widetilde{\omega}_A^\vee + 2Q\, K\,,\qquad \theta_A \mu^A =32\,.
	\end{equation}
	Its coefficients $\mu^A$ correspond to the dimension of the fundamental
	representations of the flavor symmetries, $f^I=\mathcal{F}(\mu^A)$. 

	This subset of dominant coweights, denoted $\widetilde{\mathcal{P}}^\vee_\text{LST}$,
	must furthermore satisfy the following constraints:
	\begin{enumerate}
			\item Defining $\widetilde{S}^\vee = S^A\widetilde{\omega}_A^\vee$ from the
					Frobenius--Schur indicator the weight $\widetilde{\mu}^\vee
					- 16\widetilde{S}^\vee $ is a positive coroot, with
					components: 
			\begin{equation}
					d_A = \langle\widetilde{\omega}_A, \widetilde{\mu}^\vee - 16 \widetilde{S}^\vee\rangle = 2Q\,K_A + X_{AB}(\mu^B - 16S^B)\,\in\,\mathbb{N}\,,
			\end{equation}
			and correspond to the dimensions of the fundamental representation
			of the gauge algebras in the folded base algebra $d_I=\mathcal{F}(d_A)$.
	\item So that algebras of type $\mathfrak{sp}_k$ have an even fundamental representation, if $S^A=-1$ then $\mu^A$ is even, and if $S^A=1$, then $d_A$ is even.
	\item Complex representations of the McKay-dual finite group $\Gamma$ and
			their conjugates have the same coefficients: if
			$\mathcal{R}_A=\overline{\mathcal{R}}_B$, then $\mu^A=\mu^B$.
	\end{enumerate}
\end{classification}

One can equivalently work directly in the affine root system of the base
$\widetilde{\mathfrak{b}}$, rendering the last constraint trivial, and with the
Frobenius--Schur indicator extended through Equation \eqref{f-S-folding}. It
is in practice more convenient, and it has several advantages when studying
LSTs computationally---in particular, generating all dominant coweights
associated with an LST can be considerably faster in that case. We therefore see
that a coweight of $\widetilde{\mathfrak{b}}$ completely defines an LST without
needing any additional external data.

Similarly, the triplet $\widetilde{\mu}^\vee = (\mu^\vee, 32, 2Q)$ also has
advantages, as one can work with the simpler finite coweight $\mu^\vee$. The
flavor on the affine node is then simply $\mu^0 = 32 - \theta_a \mu^a$.

\paragraph{Example:} As an example, let us consider the LST associated with the
algebra $\widetilde{\mathfrak{g}}=A_3^{(1)}$. After folding, we obtain
$\widetilde{\mathfrak{b}}=C_2^{(1)}$, from which we can easily infer the curve
configuration via equation \eqref{eta-from-G}:
\begin{equation}
		A_3^{(1)}:\quad\begin{matrix}\begin{tikzpicture}
	    \node[node, label=left:{\footnotesize $\mu^0$}, fill=red] (A0)  {};
	    \node[node, label=below:{\footnotesize $\mu^1$}, fill=white] (A1) [below right=6mm of A0] {};
	    \node[node, label=above:{\footnotesize $\mu^3$}, fill=black] (A3) [above right=6mm of A0] {};
	    \node[node, label=right:{\footnotesize $\mu^2$}, fill=red] (A2) [below right=6mm of A3] {};
	    \draw (A0.south east) -- (A1.north west);
	    \draw (A1.north east) -- (A2.south west);
	    \draw (A2.north west) -- (A3.south east);
	    \draw (A0.north east) -- (A3.south west);
	\end{tikzpicture}\end{matrix}
	\qquad\Rightarrow\qquad
	C_2^{(0)}:\quad\begin{matrix}\begin{tikzpicture}
		\node[node, label=below:{\footnotesize $f^0$}, fill=red]   (A0) {};
    	\node[node, label=below:{\footnotesize $f^1$}, fill=black] (A1) [right=6mm of A0] {};
    	\node[node, label=below:{\footnotesize $f^2$}, fill=red] (A2) [right=6mm of A1] {};
		\node[yscale=1.4] (C) [right=.2mm of A0] {$>$};
		\node[yscale=1.4] (D) [right=.2mm of A1] {$<$};
		\draw ([yshift=1.5pt]A0.east) -- ([yshift=1.5pt]A1.west);
    	\draw ([yshift=-1.5pt]A0.east) -- ([yshift=-1.5pt]A1.west);
		\draw ([yshift=1.5pt]A1.east) -- ([yshift=1.5pt]A2.west);
    	\draw ([yshift=-1.5pt]A1.east) -- ([yshift=-1.5pt]A2.west);
	\end{tikzpicture}\end{matrix}
	\qquad\leftrightarrow\qquad
	\underset{[\mathfrak{so}_{f^0}]}{\overset{\mathfrak{sp}_{k_0/2}}{1}}\,
	\underset{[\mathfrak{su}_{f^1}]}{\overset{\mathfrak{su}_{k_1}}{2}}\,
	\underset{[\mathfrak{so}_{f^2}]}{\overset{\mathfrak{sp}_{k_2/2}}{1}}\,
\end{equation}
The condition that the level of the affine coweight $\widetilde{\mu}^\vee$ is 32
corresponds to the relation
\begin{equation}
		\theta_A\mu^A = \mu^0+\mu^1+\mu^2+\mu^3 = f^0+2f^1+f^2 = 32\,,
\end{equation}
where after folding we have $\mu^1 = f^2 = \mu^3$. Cancellation of gauge
anomalies, which is equivalent to demanding that $\widetilde{\mu}^\vee$ is both
a dominant coweight and a positive coroot, is simplest in terms of the
folded-algebra quantities. Given the dimensions of the fundamental representations of flavor and gauge symmetries $f^I$, and $k_I$, respectively, we have:
\begin{equation}
		C^{IJ}k_J = f^I - 16 S^I \quad\Leftrightarrow\quad 
		\begin{pmatrix}2 & -2 &0 \\ -1 & 2 & -1\\0&-2&2\end{pmatrix}
		\begin{pmatrix}k_0 \\ k_1 \\ k_2\end{pmatrix} = 
		\begin{pmatrix}f^0 \\ f^1 \\ f^2\end{pmatrix}\, 
		-16\,\begin{pmatrix}1 \\ 0 \\ 1\end{pmatrix}\,.
\end{equation}
where $C^{IJ}$ is the Cartan matrix of $\widetilde{\mathfrak{b}}=C_2^{(1)}$,
and we have used that, following Table \ref{tbl:node-color}, the
Frobenius--Schur indicators are given by $S^I = (1,0,1)$. 

Since we have two symplectic algebras $\mathfrak{sp}_{k_0/2}$ and
$\mathfrak{sp}_{k_2/2}$, we must further impose that $k_0,k_2$ are both even. It is
then easy to show that there are 81 doublets $(f^1,f^2)$ satisfying the
constraints above, leading to 81 dominant finite coweights:
\begin{equation}
		\mu^\vee = \mu^a\omega_a^\vee = f^1\omega_1^\vee + f^2\omega_2^\vee + f^1\omega_3^\vee\,,
\end{equation}
The whole set of consistent dominant affine coweights $\mu^\vee= (\mu^\vee, 32,
2Q)$ defining the $Spin(32)/\mathbb{Z}_2$ LSTs is then obtained by
supplementing $\mu^\vee$ with the instanton number, i.e. setting $k_0=2Q$.

In Sections \ref{sec:structure-duality-orbits} and \ref{sec:example-so8}, we
give additional examples for $\widetilde{\mathfrak{g}} = E_6^{(1)} $ (or
equivalently $\widetilde{\mathfrak{b}} = F_4^{(1)} $) and
$\widetilde{\mathfrak{g}}= D_4^{(1)}$.

\paragraph{T-duality invariants and coweights:}
In the language above, we can rewrite the two-group constant $\kappa_R$, see
Equation \eqref{closed-form-kappa}, in a particularly simple form:
\begin{equation}\label{kappa-closed-form}
	\kappa_R = \frac{1}{2}\langle \widetilde{T}, \widetilde{\mu}^\vee - 16 S\rangle = \Gamma_{\mathfrak{g}}(Q-r_{\mathfrak{g}}-1) + 2 + \frac{1}{2}\mu^aC_{ab}\theta^a\,,
\end{equation}
where $\widetilde{T} = \sum_A\theta_A\widetilde{\omega}_A$ is the affine weight
whose coefficients are the marks---not to be confused with the null root
$\widetilde{\theta} = \theta_A\alpha^A$. The pairing is taken equivalently over
either $\widetilde{\mathfrak{g}}$ or $\widetilde{\mathfrak{b}}$, and we recall
that lowercase indices $a,b$ are taken over only the finite part of the
algebra---where the Cartan matrix is invertible, with $C_{ab}$ its inverse. The
properties of the involved (co)weights and (co)roots are collected in Appendix
\ref{app:roots}.

The Coulomb branch dimension of the five-dimensional theory can be similarly
expressed in closed form. Due to the fact that the rank of $\mathfrak{so}_d$ is
$\lfloor\frac{d}{2}\rfloor$, the resulting expression is a bit more involved:
\begin{equation}\label{CB-closed-form}
		\begin{aligned}
				\text{dim}(\text{CB}) &= 2r_{\mathfrak{b}}-r_{\mathfrak{g}} + \frac{1}{2}\langle \widetilde{\rho}, \widetilde{\mu}^\vee\rangle + \sum_{A\in \mathcal{P}} \text{frac}(\frac{d_I}{2})\\
				&= Qh_{\mathfrak{g}}^\vee -\text{dim}(\mathfrak{g}) + \frac{1}{2}\rho^aC_{ab}\mu^a +\sum_{a\in \mathcal{P}} \text{frac}(\frac{C_{ab}\mu^a}{2})\,,
		\end{aligned}
\end{equation}
where $\widetilde{\rho}$ is the weight with component $\rho^I = D^{II}$, and we
have used that the number of $\mathfrak{su}_n$ gauge algebras on the
six-dimensional quiver is always
$r_{\mathfrak{g}}-r_{\mathfrak{b}}$.\footnote{For ADE algebras, the finite part of $\widetilde{\rho}$
is the Weyl vector, with components $\rho^a = 1$. When working at the level of the folded algebras on the
other hand, we stress that $\rho^I = D^{II}$. This distinction is particularly
important when using the closed-form formulas, as we must ensure that folded
nodes are taken into account correctly.}

Finally, the total flavor rank also admits a simple closed-form. There, however,
the instanton number $Q$ of course cannot appear explicitly, and we need  
to restrict the coweight in Equation \eqref{LST-coweight} to omit the central
element $K$. For brevity, we only give the expression in
components:\footnote{Type-A LSTs have an additional ``baryonic''
		$\mathfrak{u}_1$ symmetry rotating all hypermultiplets simultaneously,
		which we did not include in Equation \eqref{rkf-closed-form}. However,
		every statement we will make about matching of the flavor symmetry rank
will be correct when it is included, see e.g. references \cite{Ahmed:2023lhj, Lawrie:2023uiu}.}
\begin{equation}\label{rkf-closed-form}
		\text{rk}(f) = \frac{1}{2}\rho_A\,\mu^A - \sum_{A\in\mathcal{R}}\text{frac}(\frac{\mu^A}{2})\,,
\end{equation}
where we have used---as indicated around Equation \eqref{I8-ABJ}---that a
$\mathfrak{su}_f$ flavor symmetry is always accompanied by an unbroken
Abelian factor $\mathfrak{u}_1$. These closed-form expressions are very
important as they make the search for T-dual theories much simpler. In
addition, we can now use standard results in the theory of affine root systems
to greatly simplify certain computations.

\paragraph{The $\kappa$-theorem:} in two- and four-dimensional Quantum Field
Theory, the central charges $a$ and $c$ are known to decrease monotonically
along RG flows \cite{Zamolodchikov:1986gt, Komargodski:2011vj}, and are
therefore interpreted as a non-perturbative counting of physical degrees of
freedom. For six-dimensional QFTs, a general $a$-theorem remains elusive
\cite{Elvang:2012st, Baume:2013ika, Cordova:2015fha,
Heckman:2015axa,Stergiou:2016uqq, Heckman:2021nwg}. However, through numerical
scans and due to the deep relationship between group theory and these SQFTs, it
has been established for large classes of theories using properties of the
underlying root systems \cite{Mekareeya:2016yal,Baume:2023onr, Fazzi:2023ulb}.

On the other hand, in the case of LSTs the absence of a local energy-momentum
tensor at the singular point of the tensor branch does not allow for such
theorems. However, it has been conjectured that $\kappa_R$ should give rise to
a similar version of at least the weak $a$-theorem \cite{DelZotto:2022ohj}: given an RG
flow between two LSTs $\mathcal{K}^\text{UV}$ and $\mathcal{K}^\text{IR}$, we
must have 
\begin{equation}
	\kappa_R(\mathcal{K}^\text{UV}) > \kappa_R(\mathcal{K}^\text{IR})\,.
\end{equation}

This ``$\kappa$-theorem'' was proven in reference \cite{Lawrie:2023uiu} for any
LST admitting a description in terms of a generalized quiver. Due to their
relation to coweights we can give an equivalent, but simpler, proof for any
$\mathfrak{so}_{32}$ LST. Indeed, anticipating a more complete discussion in the next
section, dominant coweights admit a partial order. One says that
$\widetilde{\nu}^\vee<\widetilde{\mu}^\vee$ if $\widetilde{\mu}^\vee -
\widetilde{\nu}^\vee$ lies in the cone of positive coroots. If
$\mathcal{K}^\text{UV}$ and $\mathcal{K}^\text{IR}$ are associated with
$\widetilde{\mu}^\vee$ and $\widetilde{\nu}^\vee$, respectively, a necessary
condition for an RG flow to exist is then that
$\widetilde{\nu}^\vee<\widetilde{\mu}^\vee$.

While the original proof describes how $\kappa_R$ decreases after performing a
Higgs mechanism, the theory of coweights shows this result immediately for
$\mathfrak{so}_{32}$ heterotic LSTs: it is a standard result that, given any
dominant weight $w$, the pairing $\langle w,\cdot \rangle$ is a monotonic
function with respect to the partial order \cite{stembridge1998partial},
and we indeed recover
\begin{equation}
		\kappa_R(\widetilde{\mu}^\vee) - \kappa_R(\widetilde{\nu}^\vee) = \frac{1}{2}\langle\widetilde{T}, \widetilde{\mu}^\vee - \widetilde{\nu}^\vee \rangle > 0\,.
\end{equation}
While this proof of the $\kappa$-theorem is of course completely equivalent to
that of reference \cite{Lawrie:2023uiu}, it does not explicitly require prior
knowledge of the quiver or the F-theory description. This illustrates how the
language of dominant coweights can greatly simplify certain computations, as we
will now see for T-dual theories.

\section{T-duality of \texorpdfstring{$Spin(32)/\mathbb{Z}_2$}{Spin(32)/Z2} Little String Theories}\label{sec:t-duality}

Little String Theories have the important property that they can exhibit
T-duality. To avoid confusion, we now spell out carefully what we mean by
such a feature, which follows from what is commonly thought of as T-duality in
ten dimensions.\\

\begin{definition}[T-duality of LSTs]\label{def:t-duality} 
	Two Little String Theories are T-dual if, upon a circle reduction, there
	exists a common subspace in their five-dimensional extended Coulomb-branch
	moduli space along which the two theories become identical.
\end{definition}

Note that two T-dual theories are generically \text{not identical} upon a
simple circle reduction. However, due to symmetry-valued Wilson lines along the
circle, the lower-dimensional moduli space is much larger than its
higher-dimensional analogue. For specific values of Wilson lines---and possibly
the tensor-branch moduli giving rise to part of the five-dimensional
Coulomb-branch moduli space---there exists a subspace where both theories are
exactly identical. A direct consequence of this requirement is that the
Coulomb-branch dimension in five dimensions---corresponding to the sum of the rank
of the full gauge symmetry and the number of dynamical tensor multiplets in
six dimensions---must be the same for the two LSTs.

For LSTs of heterotic type, we will differentiate between three different
classes of theories exhibiting T-duality. Namely, we will refer to a pair of
T-dual LSTs as:
\begin{enumerate}
		\item \textbf{internal (T-)dual}, when they are both heterotic LSTs of
				type $\mathfrak{so}_{32}$, \emph{or} both of type
				$\mathfrak{e}_8\oplus \mathfrak{e}_8$ for a given ADE algebra
				$\mathfrak{g}$;
		\item \textbf{fiber-base (T-)dual}, when one is
				of type $\mathfrak{so}_{32}$ while the other is of type
				$\mathfrak{e}_8\oplus \mathfrak{e}_8$ for a given ADE algebra
				$\mathfrak{g}$;
		\item \textbf{exotic (T)-dual} if they relate LSTs associated with
				different ADE algebras $\mathfrak{g}$.
\end{enumerate}
Fiber-base duality is the standard T-duality between the
$\mathfrak{e}_8\oplus\mathfrak{e}_8$ and $\mathfrak{so}_{32}$ heterotic string
theories \cite{Aspinwall:1997ye}, its name deriving in the geometric picture
from the exchange of the algebra associated with the base of the elliptic
fibration with that of the elliptic fibers. The term also applies to those of
Type II. There, LSTs are uniquely defined by a choice of base algebra
$\mathfrak{g}_B$ and a fiber algebra $\mathfrak{g}_F$, and fiber-base duality
simply exchanges them:
$(\mathfrak{g}_B,\mathfrak{g}_F)\leftrightarrow(\mathfrak{g}_F,\mathfrak{g}_B)$
\cite{DelZotto:2020sop, Baume:2024oqn}.

Furthermore, in the presence of discrete symmetries, a twist along the circle
can be performed, opening up the possibility of new T-dual pairs
\cite{Bhardwaj:2022ekc, Ahmed:2024wve}. While we will not discuss twisted
compactification in this work, these would fall into the class of internal
duals. 

\emph{A contrario}, pairs of exotic type often arise when an LST can be thought
of as originating through a Higgs-branch flow from theories associated with
different algebras. This therefore especially occurs when the number of curves
is low, see e.g. references \cite{DelZotto:2022xrh, Ahmed:2023lhj,
Ahmed:2024wve}. In this work, we will consider the instanton number to be large
so that there are no base-changing deformations, and will generically not
consider those cases. In Section~\ref{sec:extension}, we will however discuss
infinite families which exhibit exotic duality \cite{DelZotto:2022xrh,
Ahmed:2023lhj} which can partly be explained by a group-theoretic property
relating certain invariants of LSTs with different quiver topology.

Of course, more than two LSTs can be dual to one another, and there are many
examples of such ``T-$n$-ality'' \cite{Bhardwaj:2015oru, DelZotto:2020sop,
DelZotto:2022ohj, DelZotto:2022xrh,Bhardwaj:2022ekc,
Ahmed:2023lhj, Mansi:2023faa, Lawrie:2023uiu, Baume:2024oqn, Ahmed:2024wve}. We
will refer to such groupings as \textbf{(T-)duality orbits}. These can
contain a quite large number of theories. For instance, in reference
\cite{DelZotto:2022xrh} it was shown through a specific string-theoretic
realization that a certain geometry led to sixteen different, but T-dual,
Little String Theories.

\subsection{Structure of Duality Orbits}\label{sec:structure-duality-orbits}

Identifying LSTs falling in the same duality orbit and proving T-duality can be
an arduous task, and generally requires either a brane construction or
geometric engineering in the F-theory picture. Certain quantities that are
invariant under T-duality have however been identified and can be used to
facilitate the search for T-dual pairs \cite{DelZotto:2022ohj,
DelZotto:2022xrh, Ahmed:2023lhj, Lawrie:2023uiu}:
\begin{equation}\label{T-dual-invariants}
		\kappa_P\,,\quad
		\kappa_R\,,\quad\,
		\text{dim}(\text{CB})\,,\quad
		\text{rk}(\mathfrak{f})\,.
\end{equation}
The last two must match as they follow from the definition of T-duality, while
the two-group structure constants are related to correlators of the associated
conserved currents \cite{Cordova:2020tij}, and are therefore expected to match
across T-duality since this does not rely on a specific gauge or geometric
realization. When convenient, and for ease of notation, we will often denote
$\text{dim}(\text{CB})$ simply as $\text{CB}$.

In reference \cite{Baume:2024oqn}, it was further shown that higher-form
symmetries are preserved under T-duality, in the sense that six-dimensional
one-form symmetries and the defect group give rise to one-form symmetries in
five dimensions, and must therefore match. However, in the case of heterotic
theories the defect group is always trivial, and this will not be relevant for
our purposes.

On the $\mathfrak{so}_{32}$ side, the T-dual invariants are of course not
independent, as they are ultimately related to the coweight defining the theory
through the closed-form formulas derived in the previous section. On the other
hand, the invariants in Equation \eqref{T-dual-invariants} are extremely
powerful to carve out the space of heterotic LSTs into candidate duality
orbits. 

Consider two LSTs associated with affine coweights
$\widetilde{\mu}^\vee=(\mu^\vee, 32, 2Q)$ and $\widetilde{\nu}^\vee=(\nu^\vee,
32, 2Q')$, in the notation of Equation \eqref{LST-coweight}, with
$\mu^\vee,\nu^\vee\in \mathcal{P}_\text{LST}^\vee(\mathfrak{g})$ two dominant
finite coweights of the same algebra $\mathfrak{g}$. From the closed-form
expressions given in equations \eqref{kappa-closed-form} and
\eqref{CB-closed-form}, we see that the two invariants take the form:
\begin{equation}
		\kappa_R(\widetilde{\mu}^\vee) =  \Gamma\, (Q-r_{\mathfrak{g}}-1)+2 + \delta\kappa_R(\mu^\vee)\,,\qquad
		\text{CB}(\widetilde{\mu}^\vee) = h^\vee Q - \text{dim}(\mathfrak{g}) + \delta\text{CB}(\mu^\vee)\,,
\end{equation}
where $\delta\kappa_R$ and $\delta\text{CB}$ depend only on the finite part
$\mu^\vee$ of the affine coweight $\widetilde{\mu}^\vee$. A necessary condition
for the two associated LSTs to be T-dual is therefore that:\footnote{ We will
		not consider $\kappa_P = 2$, as it always matches for any heterotic LST
and assuming that there are no additional discrete twists along the Kaluza--Klein circle.}
\begin{equation}\label{duality-condition}
	\delta\kappa_R(\mu^\vee) - \delta\kappa_R(\nu^\vee) = \Gamma(Q-Q')\,,\qquad
	\delta\text{CB}(\mu^\vee) - \delta\text{CB}(\nu^\vee) = h_\mathfrak{g}^\vee(Q-Q')\,,
\end{equation}
The two $\mathfrak{so}_{32}$ theories are therefore putatively in the same
T-duality orbit if their part of the invariants controlled by the finite part
of the coweight coincide up to a particular multiple of $Q-Q'$. 

In this section, we will therefore restrict ourselves to dominant coweights of
the simple algebra $\mathfrak{g}$ and find those that lead to LSTs in the same
orbit up to this multiple, rather than affine coweights. We recall that
$\mathcal{P}_+^\vee$ is the set of dominant finite coweights, while
$\mathcal{P}_\text{LST}^\vee\subset\mathcal{P}_+^\vee$ denotes those associated
with an LST, as defined at the end of
Section~\ref{sec:classification-spin32-lst}. 

We can significantly simplify our work by using additional properties of the
root system. In particular, it was shown by Stembridge that dominant
(co)weights admit a natural partial order \cite{stembridge1998partial}.
Indeed, two dominant coweights $\mu^\vee, \nu^\vee\in \mathcal{P}_+^\vee$ are
partially ordered if their difference is a positive sum of coroots:
\begin{equation}\label{po-coweights}
		\nu^\vee \leq \mu^\vee\,,\quad \Leftrightarrow\quad \mu^\vee - \nu^\vee = m_a \alpha^{a\,\vee}\,,\quad m_a\in\mathbb{N}\,,
\end{equation}
with the equality satisfied only if $\mu^\vee=\nu^\vee$. This definition
extends to affine coweights $\widetilde{\mu}^\vee=(\mu^\vee, 32, 2Q)$ and
$\widetilde{\nu}^\vee=(\nu^\vee, 32, 2Q')$ if in addition $Q'\leq Q$.

Physically, we have seen in Section~\ref{sec:classification-spin32-lst} that
$\widetilde{\mu}^\vee - 16\widetilde{S}^\vee$ must be a positive coroot for the
gauge algebras to be well defined. The partial order in equation
\eqref{po-coweights} therefore simply means that an LST is ``smaller than''
another if the change in the associated dimensions $d_I$ of the
fundamental representation of a gauge symmetry are positive integers. As we
will see in the next section, this is related to the fact that there can be a
Higgs-branch Renormalization Group flow between the two.

The set of dominant coweights endowed with the partial order above is seen as the
poset $(\mathcal{P}^\vee_+, <)$, and has certain attractive properties
\cite{stembridge1998partial} with important physical consequences. First, it is
disconnected, with the number of components given by the order of the center
$Z(\mathfrak{g}) = \mathcal{P}^\vee/\mathcal{Q}^\vee$.\footnote{ The center $Z$
		is that of the simply-connected Lie group associated with
		$\mathfrak{g}$. As we work with an ADE algebra, we write
$Z(\mathfrak{g}) = \mathcal{P}^\vee/\mathcal{Q}^\vee$ to emphasize that it can be
obtained via the root system of $\mathfrak{g}$.} However, an LST with $\mu^\vee\in
\mathcal{P}_\text{LST}^\vee$ must then be in the component of the poset connected to the
trivial coweight: for any LST we must have $0<\mu^\vee$, as all $d_I$ are positive
integers.

Let us now assume that we have found all possible coweights in a putative
orbit, defined as the set of coweights leading to the same T-dual invariants up
to a shift in $Q$, i.e. satisfying the condition in equation
\eqref{duality-condition}, and the same flavor rank. The smallest coweight
$\mu^\vee$ of the putative orbit under the partial order defined in equation
\eqref{po-coweights} defines a natural representative of that orbit, as all the
other coweights are by definition obtained by adding a coroot mimicking the
shift in the instanton number matching of the T-dual invariants. We denote this
orbit by $\mathcal{O}_{\mu^\vee}$, and say that it is \emph{seeded} by
$\mu^\vee$.

Therefore, if $\nu^\vee\in\mathcal{O}_{\mu^\vee}$, then
$\lambda^\vee=\mu^\vee-\nu^\vee$ must be a positive coroot:
\begin{equation}
		\nu^\vee\in\mathcal{O}_{\mu^\vee}\,, \qquad \mu^\vee = \lambda^\vee + \nu^\vee\,,\quad \lambda^\vee\in \mathcal{Q}^\vee_+\,.
\end{equation}
This shows that the T-duality orbits are not mere subsets of
$\mathcal{P}^\vee_\text{LST}$, but they have an interesting structure: the set
of admissible $\lambda^\vee$ is constrained by the matching of the T-dual
invariants, see Equation \eqref{duality-condition}. 

Let us first consider $\kappa_R$; from Equation \eqref{closed-form-kappa}, we
must have
\begin{equation}\label{modGamma}
		\kappa_R:\qquad \frac{1}{2}\left<T, \lambda^\vee\right> = \frac{1}{2}\theta^a\, C_{ab}\, \lambda^b\equiv 0 \mod \Gamma\,,\qquad 
\end{equation}
where we recall that $T = \sum_a\theta^a\omega_a$, and $C_{ab}$ is the inverse
Cartan matrix of $\mathfrak{g}$. This congruence equation defines an integer
cone $\mathcal{C}_{\Gamma}\subset \mathcal{Q}^\vee_+$ to which $\lambda^\vee$
must belong. Its generators are easily obtained by e.g. computing the Hilbert
basis through standard linear-optimization techniques for any class of LSTs.
Since $\langle\theta, P^\vee_\text{LST}\rangle\leq32$, see Classification
\ref{class:spin32}, the elements of $\mathcal{O}_{\mu^\vee}$ form a truncated
cone pointed at $\mu^\vee$.

We note that the ordering of dominant coweights is not total: there are elements
$\mu^\vee, \nu^\vee \in \mathcal{P}^\vee_+$ satisfying neither
$\nu^\vee<\mu^\vee$ nor $\mu^\vee<\nu^\vee$. This means that given a putative
T-duality orbit, there can be multiple coweights $\mu^\vee_i$ for
which there are no $\nu^\vee<\mu^\vee_i$. In other words, the orbit is seeded
by more than one coweight, and given by $\cup_i\,\mathcal{O}_{\mu^\vee_i}$.
This phenomenon is fairly common among all possible $\mathfrak{g}$; however, we
have checked explicitly by construction that the majority of putative T-dual
orbits are seeded by a single coweight $\mu^\vee$. As an example, since
$0<\nu^\vee\in\mathcal{P}^{\vee}_\text{LST}$, the orbit containing the theory
with an $\mathfrak{so}_{32}$ flavor symmetry is only seeded by the trivial
coweight.

The constraint on $\lambda^\vee$ due to $\kappa_R$ must be supplemented by also
matching $\text{CB}$ and $\text{rk}(\mathfrak{f})$. As we need to consider
the rank $\lfloor k/2\rfloor$ of $\mathfrak{so}_k$ we need to separate cases
with odd and even dimensions of flavor and gauge symmetries, which for the Coulomb
branch is given in components by: 
\begin{equation}\label{modh-rkf}
		\text{CB}:\qquad \frac{1}{2}\rho^aC_{ab}(\nu^a-\mu^a) + \sum_{\alpha\in\mathcal{P}}\left(\text{frac}(\frac{C_{\alpha a}\nu^a}{2}) - \text{frac}(\frac{C_{\alpha a}\mu^a}{2})\right) \equiv 0 \mod h^\vee\,,
\end{equation}
and similarly for the flavor rank. The appearance of floor functions therefore
adds a layer of computational subtlety, but does not substantially change the
implementation of those constraints: instead of the matching of the flavor
rank and CB cutting out a unique integer subcone of $\mathcal{C}_{\Gamma}$, it
may rather result in a union of such subcones. Generically, we therefore find
the set of dominant coweights $\mathcal{P}^\vee_\text{LST}$ associated with a
$Spin(32)/\mathbb{Z}_2$ LST has the following properties:

\begin{classification}[Structure of putative T-duality orbits]\label{class:duality-structure}
	Given a simple Lie algebra $\mathfrak{g}$ and a fixed instanton number $Q$,
	the set of dominant finite coweights $\mathcal{P}_\text{LST}^\vee$
	associated with a $Spin(32)/\mathbb{Z}_2$ LST is partitioned into disjoint
	unions of sets of coweights distinguished by their T-dual invariants
	$\kappa_R,\text{CB}, \text{rk}(\mathfrak{f})$. Each orbit is seeded by
	(possibly multiple) minimal coweight(s) $\mu^\vee_i$, and we have
	\begin{equation}
			\mathcal{P}_\text{LST}^\vee(\mathfrak{g}) = \bigcup_{\mu^\vee_i}\, \mathcal{O}_{\mu^\vee_i}\,.
	\end{equation}

	Up to a shift in the instanton number $Q$, elements of
	$\mathcal{O}_{\mu^\vee_i}$ satisfy all the known necessary conditions to be
	internal T-duals. The orbits themselves are unions of truncated integer
	cones translated to $\mu^\vee_i$ which are obtained by solving Equations
	\eqref{modGamma} and \eqref{modh-rkf}, and matching the flavor rank.
\end{classification}

We stress that the matching of the T-dual invariants is only a necessary
condition rather than a \emph{sufficient} condition, and the properties of the
coweight lattice cannot, as is, ensure that putative pairs are in fact T-dual
without an explicit geometric construction. Heterotic LSTs are particularly
amenable to toric constructions and large classes of models have been
constructed \cite{DelZotto:2020sop, DelZotto:2022ohj, DelZotto:2022xrh,
Ahmed:2023lhj, Ahmed:2024wve}. Although these methods cannot accommodate a
description of all LSTs at any instanton number, we are not aware of a single
example where two different putative T-dual LSTs---that is two theories with
the same invariants---each with an explicit geometric realization, but can be shown to be part of
two inequivalent geometries thereby ruling out T-duality. In all known
constructions, matching of the invariants will result in T-duality. This also
extends to cases where one allows twists by discrete symmetries along the
compactification circle; putative duals with the same invariants continue to
admit a description as inequivalent fibrations of $6d$ F-theory vacua
\cite{Ahmed:2024wve}.

\paragraph{T-duality and low-rank algebras $\mathfrak{g}$:} using the
closed-form expressions in terms of the coweights, we observe that for
$\mathfrak{g}=\mathfrak{so}_{8}, \mathfrak{so}_{10}, \mathfrak{e}_6$, only two
T-duality invariants are sufficient to define a duality orbit. This can be
shown explicitly by finding a linear relation between the invariants, which
follows from the numerology of the related algebras. We have furthermore
checked that one indeed only needs two invariants by explicitly scanning over
the set of dominant coweights. The same can easily be shown for
$\mathfrak{g}=\mathfrak{su}_{K}$, since $\Gamma=K=h^\vee$. Furthermore,
the orbifold $\mathbb{C}^2/\mathbb{Z}_K$ being obtained from an Abelian
discrete group, the flavor cannot be broken and all theories have maximal
flavor rank. 

While this property does not extend to other ADE algebras $\mathfrak{g}$, it
has interesting consequences for the geometry of F-theory. The T-duality
invariants in Equation \eqref{t-dual-invariants} encode certain geometric
properties of the non-compact Calabi--Yau threefold on which F-theory is
compactified to get the LSTs. For example, the flavor rank is reflected in the
number of non-compact divisors in the geometry, whereas the
$\text{dim}(\text{CB})$ encodes compact divisors. The two-group constant
$\kappa_{R}$ encodes information about both the singularity structure as well
as the intersection data of the base curves. Our observation---at least for the
algebras above---suggests a deeper non-trivial relation between these
quantities at the geometric level. It would be interesting to study this
relation in future work, as well as explaining why it seems to fail for other
singularities.

\subsection{Orbits at Maximal Rank}

While the complexity of the orbits grows as the total flavor rank decreases
from the matching of $\text{rk}(\mathfrak{f})$ and $\text{CB}$---particularly
due to floor functions---the orbits at maximal rank are particularly
simple, which we now use as an opportunity to exemplify the procedure laid out
above.

In the sequel, we will work solely at the level of the algebra $\mathfrak{g}$
rather than that of the base. Similar arguments may be given in terms of the
Lie theory of the folded algebras, but using the simply-laced case is more
straightforward as it involves a relation between the order of the ADE discrete
group $\Gamma$ and the center $Z(\mathfrak{g})$.

Consider an affine coweight $\widetilde{\mu}^\vee = (\mu^\vee, 32, 2
Q)\in\widetilde{\mathcal{P}}_\text{LST}^\vee$. Using that its level gives the
constraint $\theta_A\mu^A=32$, from the closed-form expressions we have derived
for the flavor rank in Equation \eqref{rkf-closed-form}, it is easy to show
that the flavor symmetry of any $Spin(32)/\mathbb{Z}_2$ LST must satisfy
\begin{equation}\label{flavor-constraint}
		\mu^A(\theta_A-\rho_A) \leq 32 - 2\,\text{rk}(\mathfrak{f})\,.
\end{equation}
This expression can also be used to simplify the scans of valid dominant
coweights. The inequality is saturated only at maximal flavor rank, where we
obtain
\begin{equation}
		\text{rk}(\mathfrak{f}) = 16: \qquad \mu^A(\theta_A-\rho_A) =0\,.
\end{equation}
with $A=0,1,\dots,\text{rk}(\mathfrak{g})$. Since $\rho_A=1\leq \theta_A $ for
simply-laced algebras, this equation is extremely constraining: we can only
have non-zero values of $\mu^A$ when $\theta_A=1$. In terms of the McKay
correspondence, it means we can only decompose the fundamental of $Spin(32)/\mathbb{Z}_2$ in
terms of one-dimensional representations of $\Gamma$. 

By inspection, see Table~\ref{tab:algebra-values}, we see that such
representations have two important consequences. First, the Abelianization
$\Gamma_{ab}$ of $\Gamma$---whose order is the number of one-dimensional
representations---is given by the center: $\Gamma_\text{ab} = Z(\mathfrak{g}) =
\mathcal{P}^\vee/\mathcal{Q}^\vee$. Moreover the associated flavor symmetries
are either $\mathfrak{so}_f$ or $\mathfrak{su}_f$, which means that the
corresponding $\mu_A$ are not \emph{a priori} restricted to be even.

To solve the flavor constraints, we therefore only have to consider the indices
$\alpha$ such that $\theta_\alpha=1$. From Equation \eqref{kappa-closed-form},
in order to find the possible values of $\kappa_R$ we must consider
\begin{equation}
		\frac{1}{2}\langle T, \mu^\vee\rangle = \frac{1}{2}\theta^b \, C_{b\alpha}\, \mu^\alpha = \frac{1}{2|Z|}(\theta\, \text{Ad}(C))_\alpha \,\mu^\alpha\,,
\end{equation}
where $\text{Ad}(C) = \text{det}(C)\cdot C^{-1}$ is the adjugate of the Cartan
matrix of $\mathfrak{g}$, and $\det C=|Z|$ is the order of its center
$Z(\mathfrak{g})$. The matrix $\text{Ad}(C)$ is by definition integer valued,
and it is straightforward to further show that
$\frac{1}{2}(\theta\text{Ad}(C))_\alpha\in\mathbb{N}$. 

The number of \emph{a priori} distinct duality orbits is found by solving
Equation \eqref{duality-condition}. For the matching of $\kappa_R$, we
therefore have the congruence equation:
\begin{equation}\label{congruence-kappa}
		\frac{1}{2}\langle T, \mu^\vee\rangle \equiv\, c\, \mod |\Gamma|\,,
\end{equation}
for some integer $c$. Using the integer properties of the adjugate matrix
above, one can see that to solve this equation, there must also exist an
integer $b$ related to $c$ for which $\frac{1}{2}(\theta\,\text{Ad}(C))_\alpha
\mu^\alpha \equiv b\,\mod |Z|$. Since $|Z|$ divides $|\Gamma|$, this means that
there are \emph{at most} $|Z|$ distinguished solutions to equation
\eqref{congruence-kappa}. The \emph{actual} number of solutions is obtained by
relating the integers $b$ to $c$. A case-by-case analysis shows that the number
of putative orbits at maximal flavor rank is indeed given by the order of the
center, except for $\mathfrak{g}=\mathfrak{so}_{4k}$, where there are only two
orbits, rather than four.

A more Lie-algebraic explanation for this fact is as follows. By considering
only $\mu^\alpha$ for which $\theta_\alpha=1$, we only take into account the
generators of the center $Z$, and the congruence equation
\eqref{congruence-kappa} can be thought of as a map
$Z(\mathfrak{g})\to\mathbb{Z}_{|\Gamma|}$. Since $|Z|$ always divides
$|\Gamma|$, the map is straightforward when the center is cyclic. This is the
case for every algebra except for
$Z(\mathfrak{so}_{4n})=\mathbb{Z}_2\times\mathbb{Z}_2$ where the embedding is
made only through its diagonal subgroup $\mathbb{Z}_2$.

The same reasoning applies for the matching of the Coulomb branch, where the
condition is the same after exchanging $\Gamma$ with $h^\vee$. It is then
straightforward to see that anomaly cancellation in that case only leads to
even fundamental representations of the gauge symmetry, so that there are no
fractional parts involved, and we are led to the same conclusion as above,
namely:\\

\begin{classification}[Number of putative orbits at maximal flavor rank]\label{class:orbits}
	For any $Spin(32)/\mathbb{Z}_2$ LST of type $\mathfrak{g}$, the number
	of orbits at maximal flavor rank $\text{rk}(\mathfrak{f})$ is given by
	the order of the center $Z=\mathcal{P}^\vee/\mathcal{Q}^\vee$ related
	to $\mathfrak{g}$, except in the case
	$\mathfrak{g}=\mathfrak{so}_{4k}$, where it is given by the order of
	the diagonal subgroup
	$\mathbb{Z}_2\subset\mathbb{Z}_2\times\mathbb{Z}_2$ instead.
\end{classification}

\paragraph{Example $\fg=\fe_6$ at flavor rank 16:} let us illustrate this result for
$\mathfrak{g}=\mathfrak{e}_6$. From Table \ref{tab:algebra-values}, we see that we
have only three nodes of the affine Dynkin diagram of
$\widetilde{\mathfrak{g}} = E_6^{(1)}$ with $\theta_A=1$. Two of the
one-dimensional representations of $\Gamma$ are conjugate, and we must set
$\mu^5=\mu^1$. The finite dominant coweight is therefore constrained to be of
the form $\mu^\vee = \mu^1 (\omega_1^\vee +
\omega_5^\vee)\in\mathcal{P}^\vee_\text{LST}$, and the congruence equation is
given by
\begin{equation}
		\mathfrak{g}=\mathfrak{e}_6:\qquad\frac{1}{2}\langle T, \mu^\vee\rangle = 16\, \mu^1 \equiv c \mod 24\,.
\end{equation}
The representatives of the orbits are defined by $\mu^1=0,1,2$, which matches
the order of $Z = \mathbb{Z}_3$. We can further see that the integer cones are
simply generated by $\mathcal{C}_\Gamma = \text{span}(3\,\omega^\vee_1)$. After
folding, the flavor is located on the last node with $f^4= \mu^1 + 3n$, and the
affine node is constrained to be $f^0 = 32 - 2(\mu^1 + 3n)$ so that the level
of the affine coweight satisfies $\theta_A\mu^A = 32$; this gives an upper
bound on $n$. 

The ranks of the gauge symmetry are then fixed by anomaly cancellation,
enabling us to indeed find three orbits. From smallest to largest with respect
to the partial order, the quivers are given by:
\begin{equation}
	\begin{gathered}
			\mathcal{O}_{0\cdot(\omega^\vee_1+\omega_5^\vee)}:\qquad
			\underset{[\mathfrak{so}_{32}]}{\overset{\mathfrak{sp}_{Q}}{1}}\,\overset{\mathfrak{so}_{4Q - 16}}{4}\,\overset{\mathfrak{sp}_{3Q - 24}}{1}\,\overset{\mathfrak{su}_{4Q - 32}}{2}\,\overset{\mathfrak{su}_{2Q - 16}}{2}\,\qquad\dots\qquad
    \underset{[\mathfrak{so}_{2}]}{\overset{\mathfrak{sp}_{Q}}{1}}\,\overset{\mathfrak{so}_{4Q + 14}}{4}\,\overset{\mathfrak{sp}_{3Q + 6}}{1}\,\overset{\mathfrak{su}_{4Q + 13}}{2}\,\underset{[\mathfrak{su}_{15}]}{\overset{\mathfrak{su}_{2Q + 14}}{2}}\\
			\mathcal{O}_{1\cdot(\omega^\vee_1+\omega_5^\vee)}:\qquad
	\underset{[\mathfrak{so}_{30}]}{\overset{\mathfrak{sp}_{Q}}{1}}\,\overset{\mathfrak{so}_{4Q - 14}}{4}\,\overset{\mathfrak{sp}_{3Q - 22}}{1}\,\overset{\mathfrak{su}_{4Q - 29}}{2}\,\underset{[\mathfrak{su}_{1}]}{\overset{\mathfrak{su}_{2Q - 14}}{2}}\qquad\dots\qquad
	\overset{\mathfrak{sp}_{Q}}{1}\,\overset{\mathfrak{so}_{4Q + 16}}{4}\,\overset{\mathfrak{sp}_{3Q + 8}}{1}\,\overset{\mathfrak{su}_{4Q + 16}}{2}\,\underset{[\mathfrak{su}_{16}]}{\overset{\mathfrak{su}_{2Q + 16}}{2}}\\
	\mathcal{O}_{2\cdot(\omega^\vee_1+\omega_5^\vee)}:\qquad
    \underset{[\mathfrak{so}_{28}]}{\overset{\mathfrak{sp}_{Q}}{1}}\,\overset{\mathfrak{so}_{4Q - 12}}{4}\,\overset{\mathfrak{sp}_{3Q - 20}}{1}\,\overset{\mathfrak{su}_{4Q - 26}}{2}\,\underset{[\mathfrak{su}_{2}]}{\overset{\mathfrak{su}_{2Q - 12}}{2}}\qquad\dots\qquad
    \underset{[\mathfrak{so}_{4}]}{\overset{\mathfrak{sp}_{Q}}{1}}\,\overset{\mathfrak{so}_{4Q + 12}}{4}\,\overset{\mathfrak{sp}_{3Q + 4}}{1}\,\overset{\mathfrak{su}_{4Q + 10}}{2}\,\underset{[\mathfrak{su}_{14}]}{\overset{\mathfrak{su}_{2Q + 12}}{2}}
	\end{gathered}
\end{equation}
Recall from the discussion around Equation \eqref{ABJ-I8} that for each
non-trivial $\mathfrak{su}_f$ flavor symmetry, there is an additional
$\mathfrak{u}_1$ factor possibly delocalized over all complex hypermultiplets,
and all LSTs have $\text{rk}(\mathfrak{f})=16$. Furthermore, with the bound on
$n$ above, one finds that each orbit contains
$\{|\mathcal{O}_{i\cdot(\omega^\vee_1+\omega_5^\vee)}|, i=0,1,2\}=\{6,6,5\}$
LSTs that are putatively T-dual.

\paragraph{Lower-rank orbits:} as we have already argued above, solving the
congruence constraints for $\text{rk}(\mathfrak{f})<16$ is more involved, both
because the flavor relation in Equation \eqref{flavor-constraint} admits more
solutions, and because we need to treat $\text{dim}(\text{CB})$ and $\kappa_R$
separately due to floor functions.

In particular, we have mainly relied on the fact that $C^{-1} =
|Z|^{-1}\text{Ad}(C)$. One could therefore naively expect that at lower flavor
rank, the number of orbits can be predicted by the possible values of
$\kappa_R~\mod\Gamma$ and is again of the order of the center, compounded by the
different values that the Coulomb branch might take. However, the correct
number of orbits is difficult to predict. First, as it might be that instead of
being counted by the cyclic group $\mathbb{Z}_{|Z|}$, the numerology forces it
to be a smaller subgroup. Furthermore, a particular value of $\kappa_R$ might
correspond to two different values of the Coulomb-branch dimension. The
constraints are however easily implemented on a computer to efficiently obtain
the number of orbits for a specific algebra $\mathfrak{g}$ at any flavor rank. 

We have performed such a scan for all type-DE LSTs up to $r_{\mathfrak{g}}=8$.
At maximal flavor rank, one can see from Table \ref{tab:nbr-orbits} that we
indeed have a number of orbits corresponding to the order of the center, except
for $\mathfrak{g}=\mathfrak{so}_8$, which only has two for the reason
shown above. Note that as the dominant affine coweights realizing an LST must
have level 32, the minimal flavor rank is constrained, and depends on the
largest value the mark $\theta_A$ can take. In particular, no
$Spin(32)/\mathbb{Z}_2$ LST can be flavorless. This is in contradistinction
to LSTs of $\mathfrak{e}_8\oplus\mathfrak{e}_8$ heterotic type, where there
are theories with low-rank flavor symmetry, and even no flavor
\cite{Frey:2018vpw}. In those cases, we expect the fiber-base dual theories to
be associated with twisted compactifications. 

Furthermore, due to their low rank, LSTs of type $\mathfrak{g} =
\mathfrak{so}_8\,,\mathfrak{so}_{10}$ have an interesting structure as the
number of orbits is the same at all flavor ranks, with the exception of very low
flavor where almost no models are available. In general, we observe that the
number of orbits monotonically grows as the flavor rank decreases, possibly
plateaus, and then monotonically decreases. While this pattern continues for
$r_{\mathfrak{so}_{2k}}>8$, we however do not have an explanation, either Lie
theoretic or geometric, for this behavior. We leave a more detailed study for
future work.

\begin{table}
\centering
\begin{tabular}{|c||c|c|c|c|c||c|c|c|}\hline
$\text{rk}(\mathfrak{f})$ & $\fso_8$ & $\fso_{10}$ & $\fso_{12}$ & $\fso_{14}$ & $\fso_{16}$ & $\fe_6$ & $\fe_7$ & $\fe_8$ \\ \hline
16 & 2 & 4 & 2  & 4   & 2  & 3   & 2  & 1   \\
15 & 2 & 4 & 4  & 8   & 6  & 3   & 4  & 2   \\
14 & 2 & 4 & 6  & 12  & 12 & 6   & 8  & 5   \\
13 & 2 & 4 & 8  & 16  & 20  & 6   & 14 & 10  \\
12 & 2 & 4 & 10 & 20  & 28  & 9   & 24 & 22  \\
11 & 2 & 4 & 12 & 24  & 36  & 9   & 36 & 40  \\
10 & 2 & 4 & 14 & 28  & 44  & 12  & 50 & 73 \\
9  & 2 & 4 & 16 & 32  & 52  & 12  & 66 & 121 \\
8  & 1 & 4 & 17 & 36  & 59  & 12  & 83 & 190 \\
7  & 1 & 4 & 15  & 40  & 62  & 12  & 97 & 286 \\
6  & ---  & ---& ---  & ---   & 20  & 11  & 96 & 388 \\
5  & ---  & ---& ---  & ---   & ---  &  2  & 75 & 440 \\
4  &  --- & ---& ---  & ---   & ---  &  ---  & 27 & 396 \\
3  & ---  & ---& ---  & ---   & ---  & ---   & 2 & 191 \\
2  & ---  & ---&---   & ---   & ---  & ---   & --- &   22  \\
1  & ---  & ---& ---  & ---   & ---  &  ---  &  ---  &  ---   \\
0  & ---  & ---& ---  & ---   & ---  & ---   &  ---  &  ---   \\ \hline
\end{tabular}
\caption{\label{tab:nbr-orbits}Number of putative duality orbits per flavor rank for type-DE $\fso_{32}$ LSTs of rank up to eight.}
\end{table}

\subsection{Example: Duality Orbits of Heterotic LSTs on $\fg=\fso_{8}$}\label{sec:example-so8}

In the following we give a complete classification of the T-duality orbits of
the heterotic string appearing in Table \ref{tab:nbr-orbits} for
$\mathfrak{g}=\mathfrak{so}_8$. These theories have the convenient property that the
number of duality orbits is at most two. Moreover we will show that this
property is the same for all other heterotic LSTs and that all such theories
fall into the same two possible duality orbits at each flavor rank.

\paragraph{The \texorpdfstring{$Spin(32)/\mathbb{Z}_2$}{Spin(32)/Z2} theory:} for theories with
$\mathfrak{g}=\mathfrak{so}_8$, the base is
$\widetilde{\mathfrak{b}}=D_4^{(1)}$. It is then straightforward to show that
the anomaly constraints in Equation \eqref{quartic-cancellation} force the
quiver to take the simple form
\begin{equation}
		\overset{\displaystyle  \overset{[\mathfrak{so}_{a+4m_0}]}{\overset{\mathfrak{sp}_{Q}}{1}}}{\underset{\displaystyle \underset{[\mathfrak{so}_{a+4m_1}]}{\overset{\mathfrak{sp}_{Q-m_0+m_1}}{1}}}{\vphantom{\underset{[\mathfrak{so}_1]}{\overset{\mathfrak{so}_1}{4}}}}}\underset{[\mathfrak{sp}_{8-a-N}]}{\overset{\mathfrak{so}_{4Q+16-a-4m_0}}{4}}\overset{\displaystyle \overset{[\mathfrak{so}_{a+4m_3}]}{\overset{\mathfrak{sp}_{Q-m_0+m_3}}{1}}}{\underset{\displaystyle \underset{[\mathfrak{so}_{a+4m_4}]}{\overset{\mathfrak{sp}_{Q-m_0+m_4}}{1}}}{\vphantom{\underset{[\mathfrak{so}_1]}{\overset{\mathfrak{so}_1}{4}}}}} \qquad \qquad 
		\begin{aligned}
				\text{rk}(\ff) ~=&~ 8+4\,\lfloor\frac{a}{2}\rfloor+ N - a\,;\\
		\text{dim}(\text{CB}) ~=& ~6\,Q+ 12+N-6\,m_0 - \lceil\frac{a}{2}\rceil\,;\\
		\kappa_R ~= &~8\,Q + 18+N-8\,m_0-a\,,
		\end{aligned}
\end{equation}
with integers $a = 0\ldots 3$ and $m_i$ such that $N=\sum_{i} m_i$, and must
satisfy $a+N \leq 8$. The coefficients of the affine coweight $\mu^A\in\{a+m_0,
\dots, a+4m_4\}$ can therefore be obtained from a 4-partition of $N$. Note that the
quiver admits an $S_4$ symmetry permuting the $(-1)$-curves and one must take
care of not double-counting equivalent quivers. As the T-duality invariants
only depend on $N$ and the value $m_0$ on the affine node, this does not affect
the number of putative T-duality orbits.

As we can be very explicit in that case, we find at most two duality orbits at each
rank, labeled only by $\tilde{a}=\lfloor \frac{a}{2}\rfloor=0, 1$ and the
flavor rank $\text{rk}(\mathfrak{f})$. This shows that with
$\mathfrak{g}=\mathfrak{so}_8$, an orbit is only characterized by two T-duality
invariants. For ease of reading we will denote an orbit via this data rather
than the coweight $\mu^\vee$ seeding it:
$\mathcal{O}_{\mu^\vee}=\mathcal{O}_{\tilde{a}, \text{rk}(\mathfrak{f})}$. The other two invariants are then given by:
\begin{align}
	\label{eq:so8orbits}
	\mathcal{O}_{\tilde{a}, \text{rk}(\mathfrak{f})}:\qquad 
	\begin{pmatrix}
			\kappa_R\\\text{dim}(\text{CB}) 
	\end{pmatrix} = \text{rk}(\mathfrak{f}) + \begin{pmatrix}
			\;2\;\\4
	\end{pmatrix} + \tilde{a}
	\begin{pmatrix}
			4\\-3
	\end{pmatrix}\mod \begin{pmatrix}
			8\\6
	\end{pmatrix}\,,
\end{align}
where the congruence is in terms of $(\Gamma,h^\vee)=(8, 6)$ takes into account
the need for possible shifts in the number of five-branes $Q$ so as to
reproduce the constraints in Equation~\eqref{duality-condition}. We therefore
clearly see that there can only be at most two putative T-duality orbits at
each flavor rank labeled by $\tilde{a}=0, 1$.

Anticipating a discussion in Section~\ref{sec:extension}, these
$Spin(32)/\mathbb{Z}_2$ theories can be dual to another type of $\mathfrak{so}_{32}$
heterotic theory corresponding to an example of internal duality, as well as
fiber-base duality to LSTs arising from compactifications of the
$\mathfrak{e}_8\oplus\mathfrak{e}_8$ heterotic string.

\paragraph{The $\mathbf{(\text{SU(16)} \times \text{U(1)})/\mathbb{Z}_2}$
theory:} the first case corresponds to LSTs whose base is described by the twisted
affine algebra $D_3^{(2)}$, see Section~\ref{sec:su16-lst}. After taking into
account cancellations of quartic anomalies, it is constrained to take the form:
\begin{align}
		[\mathfrak{u}_{4n+\tilde{a}}] \overset{\fsu_{2Q}}{2} \,\overset{\fsp_{2Q-2n-\tilde{a}}} {\underset{[\displaystyle \fso_{2b} ]}{1}} \,  \overset{\fsu_{2Q-4n-2\tilde{a}-b+8}}{2} [\mathfrak{u}_{16-(4n+a)-2b}]
\qquad \text{rk}(\mathfrak{f})=16-b \,.
\end{align}
The parameter $\tilde{a}=0,1$ can again take on exactly two values, and determines
the T-duality orbits $\mathcal{O}_{\tilde{a}, \text{rk}(\mathfrak{f})}$ the
theory belongs. It is easy to see that the flavor rank is bounded by
$\text{rk}(\ff)\geq8$. For this class of theories, we therefore find the same formulas
as in Equation \eqref{eq:so8orbits} and precisely the same structure for
putative T-duality orbits as for $Spin(32)/\mathbb{Z}_2$ theories.

\paragraph{The $\mathbf{\text{E}_8 \times \text{E}_8}$ theory:} the last class
of T-duality that can occur at arbitrary number of instantons $Q$ is
``fiber-base'' duality between the $\mathfrak{so}_8$ $Spin(32)/\mathbb{Z}_2$
LSTs and those of $\mathfrak{e}_8\oplus \mathfrak{e}_8$ type. As briefly
reviewed in Section \ref{ssec:e8e8theories}, these theories are classified by a
pair of embeddings of $\Gamma_{\mathfrak{so}_8}$ into the group $E_8$: $\lambda_L, \lambda_R
\in\text{Hom}(\Gamma_{\mathfrak{so}_8}, E_8)$. Individually, they can be understood as
independent breaking of the $\mathfrak{e}_8$ factors of two SCFTs called
orbi-instantons, which have $\mathfrak{e}_8\oplus\mathfrak{so}_8$ flavor
symmetry \cite{Heckman:2018pqx}. The list of quivers for these SCFTs and all
possible choices $\lambda\in\text{Hom}(\Gamma_{\mathfrak{so}_8},E_8)$ can be found in reference
\cite{Frey:2018vpw}.

Interestingly, the contribution $\delta \kappa_R(\lambda),
\delta\text{CB}(\lambda)$ from each SCFT to the T-duality invariants of
$\mathfrak{e}_8\oplus\mathfrak{e}_8$ LSTs can be computed independently.
Furthermore, even though the SCFTs do not exhibit a two-group symmetry,
$\delta\kappa_R(\lambda)$ can still be obtained directly at the level of the
orbi-instanton since it descends from the anomaly polynomial. The key
observation is that, as for the other two classes of LSTs with
$\mathfrak{g}=\mathfrak{so}_8$, there are only two types for a
given SCFT, labeled by an integer $\tilde{a}=0,1$ and the rank of the flavor
symmetry $\mathfrak{f}$ associated with a choice of embedding
$\lambda\in\text{Hom}(\Gamma_{\mathfrak{so}_8}, E_8)$:
\begin{equation}
	\begin{pmatrix}
			\delta\kappa_R(\lambda)\\
			\delta\text{CB}(\lambda)
	\end{pmatrix} = 
	\text{rk}(\mathfrak{f}) + \begin{pmatrix}
			\;0\;\\2
	\end{pmatrix} + \tilde{a}
	\begin{pmatrix}
			\;3\;\\4
	\end{pmatrix}\mod \begin{pmatrix}
			8\\6
	\end{pmatrix}\,,
\end{equation}
In Table \ref{tab:OIorbits}, we have collated all nineteen orbi-instanton
theories and the value of $\tilde{a}$ they give rise to. We see that the
minimal flavor rank of those SCFTs is four, which in turn means that at the
level of the $\mathfrak{e}_8\oplus\mathfrak{e}_8$ LST, the minimal flavor rank
is eight. Interestingly, as opposed to what occurred on the
$\mathfrak{so}_{32}$ side of the duality, there cannot be any LST with flavor
rank seven if there is no twist along the compactification circle.

\begin{table}[t]
\begin{align*}
    \begin{array}{|c|c|c|}\hline
			\text{rk}(\ff)& \text{flavors with $\tilde{a}=0$}    &  \text{flavors with $\tilde{a}=1$} \\ \hline  8 & \fe_8, \fe_7\oplus \fsu_2, \fso_{16}, \fso_8^2, 
 \fso_{12} \oplus \fso_4& \fsu_8 \oplus \mathfrak{u}_1, \fe_6 \oplus \mathfrak{u}_1^2 \\ \hline
 7&  \fe_7,\fso_{12}\oplus \fsu_2 , \fso_8 \oplus \fsu_2^3   & \fsu_6 \oplus \mathfrak{u}_1^2 \\ \hline
6&   \fsp_2^3, \fso_9 \oplus \fsp_2, \fso_{13} & \fso_7 \oplus \fsu_2^3 \\ \hline
5&   \ff_4 \oplus \fsu_2, \fsp_4\oplus \fsu_2 & \fso_8 \oplus \fsu_2 \\ \hline
4&  \fsp_4 &\text{---} \\ \hline
\end{array}
\end{align*}
\caption{\label{tab:OIorbits}
		All nineteen $\fg=\fso_8$ orbi-instanton theories \cite{Frey:2018vpw}
		identified by their flavor ranks $\ff$ and the coefficient
		$\tilde{a}=0,1$ that defines the duality orbit of the associated
		$\mathfrak{e}_8\oplus\mathfrak{e}_8$ heterotic LST.
}
\end{table}

Starting with two orbi-instanton SCFTs of type $\mathfrak{so}_8$ whose quivers are fixed by choices
$\lambda_L, \lambda_R\in\text{Hom}(\Gamma_{\mathfrak{so}_8}, E_8)$, it is straightforward to show
that, taking into account the contribution of the vector multiplet used to
gauge the common $\mathfrak{so}_8$ flavor symmetry to produce an LST, we have
\begin{equation}
		\mathcal{O}_{\tilde{a}_\text{tot},\text{rk}(\mathfrak{f}_L\oplus\mathfrak{f}_R)}: 
\begin{pmatrix}
		\kappa_R(\lambda_L, \lambda_R)\\
		\text{CB}(\lambda_L, \lambda_R)
\end{pmatrix} =
\begin{pmatrix}
		\delta\kappa_R(\lambda_L)\\
		\delta\text{CB}(\lambda_L)
\end{pmatrix} + 
\begin{pmatrix}
		\delta\kappa_R(\lambda_R)\\
		\delta\text{CB}(\lambda_R)
\end{pmatrix} + 
\begin{pmatrix}
		\;6\;\\4
	\end{pmatrix}\mod \begin{pmatrix}
			8\\6
	\end{pmatrix}\,,
\end{equation}
with $\tilde{a}_\text{tot}=\tilde{a}_L + \tilde{a}_R\mod 2$. Therefore, while we have \emph{a
priori} more freedom in $\mathfrak{e}_8\oplus\mathfrak{e}_8$ heterotic LSTs as
we may choose two independent embeddings rather than one, in the end we once
again find only two duality orbits. 

Finally we remark that the above considerations can also be extended to twisted
compactifications following reference \cite{Ahmed:2024wve}. Although beyond the
scope of this work, it can be shown that all $\fso_8$ twisted heterotic
theories fall into at most two duality orbits as in untwisted cases. This hints
at a much deeper structure that the duality orbits may have even when including
possible discrete symmetries.

\section{Higgs Branches and the Partial Order of Coweights}\label{sec:higgs}

With a concise way of encoding many properties of $Spin(32)/\mathbb{Z}_2$ LSTs
in terms of dominant affine coweights, one may ask what can be learned about
how they are related using the partial order defined in equation
\eqref{po-coweights}. As anticipated above, it can be interpreted as a
consistent change of the gauge symmetries between two LSTs and is related to
Higgs-branch Renormalization Group (RG) flows.

Concretely, let us consider two LSTs with affine coweights
$\widetilde{\mu}^\vee = (\mu^\vee, 32, 2Q)$ and $\widetilde{\nu}^\vee =
(\nu^\vee, 32, 2Q)$ of a fixed base algebra $\widetilde{\mathfrak{b}}$, in the
notation of Equation \eqref{triplet-coweight}. We assume for now that $Q$ is
set to a fixed value for simplicity. If the two LSTs are related by a
consistent RG flow, then the dimension $d_I = \langle \widetilde{\omega}_I,
\widetilde{\mu}^\vee - 16 \widetilde{S}^\vee \rangle $ of the fundamental
representation of each gauge symmetry can only change by an integer. As we have
seen around Classification \ref{class:spin32}, in terms of the affine root
system it means that the difference $\mu^\vee-\nu^\vee$ is a positive coroot. 

For any Higgs-branch RG flow to take place, the two coweights
must therefore be ordered:
\begin{equation}\label{RG-po}
		\widetilde{\mathcal{K}}_Q(\mu^\vee; \mathfrak{g}) 
		\quad\xrightarrow{\text{RG}}\quad
		\widetilde{\mathcal{K}}_Q(\nu^\vee; \mathfrak{g}) 
		\qquad\Rightarrow\qquad \nu^\vee < \mu^\vee\,.
\end{equation}

While it is true that the partial order of dominant coweights is related to
Higgs-branch flows, there are important caveats. First, for low values of $Q$,
symmetry enhancements may occur. For instance, an undecorated $(-1)$-curve is
well known to support an $\mathfrak{e}_8$ flavor symmetry rather than a
classical algebra, which is not captured by Classification
\ref{class:spin32}---at least not in a direct way. Worse, deep inside the Higgs
branch, one ultimately encounters LSTs described by zero curves
$\overset{\mathfrak{g}}0$, which have been shown to have a very rich
Higgs-branch structure \cite{Mansi:2023faa}. There are also Higgs-branch
deformations that can change the curve configuration of the base---or even lead
to product theories \cite{Baume:2021qho, Lawrie:2023uiu,
Bourget:2024mgn,Lawrie:2024zon, Lawrie:2024wan}---which also cannot be captured
by the partial order above. While certain formulas can sometimes be
``analytically continued'' to cover such cases \cite{Mekareeya:2017sqh,
Heckman:2018pqx}, they are beyond the scope of this work.

To avoid such subtleties, we will restrict ourselves to theories where those
cases do not arise. In particular, we will always assume $Q$ to be ``large
enough''---that is arbitrary, but large. The precise numerology depends on the
type of theory, but in practice one usually requires $Q \gtrsim
h^\vee_{\mathfrak{g}}$. In the sequel, we will therefore only consider flows
between LSTs whose quivers are described by affine coweights and whose flavor
and gauge symmetry follows directly from them.

The Higgs branch of theories with at least eight supercharges---which include
$6d$ $\mathcal{N}=(1,0)$ SQFTs---can sometimes be studied via a three-dimensional
auxiliary quiver gauge theory. The strategy is to find a so-called magnetic
quiver: a $3d\, \mathcal{N}=4$ theory whose Coulomb branch is the same as the
Higgs branch of the higher-dimensional theory. This has led to advances in our
understanding of the structure of the Higgs branch of six-dimensional theories,
and has recently been applied to a host of cases, such as LSTs on a zero-curve
\cite{Mansi:2023faa}, SCFTs with classical symmetries \cite{Hanany:2018uhm,
Cabrera:2019izd, Cabrera:2019dob, Lawrie:2025exx, Bourget:2019aer}, as well as certain classes
of SCFTs and heterotic LSTs with $\mathfrak{g}=\mathfrak{su}_K$ that can be
built by ``fusing'' these SCFTs together \cite{Mekareeya:2017jgc,
Fazzi:2022yca,Fazzi:2022hal, DelZotto:2023myd, Fazzi:2023ulb, Lawrie:2023uiu,
Bourget:2024mgn, Giacomelli:2024ycb, Giacomelli:2025zqn}. 

For those SCFTs that can be constructed via
$\mathfrak{e}_8\oplus\mathfrak{e}_8$ heterotic string theory with
$\mathfrak{g}=\mathfrak{su}_K$, it has been shown that they are classified by
dominant affine coweights of $E_8^{(1)}$ at level $K$ \cite{Mekareeya:2017jgc,
Fazzi:2023ulb}. The partial order of coweights then reproduces the network
of Higgs-branch RG flows of the SCFTs, which is encoded in the geometry of a
group-theoretic object referred to as the (double) affine Grassmannian
\cite{Bourget:2021siw, Fazzi:2023ulb}.

Although we will not delve into the mathematical minutiae of affine
Grassmannians, we will argue that the partial order of coweights arising for
$Spin(32)/\mathbb{Z}_2$ LSTs encodes at least part of the Higgs branch. Our
goal in this section is to obtain an algorithm that, starting with any
$Spin(32)/\mathbb{Z}_2$ LST, obtains all other possible LSTs that can in
principle be reached via a Higgs-branch RG flow. 

Studying the partial order of these theories is in some sense an easier task
than was done for SCFTs arising from $\mathfrak{e}_8\oplus\mathfrak{e}_8$
heterotic string theory: going from an affine coweight of $E_8^{(1)}$ to
the six-dimensional generalized quiver is not as direct as in our case
\cite{Fazzi:2023ulb, Mekareeya:2017jgc}. On the other hand, for
$Spin(32)/\mathbb{Z}_2$ LSTs, all types of classical algebras may arise as
gauge or flavor symmetries, while affine-Grassmannian techniques are typically
studied only for unitary quivers \cite{Mekareeya:2017jgc, Fazzi:2023ulb,
Bourget:2024mgn}. We will nonetheless find that, despite needing to deal with
all classical algebras, there is only a reasonable number of inequivalent
minimal transitions that can occur between dominant coweights. The number of
LSTs for a given algebra being finite, these are straightforward to identify.

To make our task easier, the closed-form formulas we have found in the previous
section reveal themselves to be useful. In particular, they enable us to easily
find how the T-duality invariants change between two theories. Indeed, for $Q$
kept fixed---which we will assume---we have
\begin{equation}\label{RG-qt-change}
		\begin{aligned}
				\Delta\kappa_R &= \frac{1}{2}\left<T, \mu^\vee - \nu^\vee\right>\\
			\Delta\text{CB} &= \left<\rho, \mu^\vee - \nu^\vee\right> + \sum_{\alpha\in\mathcal{P}} 
			\left(
				\text{frac}(\frac{C_{\alpha a} \mu^a}{2})- \text{frac}(\frac{C_{\alpha a} \nu^a}{2})
			\right)\\
			\Delta\text{HB} &= \frac{1}{2}(\mu^\vee, \mu^\vee) - \frac{1}{2}(\nu^\vee, \nu^\vee) - (\mu^\vee - \nu^\vee, S)\,.\\
		\end{aligned}
\end{equation}
All pairings are taken over the finite algebra $\mathfrak{g}$, and we defined
$(\mu^\vee, \nu^\vee) = \mu^a C_{ab} \nu^b$ as the natural scalar product on the
coweight lattice, with $C_{ab}$ the inverse Cartan matrix, see equation
\eqref{def-symm}.

\subsection{Slice Subtraction}\label{sec:slice-subtraction}

An algorithm referred to as \emph{slice subtraction} has recently been devised
for a large class of six-dimensional SCFTs \cite{Lawrie:2024zon}. It enables one to
reconstruct the complete Hasse diagram encoding the network of Higgs-branch
flows of an SCFT. Even though its validity was proved
using similar quiver-subtraction algorithms for 3d $\mathcal{N}=4$ theories
\cite{Cabrera:2018ann, Cabrera:2019dob, Bourget:2022ehw} using their magnetic quivers,
the algorithm is performed directly at the level of the six-dimensional
generalized quiver. It therefore offers a very elegant way of obtaining all
possible theories that can be obtained by Higgs-branch RG flows from a given
SCFT.

Given an SCFT $\mathcal{K}$, one can reach a new theory $\mathcal{K}'$ through a
subtraction operation. One finds that there is a single \emph{minimal
transverse slice} (or slice, for brevity) $\mathcal{S}$ such that $\mathcal{K}'
= \mathcal{K} - \mathcal{S}$. This procedure does not require knowing the
magnetic quiver, and encodes the field theory data that are involved in the RG
flow, such as which operator is given a vacuum expectation value, or which
moduli are turned on when a brane picture is available. Once all allowed
minimal slices $\mathcal{S}$ have been found, the complete Hasse diagram can be
recreated by successive subtractions. In three dimensions the slices have been
shown to correspond to singular symplectic varieties, and have been used to
study the Higgs mechanism for these theories. In this work, we will not explore
their geometric properties, but rather simply use them as a tool encoding how
the generalized quiver of a given LST must change to obtain another consistent
theory. We instead refer to reference \cite{Cabrera:2016vvv} for a mathematical
introduction to the topic.

Although not quite phrased in those terms, the authors of reference
\cite{Lawrie:2024zon} relied on the properties of the partial order of
dominant coweights discussed above to determine the slice-subtraction
algorithm. It is therefore very tempting to try to generalize it to LSTs
admitting a perturbative description, as we have a straightforward way to
construct the set of all dominant coweights
$\widetilde{\mathcal{P}}^\vee_\text{LST}$ corresponding to consistent LSTs.

In the following, we will work for convenience with a fixed value of $Q$ and
directly at the level of the finite part of the algebra of the base
$\mathfrak{b}$ rather than the algebra $\mathfrak{g}$ associated with the
orbifold singularity. Before presenting the general algorithm, let us first
consider a specific example to illustrate the procedure: consider the two
dominant coweights $\mu^\vee = [0,0,0,2]$ and $\mu^\vee=[2,0,2,0]$ of
$\mathfrak{b} = \mathfrak{so}_9$. It is possible to show that there
are no other LSTs with dominant coweight $\lambda^\vee$ such that
$\mu^\vee<\lambda^\vee<\nu^\vee$, and the transition between the two must
therefore be minimal. In terms of generalized quivers, we have:
\begin{equation}
\underset{\textcolor{red}{[\mathfrak{so}_{2}]}}{\overset{\textcolor{blue}{\mathfrak{sp}_{Q - 6}}}{1}}\, \overset{\displaystyle \overset{[\mathfrak{so}_{26}]}{\overset{\mathfrak{sp}_{Q}}{1}}}{\overset{\textcolor{red}{\mathfrak{so}_{4Q - 10}}}{4}}\, \underset{\textcolor{red}{[\mathfrak{so}_{2}]}}{\overset{\textcolor{blue}{\mathfrak{sp}_{2Q - 12}}}{1}}\,\overset{\mathfrak{su}_{2Q - 12}}{2}\qquad
\overset{\resizebox{1.5cm}{!}{\input{figures/B4_example.tex}}}{\relbar\joinrel\relbar\joinrel\relbar\joinrel\longrightarrow}\qquad
\overset{\textcolor{blue}{\mathfrak{sp}_{Q - 7}}}{1}\, \overset{\displaystyle \overset{[\mathfrak{so}_{28}]}{\overset{\mathfrak{sp}_{Q}}{1}}}{\overset{\textcolor{red}{\mathfrak{so}_{4Q - 12}}}{4}}\, \overset{\textcolor{blue}{\mathfrak{sp}_{2Q - 13}}}{1}\,\underset{[\mathfrak{su}_{2}]}{\overset{\mathfrak{su}_{2Q - 12}}{2}}
\end{equation}
The curve corresponding to the affine node is the one with
$\overset{\mathfrak{sp}_Q}{1}$. For ease of reading, we have colored the
algebras that change, with $\textcolor{blue}{\mathfrak{sp}_k}$ algebras in blue
and $\textcolor{red}{\mathfrak{so}_k}$ in red, following the notation of table
\ref{tbl:node-color} as well as the standard conventions of $3d$
$\mathcal{N}=4$ quiver gauge theories. Over the arrow, we have indicated by how
much the dimensions of a fundamental representation $d_I$ and $f^I$ of either
gauge and flavor symmetries change, omitting those that remain the same. To
make it resemble a $3d$ $\mathcal{N}=4$ quiver and following the notation
introduced in reference \cite{Lawrie:2024zon}, we only show $d_I$ and $f^I$.
The quiver encoding the change of spectrum between the two LSTs is the minimal
slice $\mathcal{S}$ discussed above:
\begin{equation}
		\mathcal{S}: \qquad 
		\begin{matrix}{\input{figures/B4_example.tex}}\end{matrix}\,.
\end{equation}
In the following we will only denote the dimension of the fundamental
representations. We therefore see that as advertised, starting from the LSTs
with coweight $\mu^\vee$, the subtraction operation $\mathcal{K}-\mathcal{S}$
is simply given by changing the dimensions $d_I$ and $f^I$ by an amount given
by those of $\mathcal{S}$. Note that if we simply perform the subtraction, we
obtain a quiver with flavor symmetries that leads to a gauge-anomalous
spectrum. This is a standard feature of subtraction algorithms
\cite{Cabrera:2018ann, Cabrera:2019dob, Bourget:2022ehw}: after subtracting
$\mathcal{S}$ from the original quiver, the flavor symmetry must be
``rebalanced'' according to Equation \eqref{quartic-cancellation}.

Interpreting $\mathcal{S}$ as a $3d$ $\mathcal{N}=4$ quiver, it can be shown
that this RG flow corresponds to a so-called Kraft--Procesi transition
\cite{Cabrera:2016vvv, Cabrera:2017njm}, with a change in the dimensions of the
Coulomb and Higgs branches of $-3$ and $-1$, respectively. One can easily see
that this is precisely what is obtained from the closed-form expressions
derived in Section~\ref{sec:anomalies}.

For general LSTs, as we have given a way to construct all $\mu^\vee \in
P_\text{LST}^\vee(\mathfrak{b})$, we can easily find all the possible minimal
transverse slices $\mathcal{S}$ in the same fashion as the example above. We
say that given two coweights, $\mu^\vee$ \emph{covers} $\nu^\vee$, denoted
$\nu^\vee \lessdot \mu^\vee$, if there is no third coweight $\lambda^\vee$
such that $\nu^\vee<\lambda^\vee<\mu^\vee$. This definition also applies to
affine dominant coweights. In each case, we can always uniquely associate a
slice $\mathcal{S}$. Since, given an algebra $\mathfrak{b}$ associated with the
base of an LST, the number of possible coweights is finite, it is
straightforward to obtain an exhaustive list of all transverse minimal slices
$\mathcal{S}$.

The example above shows how, knowing the quiver of a particular slice
$\mathcal{S}$, one is able to find the next theory. For more general cases, the
following algorithm holds:\\

\begin{algorithm}[Slice Subtraction]\label{alg:slice-subtraction}
		Consider an LST $\mathcal{K}=\widetilde{\mathcal{K}}_Q(\mu^\vee; {\mathfrak{g}})$
		associated with an affine dominant coweight $\widetilde{\mu}^\vee\in
		\widetilde{P}_\text{LST}^\vee(\widetilde{\mathfrak{b}})$. A second LST
		$\mathcal{K}'$ can be reached through the following procedure:
\begin{enumerate}
		\item The quiver $\mathcal{S}$ of a transverse slice appearing in
				Tables
				\ref{tab:observed-slices-nilps}--\ref{tab:exceptional-slices}
				can be subtracted from the generalized quiver of $\mathcal{K}$
				if $\mathcal{S}$ appears as one of its subquivers, with
				matching types of flavor and gauge symmetries.
		\item The quiver corresponding to a subtraction operation $\mathcal{K}' = \mathcal{K} - \mathcal{S}$
				has gauge symmetries obtained by subtracting the dimensions of
				the fundamental representation $d_I, f^I$ of $\mathcal{K}$ with
				those of $\mathcal{S}$. The gauge algebras that do not align
				with those of $\mathcal{S}$ remain unchanged.
		\item After performing the subtraction, the dimension of fundamental
				representations of the flavor symmetries $f^I$ of
				$\mathcal{K}'$ must be ``rebalanced'' to ensure cancellation of
				quartic gauge anomalies:
		\begin{equation}
				f^I = C^{IJ}d_J + 16 S^I\,,
		\end{equation}
		with $C^{IJ}$ the Cartan matrix of $\widetilde{\mathfrak{b}}$, and $S^I$ the associated Frobenius--Schur indicator.
\end{enumerate}
The resulting LST $\mathcal{K}'$ is associated with an affine coweight
$\widetilde{\nu}^\vee\in\widetilde{P}_\text{LST}^\vee(\widetilde{\mathfrak{b}})$
such that $\widetilde{\nu}^\vee\lessdot\widetilde{\mu}^\vee$.
\end{algorithm}

Algorithm \ref{alg:slice-subtraction} therefore provides us with an easy way to
generate a large part of the set $\mathcal{P}^\vee_\text{LST}$ from a single
LST. For low values of $Q$, one might encounter the subtleties mentioned at the
beginning of this section, and the algorithm can break down.

We must now find all possible minimal slices $\mathcal{S}$ appearing in
$Spin(32)/\mathbb{Z}_2$ LSTs. Note that our task is slightly different from
the original classification of coverings of coweights by Stembridge
\cite{stembridge1998partial}. In our case, given two dominant coweights
$\mu^\vee,\nu^\vee\in\mathcal{P}^\vee_+$ satisfying $\nu^\vee \lessdot \mu^\vee$
might not correspond to LSTs. We are therefore only interested in coverings
within $\mathcal{P}^\vee_\text{LST}\subset\mathcal{P}^\vee_+$. Given the finite
part $\mathfrak{b}$ of a base algebra, we can then find all minimal slices
$\mathcal{S}$ that can occur by enumerating all the possible coverings in
$\mathcal{P}^\vee_\text{LST}$. As the level of the dominant coweights is set to
32, this can be done systematically for low-rank algebras $\mathfrak{b}$. For
infinite classical series, it is easy to convince oneself that no new types of slices
arise at large rank.

Our task is now to find all possible coverings for an algebra $\mathfrak{b}$
that are related to $Spin(32)/\mathbb{Z}_2$ LSTs. For simplicity, we focus here
only on \emph{finite} coweights, meaning that we are only considering LSTs with
a fixed rank $Q$. However we find that the algorithm extends to general
coweights, although there might be additional minimal slices. In fact, the
slices we find for finite coweights do often appear between coweights with
different values of $Q$.

\paragraph{Classical Kraft--Procesi transitions:} The simplest cases occur when
a single curve supporting a gauge symmetry $\mathfrak{su}_n, \mathfrak{so}_n,
\mathfrak{sp}_n$ has $n$ reduced by one. These correspond to the closure of the
minimal orbit of the associated flavor $\mathfrak{f}_I$, usually denoted by
$a_k, b_k, c_k, d_k$, depending on the type of the algebra of the flavor
symmetry. For instance, for $b_k$, we have:
\begin{equation}\label{nilps}
		\dots\underset{[\mathfrak{so}_{2k+1}]}{\overset{\mathfrak{sp}_n}{1}}\dots \qquad\xrightarrow{\quad b_k ~ = ~\begin{matrix}\scalebox{.7}{\input{figures/slice_bk.tex}}\end{matrix}}\qquad \dots\underset{[\mathfrak{so}_{2k-3}]}{\overset{\mathfrak{sp}_{n-1}}{2}}\dots
\end{equation}
The other types of minimal slices corresponding to the closure of minimal
nilpotent orbits of classical algebras are collated in Table
\ref{tab:observed-slices-nilps}.

These minimal slices are the simplest transitions appearing in the Higgs branch
of SQFTs with eight supercharges, and correspond to giving a particular vacuum
expectation value (vev) to the moment map operator, the gauge-invariant
composite scalar operator in the same supermultiplet as the flavor current. As
it transforms in the adjoint representation of the flavor algebra, one may choose
the vev to be the minimal nilpotent orbit, corresponding to the simplest
decomposition $\mathfrak{f}\to\mathfrak{su}_2\oplus\mathfrak{f}_\text{IR}$. The
algebra $\mathfrak{f}_\text{IR}$ is then the same as in Equation \eqref{nilps}.
It is then well known that the change in the Coulomb-branch dimension always
decreases by one, as can be seen from above, and that of the Higgs branch decreases
by $h^\vee_{\mathfrak{f}}-1$, which is the dimension of the minimal orbit of
$\mathfrak{f}$. 
\begin{table}
		\centering
\begin{tabular}{ccccc}
	\toprule
		Quiver:  & $\begin{matrix}\input{figures/slice_ak.tex}\end{matrix}$  & $\begin{matrix}\input{figures/slice_bk.tex}\end{matrix}$ & $\begin{matrix}\input{figures/slice_ck.tex}\end{matrix}$ & $\begin{matrix}\input{figures/slice_dk.tex}\end{matrix}$\\
		slice: & $a_k$ & $b_k$ & $c_k$ & $d_k$\\
	\bottomrule
\end{tabular}
\caption{Slices corresponding to the closure of minimal nilpotent orbits. The integer at
		each node of the quiver denotes the dimension of the fundamental
		representation, with $\textcolor{red}{\mathfrak{so}_k}$ in red,
		$\textcolor{blue}{\mathfrak{sp}_k}$ in blue, and $\mathfrak{su}_k$ in
		black.
}
\label{tab:observed-slices-nilps}
\end{table}

These RG flows are part of a larger class of transitions referred to as
classical Kraft--Procesi transitions, which---loosely speaking---arise in
physics when studying brane systems that can be labeled by integer partitions.
We defer to a more in-depth discussion of Kraft--Procesi transitions to
references \cite{Cabrera:2016vvv, Cabrera:2017njm}, which also contain a
mathematical introduction to the topic.

The other types of Kraft--Procesi
transitions are associated with Kleinian singularities $\mathbb{C}^2/\Lambda$
where $\Lambda\subset SU(2)$, and are denoted by the associated ADE algebra.
For instance a transition associated with the Kleinian singularity
$\mathbb{C}^2/\mathbb{Z}_n$ is denoted $A_{n-1}$. Physically, the RG flow is
triggered by a baryon-like operator getting a vev. We stress that the
singularity relates to the geometry of the Higgs branch of the theory, and
\emph{not} the type of LST we consider. These are engineered in heterotic
string theory on an orbifold singularity $\mathbb{C}^2/\Gamma$. The slices
$\mathcal{S}$ appearing in $Spin(32)/\mathbb{Z}_2$ LSTs and the associated
quivers are shown in Table \ref{tab:Kleinian}. 

For Kraft--Procesi transitions, whether associated with the closure of a
nilpotent orbit or a Kleinian singularity, there is always an RG flow
associated with those types of coverings in six dimensions, which has been
checked in a plethora of cases via magnetic quivers \cite{Mansi:2023faa,
		Lawrie:2024zon, Lawrie:2023uiu, Bourget:2024mgn, Mekareeya:2017jgc,
Cabrera:2019dob, Cabrera:2019izd, Bao:2024wls, Bao:2025pxe, Bennett:2025edk,
Bennett:2024llh}. The changes in the dimension of the Coulomb and Higgs branches are exchanged
with respect to that of nilpotent orbits: the former changes by $h^\vee-1$ and
the latter by one. There is an exception in cases involving
$\overset{\mathfrak{so}_k}{4}$ configurations: in most cases, the change in $k$ is
even and there are no issues, but when it is odd, the change in the Coulomb
branch might be different from the predicted $h^\vee-1$, as the rank of
$\mathfrak{so}_{2n+1}$ and $\mathfrak{so}_{2n}$ is the same. This difference between
the predicted and actual change in the Coulomb-branch dimension is a known
occurrence, and is related to the so-called $X$-collapse of the partition labeling
the branes \cite{Cabrera:2017njm}.

\begin{table}[t]
		\centering
		\begin{tabular}{ccccc}
	Slice type & Quiver & $\delta\text{CB}$ & $\delta\text{HB}$\\
	\toprule
	$A_k$ & $\begin{matrix}\input{figures/slice_Ak_brace.tex}\end{matrix}$ & $-k$ & $-1$  \\
	$A_1$ & $\begin{matrix}\input{figures/slice_123.tex}\end{matrix}$ & $\begin{smallmatrix}-2~ (\overset{\mathfrak{so}_k}{4}~\text{even})\\-1~ (\overset{\mathfrak{so}_k}{4}~\text{odd})\end{smallmatrix}$ & $-1$\\
$A_{2k-1}$ & $\begin{matrix}\input{figures/slice_Aos_brace.tex}\end{matrix}$ & $\begin{smallmatrix}-2k~ (\overset{\mathfrak{so}_k}{4}~\text{even})\\-(2k-1)~ (\overset{\mathfrak{so}_k}{4}~\text{odd})\end{smallmatrix}$ & $-1$ \\
	$A_{2k-1}\cup A_{2k-1}$ & $\begin{matrix}\input{figures/slice_2A2n-1_brace.tex}\end{matrix}$  & $-(2k-1)$ & $-1$ \\
	$A_3\simeq D_3$ & $\begin{matrix}\input{figures/slice_D4.tex}\end{matrix}$ & $\begin{smallmatrix}-1~ (\overset{\mathfrak{so}_k}{4}~\text{even})\\-3~ (\overset{\mathfrak{so}_k}{4}~\text{odd})\end{smallmatrix}$ &  $-1$ \\
	$D_k$ & $\begin{matrix}\input{figures/slice_Dk_braces.tex}\end{matrix}$ &  $\begin{smallmatrix}-(2k-3)~ (\overset{\mathfrak{so}_k}{4}~\text{even})\\-(2k-5)~ (\overset{\mathfrak{so}_k}{4}~\text{odd})\end{smallmatrix}$ &  $-1$ \\
	\bottomrule
	\end{tabular}
	\caption{\label{tab:Kleinian}Quivers appearing in the slice-subtraction
	algorithm that, interpreted as a $3d~\mathcal{N}=4$ theory, corresponds to
	Kraft--Procesi transitions associated with Kleinian singularities, taken
	from \cite[Tables 6, 7]{Cabrera:2017njm}. The integer at each node of the
	quiver denotes the dimension of the fundamental representation, with
	$\textcolor{red}{\mathfrak{so}_k}$ in red,
	$\textcolor{blue}{\mathfrak{sp}_k}$ in blue, and $\mathfrak{su}_k$ in
	black. $\delta\text{CB}$ and $\delta\text{HB}$ indicate the changes of the Coulomb- and Higgs-branches with respect to the UV theory. In the presence of $\mathfrak{so}_k$ algebras, if the quiver of the slice
	leads to an odd change in the dimension of the fundamental representation,
	the change in the Coulomb-branch dimension depends on the UV theory, and
	whether the associated $\mathfrak{so}_k$ symmetry is even or odd.}
\end{table}

\paragraph{Exotic transitions:} in addition to the usual Kraft--Procesi
transitions, we also find coverings where we can associate a quiver
$\mathcal{S}$ for which we do not know the corresponding type of transition. We
separate them into two classes. The first appears in theories whose base is
non-simply laced, while the second corresponds to orthosymplectic quivers whose
gauge nodes are shaped like an exceptional algebra. 

The slices related to non-simply-laced bases are collected in Table
\ref{tab:slice-non-simply-laced}. One can see that in each case, the quiver
corresponding to the slice arranges like the Dynkin diagram of a
(possibly-twisted) affine algebra, with the flavor symmetry on the affine node.
Interestingly, the change in the dimension of the fundamental representation
at each node of the slice is equal to the comarks of the algebra. Note that,
similarly to Kraft--Procesi transitions of the Kleinian type, when the quiver
of the slice has only a few gauge nodes, the flavor symmetry of the slice can
be larger what could be expected from the comarks.

\begin{table}
	\centering
	\begin{tabular}{cccc}
		Quiver & $\delta\text{CB}$ & $\delta\text{HB}$\\
		\toprule
				$\begin{matrix}\input{figures/slice_bt3.tex}\end{matrix}$ & $-2$ & $-1$\\
				$\begin{matrix}\input{figures/slice_bt3-1.tex}\end{matrix}$ & $-2$ & $0$\\
				$\begin{matrix}\input{figures/slice_Bk.tex}\end{matrix}$ & $\begin{smallmatrix}-k~ (\overset{\mathfrak{so}_k}{4}~\text{even})\\-(k-1)~ (\overset{\mathfrak{so}_k}{4}~\text{odd})\end{smallmatrix}$ & 0\\
				$\begin{matrix}\input{figures/slice_bt3i.tex}\end{matrix}$ & $-4$ & $0$\\
				$\begin{matrix}\input{figures/slice_C2.tex}\end{matrix}$ & $-2$& $0$\\
				$\begin{matrix}\input{figures/slice_C2-2.tex}\end{matrix}$ & $-2 $& $-1$\\
				$\begin{matrix}\input{figures/slice_C2-3.tex}\end{matrix}$ & $-2$ & $-1$ \\
				$\begin{matrix}\input{figures/slice_C2-4.tex}\end{matrix}$ & $-2$ & $-2$ \\
				$\begin{matrix}\input{figures/slice_Ck.tex}\end{matrix}$ & $-k$& $0$ \\
				$\begin{matrix}\input{figures/slice_sF4.tex}\end{matrix}$ & $-8$ & $0$\\
		\bottomrule
	\end{tabular}
	\caption{\label{tab:slice-non-simply-laced}Quivers appearing in the
			slice-subtraction algorithm for LSTs with bases
			$\widetilde{\mathfrak{b}}$ that are non-simply laced. The integer
			at each node of the quiver denotes the dimension of the fundamental
			representation, with $\textcolor{red}{\mathfrak{so}_k}$ in red,
			$\textcolor{blue}{\mathfrak{sp}_k}$ in blue and $\mathfrak{su}_k$
			in black. $\delta\text{CB}$ and $\delta\text{HB}$ indicate the changes of the Coulomb- and Higgs-branches with respect to the UV theory. In the presence of $\mathfrak{so}_k$ gauge algebras, if
			the quiver associated with the slice leads to an odd change in the
			dimension of the fundamental representation, the change in the
	Coulomb-branch dimension depends on the UV theory, and whether the
	fundamental representation of the corresponding $\mathfrak{so}_k$ gauge
	symmetry is even or odd.}
\end{table}

For the second class, collated in Tables \ref{tab:E6-slices} and
\ref{tab:exceptional-slices}, we distinguish two subclasses: the first
corresponds to quivers shaped like an $\mathfrak{e}_6$ Dynkin diagram. Such
quiver slices occur solely in the case of
$\mathfrak{g}=\mathfrak{so}_{2k},\mathfrak{e}_k$ due to their trivalent node.
The remaining quivers, collated in Table \ref{tab:slice-non-simply-laced} occur
only in the exceptional cases $\mathfrak{g}=\mathfrak{e}_7,\mathfrak{e}_8$.
Contrary to the other types of exceptional minimal slices, they induce a change
in the dimension of the fundamental that is greater than the comarks of the
corresponding Dynkin diagrams. To the best of our knowledge, when thought of as
$3d$ $\mathcal{N}=4$ quivers, these types of slices have not appeared in the
literature.

\begin{table}
	\centering
	\begin{tabular}{cccccc}
		 Quiver & $\delta\text{CB}$ & $\delta\text{HB}$ & Appears in $\mathfrak{g}$ \\
	\toprule
	$\begin{matrix}\input{figures/slice_E6.tex}\end{matrix}$ & $-4$ & $0$ & $\mathfrak{e}_7, \mathfrak{e}_{8}$\\
	$\begin{matrix}\input{figures/slice_E6-6.tex}\end{matrix}$ & $-7$ & $-1$ & $\mathfrak{e}_7, \mathfrak{e}_{8}$\\
	$\begin{matrix}\input{figures/slice_E6-2.tex}\end{matrix}$ & $-4$ & $0$ & $\mathfrak{e}_7, \mathfrak{e}_{8}$\\
	$\begin{matrix}\input{figures/slice_E6-4.tex}\end{matrix}$ & $\begin{smallmatrix}-6~ (\overset{\mathfrak{so}_k}{4}~\text{even})\\-4~ (\overset{\mathfrak{so}_k}{4}~\text{odd})\end{smallmatrix}$ & $-1$ & $\mathfrak{so}_{2k}, \mathfrak{e}_{7}, \mathfrak{e}_{8}$\\
	$\begin{matrix}\input{figures/slice_E6-3.tex}\end{matrix}$ & $-4$ & $0$ & $\mathfrak{so}_{2k}, \mathfrak{e}_7, \mathfrak{e}_{8}$\\
	$\begin{matrix}\input{figures/slice_E6-5.tex}\end{matrix}$ & $\begin{smallmatrix}-5~ (\overset{\mathfrak{so}_k}{4}~\text{even})\\-4~ (\overset{\mathfrak{so}_k}{4}~\text{odd})\end{smallmatrix}$ & $-1$ & $\mathfrak{so}_{2k}, \mathfrak{e}_7, \mathfrak{e}_{8}$\\
	\bottomrule
\end{tabular}
\caption{
\label{tab:E6-slices}
Slices appearing in the partial order of coweights related to D-type
algebras $\mathfrak{so}_{2k}$, and exceptional algebras $\mathfrak{e}_7$ and
$\mathfrak{e}_8$. The integer at each node of the quiver denotes the dimension
of the fundamental representation, with $\textcolor{red}{\mathfrak{so}_k}$ in
red, and $\textcolor{blue}{\mathfrak{sp}_k}$ in blue. $\delta\text{CB}$ and $\delta\text{HB}$ indicate the changes of the Coulomb- and Higgs-branches with respect to the UV theory. For the change in
Coulomb-branch dimensions, the different possibilities refer to whether the
fundamental representation of the $\mathfrak{so}_k$ gauge algebra of UV theory
on the trivalent node is even or odd.
}
\end{table}

\begin{table}
		\centering
\begin{tabular}{ccccc}
		Quiver & $\delta\text{CB}$ & $\delta\text{HB}$ & Appears in $\mathfrak{g}$ \\
	\toprule
	$\begin{matrix}\input{figures/slice_D5ppp.tex}\end{matrix}$ & $-8$ & 0 & $\mathfrak{e}_7, \mathfrak{e}_8$\\
	$\begin{matrix}\input{figures/slice_D5ppp9.tex}\end{matrix}$ & $\begin{smallmatrix}-8~ (\overset{\mathfrak{so}_k}{4}~\text{even})\\-9~ (\overset{\mathfrak{so}_k}{4}~\text{odd})\end{smallmatrix}$ & $-1$ & $\mathfrak{e}_7, \mathfrak{e}_8$\\
	$\begin{matrix}\input{figures/slice_D5ppp10.tex}\end{matrix}$ & $\begin{smallmatrix}-8~ (\overset{\mathfrak{so}_k}{4}~\text{even})\\-10~ (\overset{\mathfrak{so}_k}{4}~\text{odd})\end{smallmatrix}$ & $-1$ & $\mathfrak{e}_8$\\
	$\begin{matrix}\input{figures/slice_D5ppp7.tex}\end{matrix}$ & $-8$ & $0$ & $\mathfrak{e}_8$\\
	$\begin{matrix}\input{figures/slice_D5ppp4.tex}\end{matrix}$ & $-12$ & $-1$ & $\mathfrak{e}_8$\\
	$\begin{matrix}\input{figures/slice_D5ppp5.tex}\end{matrix}$ & $-13$ & $0$ & $\mathfrak{e}_8$\\
	$\begin{matrix}\input{figures/slice_D5ppp2.tex}\end{matrix}$ & $-16$ & $0$ & $\mathfrak{e}_7, \mathfrak{e}_8$\\
	$\begin{matrix}\input{figures/slice_D5ppp8.tex}\end{matrix}$ & $-28$ & $0$ & $\mathfrak{e}_8$\\
	\bottomrule
\end{tabular}
\caption{
\label{tab:exceptional-slices}
Slices only appearing in the partial order of coweights related to
exceptional algebras $\mathfrak{e}_7$ and $\mathfrak{e}_8$. The integer at
each node of the quiver denotes the dimension of the fundamental
representation, with $\textcolor{red}{\mathfrak{so}_k}$ in red, and
$\textcolor{blue}{\mathfrak{sp}_k}$ in blue. $\delta\text{CB}$ and $\delta\text{HB}$ indicate the changes of the Coulomb- and Higgs-branch dimensions with respect to the UV theory. For the change in Coulomb-branch
dimensions, the different possibilities refer to whether the fundamental
representation of the $\mathfrak{so}_k$ gauge algebra of UV theory on the
trivalent node is even or odd.
}
\end{table}

\paragraph{An example with $\widetilde{\mathfrak{b}}=B_4^{(1)}$:} having made
an inventory of the slices $\mathcal{S}$ that appear in the algorithm, in
Figure \ref{fig:B4-Hasse} we illustrate how Algorithm \ref{alg:slice-subtraction}
correctly reproduces the Hasse diagram of dominant coweights for fixed $Q$ for
$Spin(32)/\mathbb{Z}_2$ LSTs of type $\mathfrak{g}=\mathfrak{so}_{10}$. The
finite part of the base algebra is therefore
$\mathfrak{b}=\mathfrak{so}_9$. In the diagram we also indicate the
coweights that seed a putative duality orbit $\mathcal{O}_{\mu^\vee}$ in a blue
box, as explained in the previous section.

\begin{figure}[p]
	\centering
	\resizebox{4cm}{!}{\input{figures/hasse_D5.tex}} \caption{Part of the Hasse
			diagram of $Spin(32)/\mathbb{Z}_2$ LSTs with
			$\mathfrak{g}=\mathfrak{so}_{10}$ at fixed value of $Q$ following
			the partial order of \emph{finite} coweights. Theories in blue
			boxes indicate they are representatives of a putative duality orbit. Each arrow
			indicates the subtracted quiver corresponding to the slice
			$\mathcal{S}$. The integers associated with each LST indicate the
			total flavor rank or the change from the theory with trivial
			coweight, e.g. $\Delta_0\kappa_R(\mu^\vee) = \kappa_R(\mu^\vee) -
	\kappa_R(\varnothing)$.} \label{fig:B4-Hasse}
\end{figure}

\paragraph{Dominant coweight ordering and RG flows:} we have referred to the
slice-subtraction algorithm as a way, given an LST, to reconstruct all LSTs
with smaller coweights. As we have seen at the beginning of this section, for
two LSTs to be related, it is a necessary condition that the LST at the end of
the flow is smaller with respect to the partial order of dominant coweights
than that of the ultraviolet theory. On the other hand, this does not \emph{a
priori} provide a sufficient condition for the partial order of coweights to
be the same as the network of Higgs-branch RG flows.

However, to our knowledge, partial orders that have
been explored in the recent literature have been shown to be equivalent to
Higgs-branch RG flows \cite{Bourget:2021siw, Fazzi:2023ulb, Lawrie:2024zon}. In
those cases, up to possible subtleties at very low gauge-algebra ranks, there is
a one-to-one correspondence between each dominant coweight and the magnetic
quiver reproducing the partial order through quiver subtraction. However
LSTs differ from those works in two ways. First, we are dealing with
ortho-symplectic ``electric'' theories, while previous works have focused on
unitary quivers. Since, for instance, the dimension of the fundamental
representation of an $\mathfrak{sp}_k$ flavor symmetry must be even, not all
dominant coweights are associated with an LST. This means that the cone of
dominant coweights is not fully populated. Furthermore, anomaly cancellation
forces coweights to have a fixed level of 32, which is not the case in SCFTs.

Despite these differences, it seems natural to expect RG flows to be related to
that partial order. In particular, it has been shown that the closure of
nilpotent orbits and other Kraft--Procesi transitions discussed above indeed
correspond to RG flows in six-dimensional SCFTs \cite{Fazzi:2023ulb,
Lawrie:2023uiu, Cabrera:2019izd, Cabrera:2019dob}. However, some of the exotic
slices we have found have a peculiar property: they do not induce a change in
the dimension of the Higgs branch. As an example, consider again the
$\mathfrak{g}=\mathfrak{so}_{10}$ case, and let us focus on the last transition
in Figure \ref{fig:B4-Hasse}, involving LSTs associated to the
next-to-trivial and trivial finite coweights:
\begin{equation}\label{example-HB-0-sec4}
	\underset{[\mathfrak{so}_{2}]}{\overset{\mathfrak{sp}_{Q - 7}}{1}}\,
	 \overset{\displaystyle \overset{[\mathfrak{so}_{30}]}{\overset{\mathfrak{sp}_{Q}}{1}}}
	 {\overset{\mathfrak{so}_{4Q - 14}}{4}}\,
	 \overset{\mathfrak{sp}_{2Q - 15}}{1}\,
	 \overset{\mathfrak{su}_{2Q - 15}}{2}
	 \overset{\resizebox{2.5cm}{!}{\input{figures/example_slice_C4.tex}}}{\relbar\joinrel\relbar\joinrel\relbar\joinrel\longrightarrow}\qquad
	\overset{\mathfrak{sp}_{Q - 8}}{1}\,
	\overset{\displaystyle \overset{[\mathfrak{so}_{32}]}{\overset{\mathfrak{sp}_{Q}}{1}}}
	{\overset{\mathfrak{so}_{4Q - 16}}{4}}\,
	\overset{\mathfrak{sp}_{2Q - 16}}{1}\,
	\overset{\mathfrak{su}_{2Q - 16}}{2}
\end{equation}
One can verify that there is indeed no change in the Higgs-branch dimension in
agreement with Table \ref{tab:slice-non-simply-laced}. If there is a
Higgs-branch deformation between the two theories, one would---albeit maybe too
naively---expect that as the R-symmetry is broken along the RG flow, there must
be at least one hypermultiplet acting as its Nambu--Goldstone mode that
decouples from the UV theory. We therefore expect the change in
the Higgs-branch dimension to always be strictly negative.

In Tables \ref{tab:slice-non-simply-laced}--\ref{tab:exceptional-slices} one
can find several slices with that feature. One may therefore be tempted to say
that these slices cannot be physical. From a geometric point of view however,
it does not---again, at least naively---appear to be problematic: breaking a
gauge algebra to one of its subalgebras e.g. $\mathfrak{su}_{2Q-14}$ to
$\mathfrak{su}_{2Q-16}$ in Equation \eqref{example-HB-0-sec4} corresponds to
smoothing the fiber singularity from a Kodaira singularity $I_{2Q-14}$ to
$I_{2Q-16}$. Individually, this is always possible, see e.g. reference
\cite{DelZotto:2018tcj}. 

If the deformation cannot be done geometrically, a problem must occur due to a
conspiracy between singularities involving multiple curves. Ruling out these
transitions would require to explicitly finding the Weierstrass model associated
with both LSTs and check whether a complex-structure deformation between the
two is possible. While this has been done for LSTs of rank zero to explore the
Higgs-branch Hasse diagram and match with magnetic-quiver techniques
\cite{Mansi:2023faa}, in practice this is however notoriously difficult to
achieve for a large number of curves, and we are not aware of a reference that
has attempted to explicitly construct the deformation for a multi-curve
Weierstrass model. Furthermore, the LSTs where these curious slices appear do
not have, to our knowledge, a known description in terms of a magnetic quiver.
We further note that we have computed the change in the dimension of the Higgs
branch using the anomaly polynomial. The formula for the change given in
equation \eqref{RG-qt-change}, was originally argued to be valid for SCFTs.
Although we are not aware of any potential loophole, it may break down for LSTs
due to the zero curve. Whether these puzzles can be reconciled in either
picture appears to be an interesting avenue for future work. 

\section{Extension to Other Types of LSTs and SCFTs}\label{sec:extension}

We have so far solely focused on $Spin(32)/\mathbb{Z}_2$ heterotic LSTs. In
this section, we briefly explore possible extensions of our results to other
theories. Indeed, other types of six-dimensional quivers and the associated LSTs
or SCFTs involving only classical algebras can also be rephrased in the
language of coweights, and similar analyses can be performed.

The main reason we have focused on $Spin(32)/\mathbb{Z}_2$ heterotic LSTs stems
from the fact that they involve only bifundamental hypermultiplets, and the
type of flavor and gauge symmetries is uniquely set by the McKay
correspondence, as reviewed in Section~\ref{sec:classification-spin32-lst}.
However, as we will now see, it also applies straightforwardly to another class
of $\mathfrak{so}_{32}$ heterotic LSTs, with their data encoding the coweight
lattice of \emph{twisted} affine algebras.\footnote{In this section, we always
		mean twisted in the Lie-algebraic sense, rather than to refer to
		twisted compactification. The latter will not be discussed here, see however
		e.g. reference \cite{Ahmed:2024wve}.} It may be used to also incorporate cases
involving two-anti-symmetric representations of $\mathfrak{su}_k$, arising from a
curve configuration $\overset{\mathfrak{su}_k}{1}$ under a very mild
modification. We will further argue that certain properties relating to dominant
coweights also apply to SCFTs, offering a generalization of the
slice-subtraction algorithm also in that case, so that our results cover the
overwhelming majority of known consistent six-dimensional theories with only
classical symmetries.

\subsection{\texorpdfstring{$(SU(16)\times U(1))/\mathbb{Z}_2$}{(SU16xU1/Z2)} Little String Theories}\label{sec:su16-lst}

The $\mathfrak{so}_{32}$ heterotic Little String Theories we have discussed
above were labeled by (folded) ADE algebras through the McKay correspondence,
and fall in the classification by Blum and Intriligator \cite{Blum:1997mm}, see
Classification \ref{class:spin32}. In references \cite{DelZotto:2022xrh,
Ahmed:2024wve}, other types of $\mathfrak{so}_{32}$ LSTs not captured by the
original classification were discovered via explicit geometric construction.

Instead of the $Spin(32)/\mathbb{Z}_2$ global symmetry that descends from the
heterotic construction, they are characterized by one of its maximal
rank-preserving subgroups:
\begin{equation}
		Spin(32)/\mathbb{Z}_2\quad \longrightarrow \quad(SU(16)\times U(1))/\mathbb{Z}_2\,.
\end{equation}
While such a breaking could naively be thought of as being obtained from a
choice of non-trivial affine coweight of the ADE algebra
$\widetilde{\mathfrak{g}}$ that encodes the decomposition
$\mathfrak{so}_{32}\to\mathfrak{su}_{16}\oplus\mathfrak{u}_1$, it turns out
that the gauge and intersection pattern in the F-theory description are
different. For instance, one may obtain the following quiver:
\begin{equation}
	\underset{[\mathfrak{su}_{16}]}{\overset{\mathfrak{su}_{Q}}{2}}
	\overset{\mathfrak{su}_{2Q-16}}{2}
	\overset{\mathfrak{su}_{3Q-32}}{2}
	\overset{\mathfrak{sp}_{2Q-24}}{1}
	\overset{\mathfrak{so}_{Q-16}}{4}\,,
\end{equation}
which has the advertised flavor symmetry, the $\mathfrak{u}_1$ symmetry arising
through the arguments discussed around Equation \eqref{ABJ-I8}. Its linear
shape is reminiscent of the $Spin(32)/\mathbb{Z}_2$ LST with
$\mathfrak{g}=\mathfrak{e}_6$, see Equation \eqref{general-quiver-f4}. However
the gauge symmetries---and consequently the curve self-intersections---are
indeed different. The hypermultiplet adjacency matrix $G$ discussed in Classification
\ref{class:base-topology} is therefore distinct from that of $F_4^{(1)}$, and
corresponds instead to the symmetrized Cartan matrix of the twisted algebra
$\widetilde{\mathfrak{b}} = E_6^{(2)}$. For a generic quiver of the type above,
we can therefore make the following identification:
\begin{equation}
	\underset{[\mathfrak{su}_{f^0}]}{\overset{\mathfrak{su}_{d_0}}{2}}
	\underset{[\mathfrak{su}_{f^1}]}{\overset{\mathfrak{su}_{d_1}}{2}}
	\underset{[\mathfrak{su}_{f^2}]}{\overset{\mathfrak{su}_{d_2}}{2}}
	\underset{[\mathfrak{so}_{f^3}]}{\overset{\mathfrak{sp}_{d_3/2}}{1}}
	\underset{[\mathfrak{sp}_{f^4/2}]}{\overset{\mathfrak{so}_{d_4}}{4}}
	\qquad\longleftrightarrow\qquad
	\begin{matrix}
\begin{tikzpicture}
	\node[node, label=above:{$\vphantom{f^0}$}, label=below:{\footnotesize $f^0$}] (A0)  {};
    \node[node, label=below:{\footnotesize $f^1$}] (A1) [right=6mm of A0] {};
    \node[node, label=below:{\footnotesize $f^2$}] (A2) [right=6mm of A1] {};
    \node[node, label=below:{\footnotesize $f^3$}, fill=red] (A3) [right=6mm of A2] {};
    \node[node, label=below:{\footnotesize $f^4$}, fill=blue] (A4) [right=6mm of A3] {};
	\node[yscale=1.4] (C) [right=.2mm of A2] {$<$};
    \draw (A0.east) -- (A1.west);
    \draw (A1.east) -- (A2.west);
	\draw ([yshift=1.5pt]A2.east) -- ([yshift=1.5pt]A3.west);
    \draw ([yshift=-1.5pt]A2.east) -- ([yshift=-1.5pt]A3.west);
    \draw (A3.east) -- (A4.west);
\end{tikzpicture}
	\end{matrix}\,.
\end{equation}
As before, we have labeled the Dynkin diagrams with the dimension of the
fundamental representation of the associated flavor symmetries, coloring them
in the same fashion as for $Spin(32)/\mathbb{Z}_2$ LSTs, see Table
\ref{tbl:node-color}. There are only three classes of LSTs of that type, all
with a base corresponding to a twisted algebra:\footnote{In the first two
cases, there are also two analogous classes involving a
$\overset{\mathfrak{su}_k}{1}$ configuration. These are discussed in
Section~\ref{sec:antisymmetric}.}
\begin{equation}
		(SU(16)\times U(1))/\mathbb{Z}_2\quad \text{LST}:\qquad
		\widetilde{\mathfrak{b}} \in \{A_{2k-1}^{(2)}, D_{3}^{(2)}, E_6^{(2)} \}\,,
\end{equation}
The corresponding generalized quivers are given in Table \ref{tab:twisted},
along with useful quantities.

\begin{table}
    \centering
    \begin{threeparttable}
        \begin{tabular}[t]{cccc}
				\toprule
				${\widetilde{\mathfrak{b}}}$ & $h^\vee_{\mathfrak{b}}$ & Twisted Dynkin Diagram & $(SU(16)\times U(1))/\mathbb{Z}_2$ LST \\\midrule
		$D_3^{(2)}$ & $4$ &  $\begin{matrix}\input{figures/D_3_2.tex}\end{matrix}$ &
    $
		\underset{[\mathfrak{su}_{f^0}]}{\overset{\mathfrak{su}_{d_0}}{2}}
		\underset{[\mathfrak{so}_{f^1}]}{\overset{\mathfrak{sp}_{d_1/2}}{1}}
		\underset{[\mathfrak{su}_{f^2}]}{\overset{\mathfrak{su}_{d_2}}{2}}
	$ \\\midrule
	$A_{2K-1}^{(2)}$ & $2K$ &$\begin{matrix}\input{figures/Dk_2.tex}\end{matrix}$ &
	$ 
		\underset{[\mathfrak{su}_{f^1}]}{\overset{\mathfrak{su}_{d_1}}{2}}
		\overset{\displaystyle\overset{\overset{\left[\mathfrak{su}_{f^0} \right]}{\mathfrak{su}_{d_0}}}{2}}
		{\underset{[\mathfrak{su}_{d_1}]}{\overset{\mathfrak{su}_{d_2}}{2}}}
		\underset{[\mathfrak{su}_{f_3}]}{\overset{\mathfrak{su}_{d_3}}{2}}
		\cdots
		\underset{[\mathfrak{su}_{f_{K-1}}]}{\overset{\mathfrak{su}_{d_{K-1}}}{2}}
		\underset{[\mathfrak{so}_{f_{K}}]}{\overset{\mathfrak{sp}_{d_K/2}}{1}}
	$  \\\midrule

	$E_6^{(2)}$ & $12$ & $\begin{matrix}\input{figures/E6_2.tex}\end{matrix}$ & $
	\underset{[\mathfrak{su}_{f^0}]}{\overset{\mathfrak{su}_{d_0}}{2}}
	\underset{[\mathfrak{su}_{f^1}]}{\overset{\mathfrak{su}_{d_1}}{2}}
	\underset{[\mathfrak{su}_{f^2}]}{\overset{\mathfrak{su}_{d_2}}{2}}
	\underset{[\mathfrak{so}_{f^3}]}{\overset{\mathfrak{sp}_{d_3/2}}{1}}
	\underset{[\mathfrak{sp}_{f^4/2}]}{\overset{\mathfrak{so}_{d_4}}{4}}
	$ 
	\\\midrule
        \bottomrule
        \end{tabular}
    \end{threeparttable}
	\caption{\label{tab:twisted}LSTs of type $(SU(16)\times U(1))/\mathbb{Z}_2$
	and group-theoretic quantities associated with the twisted Lie algebra of
	its base $\widetilde{\mathfrak{b}}$. The labels of the Dynkin diagram
	correspond to the marks $\theta_I$ of $\widetilde{\mathfrak{b}}$.}
\end{table}

In the same fashion as in previous cases, one can show that the quartic
anomaly-cancellation condition takes precisely the expected form:
\begin{equation}\label{twisted-quartic}
		C^{IJ}d_J = f^I - 16S^I\,,
\end{equation}
where $C^{IJ}$ is now the Cartan matrix of the twisted algebra
$\widetilde{\mathfrak{b}}$. $S^I$ is the generalization of the
Frobenius--Schur indicators to twisted algebras. While the McKay
correspondence applies only to ADE algebras, we can proceed similarly as we did
for folding in Equation \eqref{f-S-folding}: if $\mathcal{T}$ is the map
transforming the root system of $\widetilde{\mathfrak{g}}$ to that of its
twisted version $\widetilde{\mathfrak{b}}$, we simply define
$S^I=\mathcal{T}(S^A)$ with $S^A$ the Frobenius--Schur indicator of the $A$-th
node of the Dynkin diagram of $\widetilde{\mathfrak{g}}$. For instance, for
$\widetilde{\mathfrak{b}}=E_6^{(2)}$, we have $S^I=(0,0,0,-1,1)$. As before,
the values of $S_I$ are a Lie-algebraic property of $\widetilde{\mathfrak{b}}$,
which can be obtained directly from the colored Dynkin diagram, where we have
used the same conventions as in Table \ref{tab:twisted}.

There is however a significant difference between the $Spin(32)/\mathbb{Z}_2$
and $(SU(16)\times U(1))/\mathbb{Z}_2$ LSTs: it is easily shown that for the
latter, Equation \eqref{twisted-quartic} leads to the constraints
\begin{equation}
		(SU(16)\times U(1))/\mathbb{Z}_2:\qquad f^I\theta_I = 16\,,\qquad	\theta_I S^I = 1\,,\qquad d_I = Q\, K_I + X_{IJ} (f^J-16S^J)\,.
\end{equation}
We recall that the coefficients $\theta_I$ and $K_I$ are the marks and comarks
of $\widetilde{\mathfrak{b}}$, and $X_{IJ}$ inverts the Cartan matrix up to an
integer factor $Q$ parametrizing its nullspace, see Equation \eqref{comarks-definition}.

These constraints are similar to those we obtained for the folded algebras, see
equations \eqref{sol-dI} and \eqref{BRso32}. In terms of coweights, they imply
that for these theories, the dimensions $f^I$ of the fundamental representation
of flavor symmetries define an affine coweight $\widetilde{\mu}^{\vee}$ of
$\widetilde{\mathfrak{b}}$ at level 16 rather than 32. Furthermore the value of
the scaling element---the coefficient $n$ in Equation \eqref{def-coweight}---is
now $Q$ rather than $2Q$. This is explained by the fact that for these twisted
algebras, on the zeroth node the fundamental representation of the
$\mathfrak{su}_Q$ gauge symmetry has a dimension $d_0=Q$, rather than $d_0=2Q$
for $\mathfrak{sp}_Q$.

The T-duality invariants of these theories are then easily computed. For the
two-group constant $\kappa_R$, one finds:
\begin{equation}
		\kappa_R = (\theta_IK^I) (Q-2r_{\mathfrak{b}}-2) + 2 + \theta^i C_{ij} f^j = \langle\widetilde{T}, \widetilde{\mu}^\vee-16\widetilde{S}^\vee\rangle\,.
\end{equation}
with, as before, $\widetilde{T}=\sum_I\theta_I\widetilde{\omega}_I$ and
$C_{ij}$ the inverse Cartan matrix when omitting the zeroth node. This form is
very reminiscent of the one we found in Equation \eqref{kappa-closed-form}. A
similar expression can be found for the Coulomb-branch dimension, albeit
slightly more complicated due to the floor function related to the rank of
$\mathfrak{so}_k$ gauge algebras, but is of analogous form to Equation
\eqref{CB-closed-form}.

We therefore reach the conclusion that all $\mathfrak{so}_{32}$ heterotic
LSTs involving only bifundamental hypermultiplets, either of type
$Spin(32)/\mathbb{Z}_2$ or $(SU(16)\times U(1))/\mathbb{Z}_2$, are classified
by affine dominant coweights of their base $\widetilde{\mathfrak{b}}$. While we will not
go over a complete analysis of the T-duality orbits and the slice subtraction
algorithm, one may proceed in a similar fashion.

To close this section, we note that the coefficients multiplying $Q$ in the
T-dual invariants are given in terms of the marks of
$\widetilde{\mathfrak{b}}$, and satisfy an interesting numerology. As was
observed in references \cite{DelZotto:2022xrh, Ahmed:2024wve}, for $D_3^{(2)},
A_{2K-1}^{(2)}, E_6^{(2)}$ these coefficients are \emph{half} those of
$Spin(32)/\mathbb{Z}_2$ LSTs with $\mathfrak{g}=\mathfrak{so}_8, \mathfrak{so}_{4K},
\mathfrak{e}_7$, respectively. For instance, with
$\widetilde{\mathfrak{b}}=E_6^{(2)}$, we have $K^I\theta_I =
\Gamma_{\mathfrak{e}_6} = 24 = \frac{1}{2}\Gamma_{\mathfrak{e}_7}$. In fact,
for all three types of algebras, not only do these quantities satisfy this
numerical curiosity, but when $Q=2Q'$, the T-dual invariants exactly match those
of the $Spin(32)/\mathbb{Z}_2$ theories with instanton number $Q'$ for
appropriate choices of finite coweights. The T-duality can be verified
geometrically \cite{DelZotto:2022xrh, Ahmed:2024wve}, and in the terminology
described at the beginning of Section~\ref{sec:t-duality}, corresponds to
infinite families of exotic T-dualities. For trivial coweights, these theories
can be found in Table \ref{tab:het-so32-unbroken-withoutvs} in Appendix
\ref{app:unbroken-LST}.

\subsection{Including Anti-symmetric Representations of \texorpdfstring{$\mathfrak{su}_k$}{su(k)}}\label{sec:antisymmetric}

Throughout this work, we have always carefully excluded the case of
$Spin(32)/\mathbb{Z}_2$ with $\mathfrak{g}=\mathfrak{su}_{2n+1}$, as one of the
nodes of its quiver includes an additional hypermultiplet transforming in the
two-anti-symmetric representation of an $\mathfrak{su}_k$ algebra, and
therefore does not fall in Classification \ref{class:base-topology}, where we
have considered only bifundamental hypermultiplets.

Geometrically, the inclusion of a single hypermultiplet in the
two-anti-symmetric representation is well known to be described by a curve
configuration $\overset{\mathfrak{su}_k}{1}$. Field theoretically, this is once
again explained at the level of cancellation of quartic anomalies: one
hypermultiplet in the anti-symmetric, in addition to those in the fundamental
representation, mimics the contribution to the quartic gauge anomaly of
hypermultiplets transforming in the fundamental representation of an
$\mathfrak{sp}_k$ gauge symmetry. 

The absence of quadratic gauge anomalies is then shown to indeed reproduce a
Dirac pairing corresponding to that of a $(-1)$-curve. This conspiracy allows
us to treat the anti-symmetric representation as if that node was associated
with an $\mathfrak{sp}_k$ symmetry. Indeed, from the F-theory classification
\cite{Bhardwaj:2015oru}, the LST with $\mathfrak{g}=\mathfrak{su}_{2n+1}$ is
the collection of $n$ curves forming the following generalized quiver:
\begin{equation}
		\underset{[\mathfrak{so}_{f^0}]}{\overset{\mathfrak{sp}_{k_0}}{1}}\underset{[\mathfrak{su}_{f^1}]}{\overset{\mathfrak{su}_{k_1}}{2}}\,\cdots\,\underset{[\mathfrak{su}_{f^{n-1}}]}{\overset{\mathfrak{su}_{k_{n-1}}}{2}}\underset{[\mathfrak{su}_{f^n}]}{\overset{\mathfrak{su}_{k_n}}{1}}\,.
\end{equation}
Using the same techniques used in Section~\ref{sec:anomalies} to find the anomaly
polynomial, it is straightforward to see that requiring cancellation of quartic
gauge anomalies imposes the condition
\begin{equation}
		C^{IJ}d_J = \mu^I - 16 S^I\,,\qquad
		\mu^J = (f^0, f^1,\dots, 2f^n)\,,\qquad
		d_I = (2k_0, k_1,\dots,k_n)\,.
\end{equation}
where $C^{IJ}$ is the Cartan matrix of $\widetilde{\mathfrak{b}}=C_n^{(1)}$,
and $S^{I}=1$ if $I=0,n$ and zero otherwise. 

We therefore see that we have a similar condition as for
$\mathfrak{g}=\mathfrak{su}_{2n}$: the Cartan matrix and the
Frobenius--Schur indicators are those of $C_n^{(1)}$ in both cases. We
therefore expect $\mu^I$ to be associated with an affine coweight of that
algebra. However the sets $\widetilde{\mathcal{P}}^\vee_\text{LST}$ leading to
those two types of LSTs are different: $\mu^n$ must be even to ensure
$f^n=\mu^n/2$ is a positive integer, while the coefficient $d_I$ of the last
coroot must not necessarily be even. For theories with
$\mathfrak{g}=\mathfrak{su}_{2n}$, these two constraints are reversed. 

We therefore conclude that even in the presence of a single anti-symmetric
representation of $\mathfrak{su}_{k_n}$, all the arguments described in the
previous sections go through with only minor modifications. 

A similar analysis also applies to the two generalized quivers corresponding to
$(SU(16)\times U(1))/\mathbb{Z}_2$ LSTs shown in Table \ref{tab:twisted}, where
we can again replace the $\mathfrak{sp}_k$ algebra on the $(-1)$-curve with an
$\mathfrak{su}_k$ symmetry to introduce an anti-symmetric representation:
\begin{equation}
		\underset{[\mathfrak{su}_{\mu^0}]}{\overset{\mathfrak{su}_{d_0}}{2}}
		\underset{[\mathfrak{su}_{\mu^1/2}]}{\overset{\mathfrak{su}_{d_1}}{1}}
		\underset{[\mathfrak{su}_{\mu^2}]}{\overset{\mathfrak{su}_{d_2}}{2}}\,,
		\qquad 
		\qquad 
		\underset{[\mathfrak{su}_{\mu^1}]}{\overset{\mathfrak{su}_{d_1}}{2}}
		\overset{\displaystyle\overset{\overset{\left[\mathfrak{su}_{\mu^0} \right]}{\mathfrak{su}_{d_0}}}{2}}
		{\underset{[\mathfrak{su}_{\mu^1}]}{\overset{\mathfrak{su}_{d_2}}{2}}}
		\underset{[\mathfrak{su}_{\mu^3}]}{\overset{\mathfrak{su}_{d_3}}{2}}
		\cdots
		\underset{[\mathfrak{su}_{\mu^{K-1}}]}{\overset{\mathfrak{su}_{d_{K-1}}}{2}}
		\underset{[\mathfrak{su}_{\mu^{K}/2}]}{\overset{\mathfrak{su}_{d_K}}{1}}\,.
\end{equation}
Together with an integer $Q$, the coefficients $\mu^I$ define an affine
coweight at level 16 of $D_3^{(2)}$ and $A_{2k-1}^{(2)}$ respectively.

Both LSTs with an $\mathfrak{su}$ or $\mathfrak{sp}$ on a $(-1)$-curve are
therefore described by coweights of the same algebra, and we can apply the
techniques described in Section~\ref{sec:t-duality} to find putative T-duality
orbits in a similar way, we expect the observations made in Section~\ref{sec:higgs} to be different. The slice-subtraction algorithm should still
hold, but since the set of coweights and gauge algebras changed, so will the
involved slices. We therefore possibly expect new minimal slices to appear.

Furthermore, the presence of a hypermultiplet in the anti-symmetric
representation opens up new transitions between e.g. $Spin(32)/\mathbb{Z}_2$
LSTs of type $\mathfrak{g}=\mathfrak{su}_{2k+1}$ and $\mathfrak{su}_{2k}$.
Indeed, giving a vacuum expectation value to this field leads to the
decomposition $\mathfrak{su}_{2k}\to\mathfrak{sp}_k$. For LSTs on a $0$-curve,
this was shown to be associated with the exotic slice $h_{2,k}$
\cite{Mansi:2023faa}. This type of deformation is indeed always possible
\cite{Johnson:2016qar}, as expected from field theory, and we have: 
\begin{equation}
		\cdots
		\underset{[\mathfrak{su}_{f^{n-1}}]}{\overset{\mathfrak{su}_{{k_{n-1}}}}{2}} 
		\underset{[\mathfrak{su}_{f^n}]}{\overset{\mathfrak{su}_{2k_n}}{1}}
		\qquad\longrightarrow\qquad
		\cdots
		\underset{[\mathfrak{su}_{f^{n-1}}]}{\overset{\mathfrak{su}_{{k_{n-1}}}}{2}} 
		\underset{[\mathfrak{so}_{\mu^n}]}{\overset{\mathfrak{sp}_{k_n}}{1}}\,.
\end{equation}
We leave a more extensive analysis of the space of dominant coweights involving
anti-symmetric representations and the possible transitions between different
types of LSTs for future work.

\subsection{\texorpdfstring{$\mathfrak{e}_8\oplus\mathfrak{e}_8$}{e8+e8} Heterotic Little String Theories}
\label{ssec:e8e8theories}

Due to their perturbative nature in the low-energy regime, we have only
considered $\mathfrak{so}_{32}$ heterotic LSTs in this work. Via the standard
duality between the $\mathfrak{so}_{32}$ and
$\mathfrak{e}_8\oplus\mathfrak{e}_8$ heterotic strings, we can have access to a
larger landscape of theories, which can exhibit what we referred to as ``fiber-base''
T-dualities in Section~\ref{sec:t-duality}. These LSTs are constructed
similarly through stacks of NS5-branes probing a $\mathbb{C}^2/\Gamma$
orbifold. 

They are labeled by a pair of embeddings of the finite group into the
\emph{group} $E_8$: $\mu_L^\vee,\mu_R^\vee \in\text{Hom}(\Gamma, E_8)$
corresponding to choices of connection at infinity for the ``left'' and ``right''
$\mathfrak{e}_8$ factors. In those cases, the presence of curves with
self-intersections different from what we have encountered in the rest of this work
implies the existence of non-perturbative states. While field-theoretic
techniques can be employed to study some of their properties, in particular
their anomalies \cite{Ohmori:2014pca, Ohmori:2014kda, Shimizu:2016lbw,
Shimizu:2017kzs,Baume:2023onr, Fazzi:2023ulb}, one usually resorts to geometric
engineering.

When both embeddings are trivial, $\mu_L^\vee=\varnothing=\mu_R^\vee$, one
obtains an $\mathfrak{e}_8\oplus\mathfrak{e}_8$ flavor symmetry, with
generalized quivers given in Table \ref{tab:het-e8-e8-unbroken} of Appendix
\ref{app:unbroken-LST}. One observes, starting with $Q$ or $M$ NS5-branes
probing the same $\mathbb{C}^2/\Gamma$ orbifold in the $\mathfrak{so}_{32}$ and
$\mathfrak{e}_8\oplus\mathfrak{e}_8$ picture, respectively, that the theories
with trivial embedding can only be T-dual if
\begin{equation}
		\kappa_R = \Gamma (M + r_{\mathfrak{g}} + 2) + 2\,,\qquad \text{dim}(\text{CB}) = h^\vee M-\text{dim}(\mathfrak{g})\,,\qquad M=Q+n_{\mathfrak{g}}-2(r_{\mathfrak{g}}+2)\,,
\end{equation}
where $n_{\mathfrak{g}}=2,4,6,8,12$ for
$\mathfrak{g}=\mathfrak{su}_K,\mathfrak{so}_{2K},\mathfrak{e}_6,\mathfrak{e}_7,\mathfrak{e}_8$.
The difference between the two instanton numbers $Q$ and $M$ on each side of
the duality is related to the fractionalization of the five-branes, see e.g.
references \cite{DelZotto:2014hpa,DelZotto:2022xrh, DelZotto:2022ohj, Mekareeya:2017sqh}. In the
F-theory picture, $n_{\mathfrak{g}}$ corresponds to (minus) the smallest
self-intersection appearing in the quiver, as can be seen in Table
\ref{tab:het-e8-e8-unbroken}.

\paragraph{A subtlety with T-duality:} the discussion above matches the
T-duality invariants at a generic point of the tensor-branch moduli space. As
was first observed in reference \cite{Aspinwall:1997ye}, on the
$\mathfrak{e}_8\oplus\mathfrak{e}_8$ side when reaching the singular point
corresponding to the actual Little String Theory, geometrically, one must
supplement this data with a choice of a contraction map. Given a brane system,
this is equivalent to a choice of partition $M = M_L + M_R$ indicating how many
five-branes are collapsed to the ``left'' and ``right'' nine-branes. This was
studied in detail for $\mathfrak{g}=\mathfrak{su}_K$ in reference
\cite{Lawrie:2023uiu}, where it was explicitly shown that different choices of
$M_L$ and $M_R$ lead to different Higgs-branches or equivalently different
magnetic quivers. The Coulomb branches of theories with different $(M_L, M_R)$
are therefore disconnected, but related geometrically by flop transitions as
those are simply contractions of the same underlying theory. 

This is why we must be allowed to move in the full \emph{extended} Coulomb branch 
moduli space to call two theories T-dual in Definition \ref{def:t-duality}
which in geometry corresponds to the 
\textit{extended} K\"ahler moduli space. Hence the different 6d F-theory models must come from the same 
\textit{birationally} equivalent CY-threefolds.

\paragraph{Non-trivial embeddings:} more general theories correspond to
generic choices of flat connections. The embeddings $\text{Hom}(\Gamma, E_8)$
and the possible quivers are in principle classified \cite{Heckman:2015bfa,
Frey:2018vpw}; however the precise mapping between the two is not known in full
generality. For $\mathfrak{g}=\mathfrak{su}_K$, embeddings
$\text{Hom}(\mathbb{Z}_K, E_8)$ are known to be classified by affine coweights
of $E^{(1)}_8$ at level $K$ \cite{Kac:1990gs}, see also reference
\cite{Fazzi:2023ulb} for recent work in the context of type-A SCFTs. An
algorithm to pass from the coweight to the generalized quiver has been proposed 
in reference \cite{Mekareeya:2017jgc}.

For other types of algebras, one also expects to associate a precise affine coweight
of $E_8^{(1)}$ to an SCFT as they are the natural object to classify the
embeddings. Two coweights $\mu^\vee_L$ and $\mu^\vee_R$ can then be used to
``glue'' two such SCFTs to reach any $\mathfrak{e}_8\oplus\mathfrak{e}_8$ LST
\cite{Bhardwaj:2015oru, Heckman:2018pqx}. Using the formula for $\kappa_R$,
this suggests that one simply sums the contributions from both theories
independently. 

Considering the various results in this work for $\mathfrak{so}_{32}$ LSTs, it
is very tempting to conjecture that one should be able to always find two
affine coweights such that $\kappa_R$ is obtained as a purely Lie-theoretic
quantity 
\begin{equation}
		\kappa_R\,\overset{?}{\propto}\,\langle \widetilde{T}, \widetilde{\mu}_L^\vee + \widetilde{\mu}_R^\vee\rangle\,,\qquad 
\end{equation}
for some affine weight $\widetilde{T}$. Note however that the precise mapping
is not clear. For $\mathfrak{so}_{32}$ LSTs, the coweight is associated with the
algebra corresponding to the orbifold singularity rather than
$\mathfrak{so}_{32}$, while type-A LSTs are associated with coweights of
$E_8^{(1)}$. It would therefore be interesting to investigate whether a similar
classification to that of $\mathfrak{so}_{32}$ LSTs is possible. This would
have important ramifications for both LSTs and the SCFTs that they are made out
of. Not only could certain quantities be efficiently computed without needing
to explicitly know the generalized quiver \cite{Baume:2023onr, Fazzi:2023ulb},
but we also expect the induced partial order to hold information on the
associated network of RG flows, as well as the network of internal T-duality
with similar tools developed in this work.

\paragraph{RG flows:} the duality between $\mathfrak{so}_{32}$ and
$\mathfrak{e}_8\oplus\mathfrak{e}_8$ could in principle also shed some light on
whether some of the transitions discussed in Section~\ref{sec:higgs} are indeed
associated with RG flows although there is no change in the Higgs-branch
dimension. For type-A LSTs, it has been argued that if there is a Higgs-branch
flow between two LSTs, there must be a similar Higgs-branch flow on the other
side of the T-duality, and vice versa \cite{Lawrie:2023uiu}. 

As an example, consider the following potential RG flow between two
$Spin(32)/\mathbb{Z}_2$ LSTs with $\mathfrak{g}=\mathfrak{e}_6$, corresponding
to the covering between the next-to-minimal and the trivial coweights:
\begin{equation}\label{RG-e6-min}
	\underset{[\mathfrak{so}_{30}]}{\overset{\mathfrak{sp}_{Q}}{1}}\,\overset{\mathfrak{so}_{4Q - 14}}{4}\,\overset{\mathfrak{sp}_{3Q - 22}}{1}\,\overset{\mathfrak{su}_{4Q - 29}}{2}\,\underset{[\mathfrak{su}_{1}]}{\overset{\mathfrak{su}_{2Q - 14}}{2}}
	 \qquad\overset{\resizebox{2.5cm}{!}{\input{figures/slice_sF4.tex}}}{\relbar\joinrel\relbar\joinrel\relbar\joinrel\longrightarrow}\qquad
	\underset{[\mathfrak{so}_{32}]}{\overset{\mathfrak{sp}_{Q}}{1}}\,\overset{\mathfrak{so}_{4Q - 16}}{4}\,\overset{\mathfrak{sp}_{3Q - 24}}{1}\,\overset{\mathfrak{su}_{4Q - 32}}{2}\,\overset{\mathfrak{su}_{2Q - 16}}{2}
\end{equation}
The minimal slice $\mathcal{S}$ is depicted above; the change in the dimension
of the fundamental representation of the gauge group arranges like the comarks
of $E_6^{(2)}$. This is an example of a minimal slice where there is no change
in the dimension of the Higgs branch. It is therefore natural to ask what
happens on the other side of a fiber-base duality. Focusing for simplicity only
on $\mathfrak{e}_8\oplus\mathfrak{e}_8$ duals with one of the embeddings
trivial, say $\widetilde{\mu}_L^\vee=\varnothing$, we find only one candidate:
\begin{equation}\label{RG-UV-e6-min}
	\underset{[\mathfrak{so}_{30}]}{\overset{\mathfrak{sp}_{Q}}{1}}\,\overset{\mathfrak{so}_{4Q - 14}}{4}\,\overset{\mathfrak{sp}_{3Q - 22}}{1}\,\overset{\mathfrak{su}_{4Q - 29}}{2}\,\underset{[\mathfrak{su}_{1}]}{\overset{\mathfrak{su}_{2Q - 14}}{2}}
		\quad\overset{\text{T-dual}}{\longleftrightarrow}\quad
		\underset{[\mathfrak{e}_8]}{1} \, 2 \overset{\mathfrak{su}_2}{2} \overset{\mathfrak{g}_2}{3} 1 \overset{\mathfrak{f}_4}{5} 1 \overset{\mathfrak{su}_3}{3} 1 \underbrace{\overset{\mathfrak{e}_6}{6}1 \overset{\mathfrak{su}_3}{3} 1 \cdots 1 \overset{\mathfrak{su}_3}{3} 1}_{Q-10}\overset{\mathfrak{e}_6}{6}1 \overset{\mathfrak{su}_3}{3} 1\overset{\mathfrak{e}_6}{6} 1 \overset{\mathfrak{su}_3}{3} \underset{[\mathfrak{u}_1]}{1} \overset{\mathfrak{so}_{10}}{4} \underset{[\mathfrak{so}_{14}]}{\overset{\mathfrak{sp}_2}{1}}\,,
\end{equation}
with $\kappa_R = 24Q - 150$ and $\text{dim}(\text{CB}) = 12Q - 70$. On the other hand for
the second $Spin(32)/\mathbb{Z}_2$ theory, we find two candidates with the property above:
\begin{equation}\label{RG-IR-e6-min}
		\begin{aligned}
		\underset{[\mathfrak{so}_{32}]}{\overset{\mathfrak{sp}_{Q}}{1}}\,\overset{\mathfrak{so}_{4Q - 16}}{4}\,\overset{\mathfrak{sp}_{3Q - 24}}{1}\,\overset{\mathfrak{su}_{4Q - 32}}{2}\,\overset{\mathfrak{su}_{2Q - 16}}{2}
		\quad\overset{\text{T-dual}}{\longleftrightarrow}\quad & 
		\underset{[\mathfrak{e}_8]}{1} \, 2 \overset{\mathfrak{su}_2}{2} \overset{\mathfrak{g}_2}{3} 1 \overset{\mathfrak{f}_4}{5} 1 \overset{\mathfrak{su}_3}{3} 1 \underbrace{\overset{\mathfrak{e}_6}{6}1 \overset{\mathfrak{su}_3}{3} 1 \cdots 1 \overset{\mathfrak{su}_3}{3} 1}_{Q-10}\overset{\mathfrak{e}_6}{6} 1 \overset{\mathfrak{su}_3}{3} 1 \overset{\mathfrak{f}_4}{5} 1 \overset{\mathfrak{g}_2}{3} \overset{\mathfrak{su}_2}{2} 2 \, \underset{[\mathfrak{e}_8]}{1}\\
		& \quad\qquad\qquad\qquad\qquad\updownarrow\text{\scriptsize T-dual}\\
		& \underset{[\mathfrak{e}_8]}{1} \, 2 \overset{\mathfrak{su}_2}{2} \overset{\mathfrak{g}_2}{3} 1 \overset{\mathfrak{f}_4}{5} 1 \overset{\mathfrak{su}_3}{3} 1 \underbrace{\overset{\mathfrak{e}_6}{6}1 \overset{\mathfrak{su}_3}{3} 1 \cdots 1 \overset{\mathfrak{su}_3}{3} 1}_{Q-10}\overset{\mathfrak{e}_6}{6} 1\overset{\mathfrak{su}_{3}}{3}1\overset{[\mathfrak{su}_3]}{\overset{\displaystyle 1}{\overset{\mathfrak{e}_{6}}{6}}}1\overset{\mathfrak{su}_{3}}{3}\underset{[\mathfrak{e}_6]}{1} 
		\end{aligned}
\end{equation}
with $\kappa_R = 24Q - 166$ and $\text{dim}(\text{CB}) = 12Q - 78$. We have
checked that the theories in equations \eqref{RG-UV-e6-min} and
\eqref{RG-IR-e6-min} are indeed T-dual: using the methods of reference
\cite{Ahmed:2023lhj} we have found explicit toric realizations from which these
theories originate.

Furthermore, on the $\mathfrak{e}_8\oplus\mathfrak{e}_8$ side, no RG flow can take
place between these theories. Generically, in the generalized quiver picture an
RG flow corresponds to either smoothing fiber singularities, corresponding to
breaking a gauge algebra to one of its subgroups, blowing down an undecorated
$(-1)$-curve with an appropriate change in the self-intersection of the
neighboring curves, or a combination of both, see e.g. reference
\cite{Bao:2024wls}.

One may therefore be tempted to say that the minimal slice in equation
\eqref{RG-e6-min} is unphysical as there is no possible RG flow on the
$\mathfrak{e}_8\oplus\mathfrak{e}_8$ heterotic side. However, we have found
examples where, while a minimal slice of $\mathfrak{so}_{32}$ heterotic LSTs
induces no change in the Higgs-branch dimension, an RG flow is possible for at
least one of the T-dual $\mathfrak{e}_8\oplus\mathfrak{e}_8$ heterotic LSTs.

Indeed, let us come back to the $\mathfrak{g}=\mathfrak{so}_{10}$ example we
have already encountered in Section~\ref{sec:slice-subtraction}:
\begin{equation}\label{example-HB-0}
	\underset{[\mathfrak{so}_{2}]}{\overset{\mathfrak{sp}_{Q - 7}}{1}}\,
	 \overset{\displaystyle \overset{[\mathfrak{so}_{30}]}{\overset{\mathfrak{sp}_{Q}}{1}}}
	 {\overset{\mathfrak{so}_{4Q - 14}}{4}}\,
	 \overset{\mathfrak{sp}_{2Q - 15}}{1}\,
	 \overset{\mathfrak{su}_{2Q - 15}}{2}
	 \overset{\resizebox{2.5cm}{!}{\input{figures/example_slice_C4.tex}}}{\relbar\joinrel\relbar\joinrel\relbar\joinrel\longrightarrow}\qquad
	\overset{\mathfrak{sp}_{Q - 8}}{1}\,
	\overset{\displaystyle \overset{[\mathfrak{so}_{32}]}{\overset{\mathfrak{sp}_{Q}}{1}}}
	{\overset{\mathfrak{so}_{4Q - 16}}{4}}\,
	\overset{\mathfrak{sp}_{2Q - 16}}{1}\,
	\overset{\mathfrak{su}_{2Q - 16}}{2}
\end{equation}
This transition also does not change the dimension of the Higgs branch. As
above, for the first theory we find the following putative T-dual
$\mathfrak{e}_8\oplus\mathfrak{e}_8$ heterotic LSTs: 
\begin{equation}\label{example-hb-0-so32-UV}
		\underset{[\mathfrak{so}_{2}]}{\overset{\mathfrak{sp}_{Q - 7}}{1}}\, \overset{\displaystyle \overset{[\mathfrak{so}_{30}]}{\overset{\mathfrak{sp}_{Q}}{1}}}{\overset{\mathfrak{so}_{4Q - 14}}{4}}\, \overset{\mathfrak{sp}_{2Q - 15}}{1}\,\overset{\mathfrak{su}_{2Q - 15}}{2}
		\quad\overset{\text{T-dual?}}{\longleftrightarrow}\quad
		\underset{[\mathfrak{e}_8]}{1} \, 2 \overset{\mathfrak{su}_2}{2} \overset{\mathfrak{g}_2}{3} 1 \overset{\mathfrak{so}_9}{4} \overset{\mathfrak{sp}_{1}}{1}\underbrace{\overset{\mathfrak{so}_{10}}{4}\overset{\mathfrak{sp}_{1}}{1}\cdots \overset{\mathfrak{so}_{10}}{4}\overset{\mathfrak{sp}_{1}}{1}}_{Q-9} \overset{\mathfrak{so}_{10}}{4}\overset{\mathfrak{sp}_{1}}{1}\underset{[\mathfrak{u}_1]}{\overset{\mathfrak{so}_{10}}{3}}\underset{[\mathfrak{so}_{14}]}{\overset{\mathfrak{sp}_{2}}{1}}
\end{equation}
with $\kappa_R = 12Q - 64 $ and $\text{CB} = 8Q - 41$. On the other hand, the IR theory has four putative T-dual $\mathfrak{e}_8\oplus\mathfrak{e}_8$ heterotic LSTs:
\begin{equation}\label{Example-HB-0}
		\begin{aligned}
		\overset{\mathfrak{sp}_{Q - 8}}{1}\, \overset{\displaystyle \overset{[\mathfrak{so}_{32}]}{\overset{\mathfrak{sp}_{Q}}{1}}}{\overset{\mathfrak{so}_{4Q - 16}}{4}}\, \overset{\mathfrak{sp}_{2Q - 16}}{1}\,\overset{\mathfrak{su}_{2Q - 16}}{2}
		\quad\overset{\text{T-dual?}}{\longleftrightarrow}\quad & 
		\underset{[\mathfrak{e}_8]}{1} \, 2 \overset{\mathfrak{su}_2}{2} \overset{\mathfrak{g}_2}{3} 1 \overset{\mathfrak{so}_9}{4} \overset{\mathfrak{sp}_{1}}{1}\underbrace{\overset{\mathfrak{so}_{10}}{4}\overset{\mathfrak{sp}_{1}}{1}\cdots \overset{\mathfrak{so}_{10}}{4}\overset{\mathfrak{sp}_{1}}{1}}_{Q-9}\overset{\mathfrak{so}_{9}}{4}1\overset{\mathfrak{g}_{2}}{3}\overset{\mathfrak{su}_{2}}{2}2\,\underset{[\mathfrak{e}_8]}{1} \\
		& \quad\qquad\qquad\qquad\qquad\updownarrow\text{\scriptsize T-dual?}\\
		& \underset{[\mathfrak{e}_8]}{1} \, 2 \overset{\mathfrak{su}_2}{2} \overset{\mathfrak{g}_2}{3} 1 \overset{\mathfrak{so}_9}{4} \overset{\mathfrak{sp}_{1}}{1}\underbrace{\overset{\mathfrak{so}_{10}}{4}\overset{\mathfrak{sp}_{1}}{1}\cdots\overset{\mathfrak{so}_{10}}{4}\overset{\mathfrak{sp}_{1}}{1}}_{Q-9}\overset{\mathfrak{so}_{9}}{4}1\underset{[\mathfrak{sp}_1]}{\overset{\mathfrak{so}_{7}}{3}}\overset{\mathfrak{su}_{2}}{2}\underset{[\mathfrak{e}_7]}{1} \\
		& \quad\qquad\qquad\qquad\qquad\updownarrow\text{\scriptsize T-dual?}\\
&\underset{[\mathfrak{e}_8]}{1} \, 2 \overset{\mathfrak{su}_2}{2} \overset{\mathfrak{g}_2}{3} 1 \overset{\mathfrak{so}_9}{4} \overset{\mathfrak{sp}_{1}}{1}\underbrace{\overset{\mathfrak{so}_{10}}{4}\overset{\mathfrak{sp}_{1}}{1}\cdots\overset{\mathfrak{so}_{10}}{4}\overset{\mathfrak{sp}_{1}}{1}}_{Q-9} \overset{\mathfrak{so}_{9}}{4}1\overset{\mathfrak{so}_{7}}{3}\underset{[\mathfrak{so}_{16}]}{\overset{\mathfrak{sp}_{2}}{1}}\\
		& \quad\qquad\qquad\qquad\qquad\updownarrow\text{\scriptsize T-dual?}\\
& \underset{[\mathfrak{e}_8]}{1} \, 2 \overset{\mathfrak{su}_2}{2} \overset{\mathfrak{g}_2}{3} 1 \overset{\mathfrak{so}_9}{4} \overset{\mathfrak{sp}_{1}}{1}\underbrace{\overset{\mathfrak{so}_{10}}{4}\overset{\mathfrak{sp}_{1}}{1}\cdots\overset{\mathfrak{so}_{10}}{4}\overset{\mathfrak{su}_{2}}{1}}_{Q-9}\overset{\mathfrak{so}_{10}}{4}\overset{\mathfrak{sp}_{1}}{1}\underset{[\mathfrak{su}_8]}{\overset{\mathfrak{su}_{5}}{2}}\,\oplus[\mathfrak{u}_1]
		\end{aligned}
\end{equation}
with $\kappa_R=12Q - 70$ and $\text{CB}= 8Q - 45$. We unfortunately did not
find a toric realization of these theories; they however have the same T-dual
invariants. Contrary to the $\mathfrak{g}=\mathfrak{e}_6$ example discussed
above, one can reach the last two $\mathfrak{e}_8\oplus\mathfrak{e}_8$
heterotic LSTs through an RG flow starting from the theory in equation
\eqref{example-hb-0-so32-UV}. On the other hand there cannot be a deformation
leading to the first two, as the number of curves is larger than that of the UV
theory. This is peculiar, as it contradicts the observation that was made for
type-A heterotic LSTs, namely that if there is an RG flow on one side of the
duality, there should be one in the other.

In the absence of a complete classification of
$\mathfrak{e}_8\oplus\mathfrak{e}_8$ in terms of coweights, we cannot perform
an exhaustive analysis of how Hasse diagrams behave under (putative) T-duality.
These two examples however show that the mapping is subtle. Geometrically, if
two theories are related by an RG flow, there should be a complex-structure
deformation from one to the other. On the other hand, as T-dual models
correspond to different elliptic fibrations of the same Calabi--Yau, it is not
guaranteed that the deformation will preserve the fibration structure. It would
therefore be interesting to study how T-duality acts on the Higgs-branch of
specific models, and whether new patterns arise.

\subsection{Type-II LSTs}

Another class of Little String Theories can be obtained from Type-II string
theory, giving them their name \cite{DelZotto:2020sop}. In the F-theory
construction, they are engineered from an elliptically-fibered Calabi--Yau over
a base birational to $(T^{2}\times \mathbb{C})/\Lambda$ with $\Lambda\subset
SU(2)$ \cite{Bhardwaj:2015oru}. Both the fiber and the base are therefore described by a Kodaira
singularity $(\mathfrak{g}_F, \mathfrak{g}_B)$. T-duality corresponds to
exchanging the two singularities---and is an example of fiber-base duality in
the nomenclature of Section~\ref{sec:t-duality}---and was established through
explicit geometric constructions \cite{Baume:2024oqn}. Such theories allow for
more complicated intersection patterns. However, when the singularity
$\mathfrak{g}_F$ corresponds to one of the classical algebras, the adjacency
pattern follows once again Classification \ref{class:base-topology}, and
corresponds to an affine algebra $\widetilde{\mathfrak{b}}$. The precise gauge
algebras are uniquely fixed by the pair $(\mathfrak{g}_F, \mathfrak{g}_B)$.
However, contrary to $\mathfrak{so}_{32}$ heterotic LSTs, the gauge content is
not fixed by the McKay correspondence, in the sense that on a given node, the
algebra does not follow directly from the type of representation of the
corresponding finite subgroup of $SU(2)$.

On the other hand, they have similar properties to the $\mathfrak{so}_{32}$
LSTs. For instance, the cancellation of quartic anomalies still takes the form
\begin{equation}
		C^{AB} d_B = f^A - 16 S^A\,,
\end{equation}
where $S^A$ follows more generally from Equation \eqref{classical-relations}
rather than the Frobenius--Schur indicator of the algebra
$\widetilde{\mathfrak{b}}$ associated to the Cartan matrix $C^{AB}$. Contrary
to the LSTs we have discussed throughout this work, they however always
satisfies $\theta_A S^A=0$. From what we have argued several times, it means
that type-II LSTs whose fiber singularity $\mathfrak{g}_F$ allows for a high
rank are classified by an affine coweight of $\widetilde{\mathfrak{b}}$ at
level zero:
\begin{equation}
		\text{type-II LSTs:}\qquad \widetilde{\mu}^\vee = (0, 0, Q)\in \widetilde{\mathcal{P}}^\vee(\widetilde{\mathfrak{b}})\,,
\end{equation}
and as before $\widetilde{\mu}^\vee - 16\widetilde{S}^\vee$ must be a positive
coroot to ensure that gauge algebras have integer rank.

It follows that there cannot be any continuous flavor symmetry associated with
such LSTs, up to Abelian factors delocalized over the whole quiver. This gives
an alternative field-theoretic proof of a more general bound: for any type-II
LST, an analysis of the worldsheet theory of the Little String shows that the
total flavor rank can be at most two \cite{Baume:2024oqn}.

In each case, the partial order of affine coweights suggests a simple RG flow
lowering $Q$, as the allowed affine coweights can only be
of the form $\widetilde{\nu}^\vee = (0, 0, Q-c)\lessdot \widetilde{\mu}^\vee =
(0, 0, Q)$ for a given constant $c$ depending on the type of algebras involved.

\subsection{SCFTs and Coweights}

In this work, we have only discussed Little String Theories. However, most of
our analysis of the cancellation of quartic anomalies and RG flows also applies
to six-dimensional SCFTs with classical gauge algebras. Indeed, in our analysis
of the anomaly polynomial, we remained for the most part agnostic on whether we
were considering an LST or an SCFT, and most of the discussion in Section~\ref{sec:anomalies} applies to SCFTs as well. 

In fact, SCFTs are arguably simpler than LSTs: the cancellation of quartic
anomalies given in Equation \eqref{quartic-cancellation} can be inverted, as
$C^{IJ}$ is the Cartan matrix of a \emph{finite} Lie algebra $\mathfrak{b}$. We
therefore expect all SCFTs with classical gauge algebras of high-enough rank to
be similarly labeled uniquely by a coweight $\mu^\vee$ of $\mathfrak{b}$.

For gauge algebras of type $\mathfrak{su}_k$, requiring that $\mu^\vee -
16S^\vee$ is a coroot so as to obtain well-defined gauge algebras is precisely the
condition demanded in reference \cite{Bourget:2021siw} to reconstruct the Hasse
diagram of RG flows associated with unitary three-dimensional $\mathcal{N}=4$
quiver from slices in the corresponding affine Grassmannian. This therefore
explains why the original slice-subtraction algorithm \cite{Lawrie:2024zon}
correctly reproduced the Hasse diagram obtained via
magnetic-quiver methods: it simply corresponds to the partial order of coweights,
with (in the case considered there) only Kraft--Procesi transitions allowed.

We therefore expect our generalization of this algorithm to also extend to
type-D conformal matter in the same way we found the minimal slices in
Section~\ref{sec:higgs}, possibly with additional slices beyond those we have
found here.

Furthermore, the overwhelming majority of SCFTs fall into infinite families
labeled by a choice of finite ADE algebra $\mathfrak{b}$ describing the base,
and a choice of algebra $\mathfrak{g}_F$ describing the gauge symmetry---which
is associated with the fiber singularity at the singular point in the F-theory
description. As the base corresponds to a finite, rather than affine, algebra,
SCFTs allow for a rich network of flavor symmetries. A choice of tuple
$(\mathfrak{b}, \mathfrak{g}_F)$ then defines a ``parent'' theory, from which
all other SCFTs falling in the same class can be reached through Higgs-branch
RG flows. As hinted above, these RG flows are labeled by quantities related to
the group theory of $\mathfrak{g}_F$.

For type-A algebras $\mathfrak{b}$ describing the base, one describes the RG
flows by breaking the flavor symmetry on either side of the quiver: for
instance, in the context of conformal matter SCFTs \cite{DelZotto:2014hpa}, one
makes a choice of a pair of nilpotent orbits of $\mathfrak{g}_F$
\cite{Heckman:2016ssk}. From the perspective of this work, it seems more
natural to encode the data of those two nilpotent orbits directly in a dominant
coweight of $\mathfrak{b}$ rather than treating the breaking on both sides
independently. In addition to a simplification of the description, it also
enables one to consider more types of breaking, and simultaneously describe all
possible choices of classical algebras $\mathfrak{g}_F$ in the same framework:
RG flows between theories with $\mathfrak{g}_F=\mathfrak{su}_k$ and
$\mathfrak{g}_F=\mathfrak{su}_{k-n}$ are both described by a dominant coweight
of $\mathfrak{b}$.\footnote{The authors are grateful to Craig Lawrie for
discussions on this topic and for sharing an advanced copy of the
manuscript of reference \cite{nilps-to-nilps} with us, where it is shown through
Class-$\mathcal{S}$ and 6d SCFT techniques that a breaking corresponding to two
nilpotent orbits can be understood in terms of a single nilpotent orbit
of a larger algebra.}

Furthermore, it can be shown that by working at a certain point of the tensor
branch, questions related to anomalies can be treated in a uniform manner for
any given algebra, including those of exceptional type \cite{Baume:2023onr}.
Similarly to what we have found for classical algebras, the complete anomaly
polynomial only depends on the choice of the base and group-theory data of the
nilpotent orbits. In particular, as in this work, most of the results do not
depend on the minutiae of the F-theory description but only necessitate the
Lie-theoretic quantities related to $\mathfrak{g}_F$. It would therefore be
interesting to see whether the language of coweights can be applied to
exceptional algebras using some of the ideas developed here.

\section{Conclusions and Summary of Results}\label{sec:conclusion}

Six-dimensional $\mathfrak{so}_{32}$ LSTs with minimal supersymmetry admit only
classical flavor and gauge zero-form symmetries. This fact enabled us to
understand several of their field-theoretic properties on the tensor branch
in terms of dominant coweights of the affine algebra $\widetilde{\mathfrak{b}}$
defining the topology of their generalized quiver independent of their
string/F-theory realization. For instance, in Classification
\ref{class:base-topology} we have shown that the topology of the quiver of any
$6d$ $\mathcal{N}=(1,0)$ SQFT with hypermultiplets only in the bifundamental
representations of classical algebras corresponds to the Dynkin diagram of a Lie
algebra.

The main constraint that we have used throughout this work is the cancellation
of quartic gauge anomalies, which allows for an interpretation in terms of
coweights and enabled us to use properties of root systems. The flavor and
gauge algebras are then uniquely determined by properties of
$\widetilde{\mathfrak{b}}$, rather than being an additional input. The
remaining freedom then corresponds to the rank of these algebras, which is
encoded in an affine dominant coweight of $\widetilde{\mathfrak{b}}$. We have
checked this for all known $\mathfrak{so}_{32}$ heterotic LSTs, including those
whose spectrum includes hypermultiplets transforming in the anti-symmetric
representation of an $\mathfrak{su}_k$ gauge algebra. This therefore generalizes the original
classification \cite{Blum:1997mm} (see Classification \ref{class:spin32}), to
all known cases:\\ 

\begin{classification}[all $\mathfrak{so}_{32}$ LSTs] All known
		$\mathfrak{so}_{32}$ heterotic LSTs can be labeled by a choice
		of affine dominant coweight of the allowed affine algebras
		$\widetilde{\mathfrak{b}}$ given in Table \ref{tab:dynkin-bases}. We differentiate between two classes of LSTs:
		\begin{enumerate}
				\item $Spin(32)/\mathbb{Z}_2$ LSTs: the coweight is of the form $\widetilde{\mu}^\vee = (\mu^\vee, 32, 2Q)$
				\item $(SU(16)\times U(1))/\mathbb{Z}_2$ LSTs: the coweight is of the form $\widetilde{\mu}^\vee = (\mu^\vee, 16, Q)$
		\end{enumerate}
		The components $\mu^I$ are the dimensions of the fundamental
		representations of the flavor symmetry on the $I$-th node of the Dynkin
		diagram, and must satisfy the following constraints.
		\begin{itemize}
				\item Given the affine coweight $\widetilde{S}^\vee= S^I\widetilde{\omega}_I^\vee$ whose components 
						$S^I\in\{-1,0,1\}$ can be read off the color of a
				node of the Dynkin diagram,
				$\widetilde{\mu}^\vee-16\widetilde{S}^\vee$ must be a
				positive coroot. The components $S^I$ define the type of gauge
				algebras, and $d_I=\langle\widetilde{\omega}_I, \widetilde{\mu}^\vee-16\widetilde{S}^\vee\rangle$ sets their fundamental representations.
		\item If $S^I=-1$ then $\mu^I$ is even, and if $S^I=1$ then $d_I$ is even.
		\item If the LST hosts a configuration $\underset{[\mathfrak{su}_{f^I}]}{\overset{\mathfrak{su}_{d_I}}{1}}$, then $\mu^I=2f^I$.
		\end{itemize}
\end{classification}
This labeling is unique except in cases with a curve configuration
$\overset{\mathfrak{su}_{d_I}}{1}$, as the spectrum includes an anti-symmetric
representation that may mimic another configuration. Passing from a particular
generalized quiver describing an LST to its coweight and vice versa is
explained in more detail in Sections \ref{sec:classification-spin32-lst},
\ref{sec:su16-lst}, and \ref{sec:antisymmetric}. 

This Lie-algebraic description of $\mathfrak{so}_{32}$ heterotic LSTs enables us
to simplify the search for T-dual pairs. Each of the known T-duality invariants
can be written in terms of closed-form expressions involving only
$\widetilde{\mu}^\vee$. These can then be used to efficiently partition the
cone of \emph{finite} dominant coweights, where each block of the partition is
made out of coweights that lead to LSTs with the same T-dual invariants up to a
shift in the instanton number $Q$, and is explained in Classification
\ref{class:duality-structure}.

This therefore reduces the search for putative T-dual pairs to a simple
linear-optimization problem that is amenable to numerical scans given a
particular algebra $\widetilde{\mathfrak{b}}$. For instance, in Classification
\ref{class:orbits}, we have shown that at maximal flavor rank, the number of
putative T-duality orbits is given by the order of the center of the algebra
$\mathfrak{g}$---except for $\mathfrak{g}=\mathfrak{so}_{4k}$, where there are
only two orbits.
\begin{table}[t]
    \centering
    \begin{threeparttable}
        \begin{tabular}[t]{cccc}
				\toprule
			& $Spin(32)/\mathbb{Z}_2$ LST & & $(SU(16)\times U(1))/\mathbb{Z}_2$ LST \\\midrule
			$B_n^{(1)}:$ & $\begin{matrix}\input{figures/B.tex}\end{matrix}$ & $A_{2l-1}^{(2)}:$  &$\begin{matrix}\input{figures/Dk_2.tex}\end{matrix}$ \\
			$C_n^{(1)}:$ & $\begin{matrix}\input{figures/C.tex}\end{matrix}$ & & \\
			$D_{2n}^{(1)}:$ & $\begin{matrix}\input{figures/Deven.tex}\end{matrix}$ & $D_3^{(2)}:$ & $\begin{matrix}\input{figures/D_3_2.tex}\end{matrix}$\\
			$F_4^{(1)}:$ & $\begin{matrix}\input{figures/F4.tex}\end{matrix}$ & $E_6^{(2)}:$ & $\begin{matrix}\input{figures/E6_2.tex}\end{matrix}$ \\
			$E_7^{(1)}:$ & $\begin{matrix}\input{figures/E7.tex}\end{matrix}$ & & \\
			$E_8^{(1)}:$ & $\begin{matrix}\input{figures/E8.tex}\end{matrix}$ & & \\
	\\\midrule
        \bottomrule
        \end{tabular}
    \end{threeparttable}
	\caption{Algebras $\widetilde{\mathfrak{b}}$ corresponding to the base of
			an $\mathcal{N}=(1,0)$ six-dimensional Little String Theory. The
			integers on each node indicate the marks $\theta_A$ of the affine algebra, and its color
			the associated flavor symmetry: $\begin{tikzpicture}\node[node,
					fill=red] (A0) at (0,0) {};\end{tikzpicture}
					\leftrightarrow\mathfrak{so}_k$;
					$\begin{tikzpicture}\node[node, fill=blue] (A0) at (0,0)
					{};\end{tikzpicture} \leftrightarrow \mathfrak{sp}_k$;
					$\begin{tikzpicture}\node[node, fill=white] (A0) at (0,0)
					{};\end{tikzpicture} \leftrightarrow \mathfrak{su}_k$. A
					choice of $\widetilde{\mathfrak{b}}$ must then be supplemented by an
					affine coweight representing the dimensions of the fundamental
					representations of the flavor symmetries. The complete data
			of the LST needed to study T-duality can then be recovered from the
	coweight. The other affine or twisted algebras do not lead to consistent
	LSTs.}
    \label{tab:dynkin-bases}
\end{table}

In the absence of a way to easily geometrically engineer all possible
$\mathfrak{so}_{32}$ LSTs, the partitioning of the cone of dominant coweights
gives a necessary condition for T-duality. It is not known whether there are
additional invariants that call for a refinement of the methods developed in
this work. However, we have shown that for certain low-rank algebras, they
satisfy linear relations, and only two out of three of the invariants we
considered here are needed. As we are not aware of any geometric construction
where the known invariants fail to label T-duality orbits, it is tempting to
conjecture that our method leads to \emph{sufficient} conditions to determine
T-duality. 

We have furthermore studied the partial order associated with dominant
coweights. This has led us to put forward slice-subtraction Algorithm
\ref{alg:slice-subtraction} that generalizes the one recently put forward for
six-dimensional generalized unitary quivers associated with SCFTs
\cite{Lawrie:2024zon}. Our generalization should apply to all
unitary-orthosymplectic generalized quivers. Given an LST, the approach
efficiently reconstructs the Hasse diagrams of all theories lower in the
partial order. We show an example of the algorithm for $Spin(32)/\mathbb{Z}_2$
LSTs with $\mathfrak{g}=\mathfrak{so}_{10}$ in Figure \ref{fig:B4-Hasse}.

In many cases, minimal changes from one coweight to another correspond to a
Kraft--Procesi transition \cite{Cabrera:2016vvv, Cabrera:2017njm}, and
reproduce the expected change in the dimension of the Higgs and Coulomb branches.
We however find several ``exotic'' slices, collated in Tables
\ref{tab:slice-non-simply-laced}--\ref{tab:exceptional-slices} for which we do
not know the associated symplectic singularity. It would be interesting to
construct the magnetic quivers for these models, and investigate if they
correspond to known cases, or whether new transitions occur.

While we expect this partial order to reproduce the network of Higgs-branch
Renormalization Group flows, we find certain slices that do not change the
dimension of the Higgs-branch. To the best of our knowledge, such minimal
slices have not appeared previously in the literature, and call for further
investigation through quiver methods. Geometrically, these deformations seem
naively possible, but F-theory geometries that include deformations involving
configurations with multiple curves have so far not been investigated
explicitly in the literature, and both techniques offer interesting avenues to
settle the question of whether these correspond to \emph{bona fide} RG flows.

We have also explored possible generalizations of our methods to
$\mathfrak{e}_8\oplus\mathfrak{e}_8$ heterotic and Type-II LSTs, as well as
SCFTs, where our results apply in certain cases. For general six-dimensional
theories with arbitrary gauge algebras, the biggest hindrance to a
generalization is the absence of a quartic Casimir for exceptional algebras.
Indeed, the main constraint we have used follows from the cancellation of
quartic gauge anomalies, and therefore does not apply to exceptional algebras.
However, $\mathfrak{e}_8\oplus\mathfrak{e}_8$ heterotic LSTs as well as
conformal matter suggest similar relationships with Lie theory. Furthermore,
we have not considered dualities with twisted compactifications
\cite{Bhardwaj:2019fzv, Anderson:2023wkr, Bhardwaj:2022ekc}, or those involving
frozen singularities and the corresponding heterotic compactification
without vector structure
\cite{Tachikawa:2015wka,Bhardwaj:2018jgp,Morrison:2023hqx, Oehlmann:2024cyn}.
It would be interesting to check whether similar group-theoretic considerations
can also be found in those cases, as the geometry of frozen F-theory vacua is
still not fully understood.

Having similar classifications in terms of explicit coweights would therefore
allow for an extension of our results not only to ``internal'' T-duality
between $\mathfrak{so}_{32}$ theories, but fiber-base dualities as well,
opening up a systematic analysis of a larger part of the landscape of dualities
of Little String Theories. 

\subsection*{Acknowledgments}

We thank Rafael \'{A}lvarez-Garc\'ia, Christian Ferko, Darrin Frey, Jonathan
Heckman, Lukas Kaufmann, Vittorio Larotonda, Craig Lawrie, Lorenzo Mansi,
Noppadol Mekareeya, Dave Morrison, and Marcus Sperling for helpful discussions.

F.B. is partially supported by the Deutsche Forschungsgemeinschaft under
Germany’s Excellence Strategy -- EXC 2121 “Quantum Universe” -- 390833306, the
Collaborative Research Center --- SFB 1624 “Higher Structures, Moduli Spaces,
and Integrability” --- 506632645, and by the German Research Foundation through
a German-Israeli Project Cooperation (DIP) grant ``Holography and the
Swampland''. H.A. is supported by startup funding from Northeastern
University. P.K.O. is supported by startup funding from UCSB. Part of this
work was performed at the Aspen Center for Physics, which is supported by
National Science Foundation grant PHY-2210452. 

\appendix
\section{Heterotic Little String Theories with Trivial Coweights}\label{app:unbroken-LST}

In this Appendix, we tabulate all heterotic LSTs with a trivial coweight,
that is with flavor symmetry $\mathfrak{e}_8\oplus \mathfrak{e}_8$,
$\mathfrak{so}_{32}$, or $\mathfrak{su}_{16}\oplus \fu_1$. We additionally give
the T-duality invariants for these theories, matching the closed-form
expression found in Section~\ref{sec:classification-spin32-lst}.
\begin{landscape}

\begin{table}
    \centering
    \begin{threeparttable}
        \begin{tabular}[t]{cccc}
				\toprule
				$\mathfrak{g}$ & LST & $\kappa_R$ & $\text{dim}(\text{CB})$ \\\midrule
				$\mathfrak{su}_{K=2k}$ & $\underset{[\mathfrak{so}_{32}]}{\overset{\mathfrak{sp}_{Q}}{1}}\underbrace{\overset{\mathfrak{su}_{2Q-8}}{2}\cdots\overset{\mathfrak{su}_{2Q-8\cdot i}}{2}\cdots\overset{\mathfrak{su}_{2Q-8(k-1)}}{2}}_{k-1}\overset{\mathfrak{sp}_{Q-4k}}{1} ~\times~\mathfrak{u}_1$ 
				& $\Gamma (Q - r_{\mathfrak{g}} - 1) + 2$ & $h^\vee Q  - \text{dim}(\mathfrak{g})$ \\\midrule
				$\mathfrak{su}_{K=2k+1}$ & $\underset{[\mathfrak{so}_{32}]}{\overset{\mathfrak{sp}_{Q}}{1}}\underbrace{\overset{\mathfrak{su}_{2Q-8}}{2}\cdots\overset{\mathfrak{su}_{2Q-8\cdot i}}{2}\cdots\overset{\mathfrak{su}_{2Q-8(k-1)}}{2}}_{k-1}\overset{\mathfrak{su}_{2Q-8k}}{1} ~\times~\mathfrak{u}_1$
				& $\Gamma (Q - r_{\mathfrak{g}} - 1) + 2$ & $h^\vee Q  - \text{dim}(\mathfrak{g})$ \\\midrule
				$\mathfrak{so}_{8}$ & $\overset{\displaystyle \overset{[\mathfrak{so}_{32}]}{\overset{\mathfrak{sp}_{Q}}{1}}}{\underset{\displaystyle \overset{\mathfrak{sp}_{Q - 8}}{1}}{\vphantom{\underset{[\mathfrak{so}_1]}{\overset{\mathfrak{so}_1}{4}}}}}\overset{\mathfrak{so}_{4Q - 16}}{1}\overset{\displaystyle \overset{\mathfrak{sp}_{Q - 8}}{1}}{\underset{\displaystyle \overset{\mathfrak{sp}_{Q - 8}}{1}}{\vphantom{\underset{[\mathfrak{so}_1]}{\overset{\mathfrak{so}_1}{4}}}}}$ & $8Q - 38$ & $6Q - 28$ \\\midrule
				$\mathfrak{so}_{2K=2(2k)}$ & {\scriptsize$
				[\mathfrak{so}_{32}]\overset{\mathfrak{sp}_{Q}}{1}
				\overset{\displaystyle \overset{\mathfrak{sp}_{Q-1}}{1}}{\overset{\mathfrak{so}_{4Q-16}}{4}}
				\underbrace{
					\overset{\mathfrak{sp}_{2Q-16}}{1}\,
					\overset{\mathfrak{so}_{4Q-32}}{4}
				    \cdots
					\overset{\mathfrak{sp}_{2Q - 8(k-i)}}{1}
					\overset{\mathfrak{so}_{4Q - 16(k-i)}}{4}
					\cdots
					\overset{\mathfrak{sp}_{2Q-8(k-1)}}{1}
					\overset{\displaystyle\overset{\mathfrak{sp}_{Q-4k}}{1}}{\overset{\mathfrak{so}_{4Q-16(k-1)}}{4}}
				}_{2(k-2)\,\text{curves}}
		\overset{\mathfrak{sp}_{Q-4k}}{1}
		$}  & $\Gamma (Q - r_{\mathfrak{g}} - 1) + 2$ & $h^\vee Q  - \text{dim}(\mathfrak{g})$ \\\midrule
				$\mathfrak{so}_{2K=2(2k+1)}$ & {\scriptsize$
				[\mathfrak{so}_{32}]\overset{\mathfrak{sp}_{Q}}{1}
				\overset{\displaystyle \overset{\mathfrak{sp}_{Q-1}}{1}}{\overset{\mathfrak{so}_{4Q-16}}{4}}
				\underbrace{
					\overset{\mathfrak{sp}_{2Q-16}}{1}\,
					\overset{\mathfrak{so}_{4Q-32}}{4}
				    \cdots
					\overset{\mathfrak{sp}_{2Q - 8(k-i)}}{1}
					\overset{\mathfrak{so}_{4Q - 16(k-i)}}{4}
				    \cdots
					\overset{\mathfrak{sp}_{2Q - 8(k-1)}}{1}
					\overset{\mathfrak{so}_{4Q-16(k-1)}}{4}
			}_{2(k-2)\,\text{curves}}
				\overset{\mathfrak{sp}_{2Q-8k}}{1}
				\overset{\mathfrak{su}_{2Q-8k}}{2}
		$}  & $\Gamma (Q - r_{\mathfrak{g}} - 1) + 2$ & $h^\vee Q  - \text{dim}(\mathfrak{g})$ \\\midrule
$\mathfrak{e}_{6}$ & $\underset{[\mathfrak{so}_{32}]}{\overset{\mathfrak{sp}_{Q}}{4}} \overset{\mathfrak{so}_{4Q - 16}}{1} \overset{\mathfrak{sp}_{3Q - 24}}{4} \overset{\mathfrak{su}_{4Q - 32}}{2} \overset{\mathfrak{su}_{2Q - 16}}{2}$ & $24Q - 166$ & $12Q - 78$ \\
$\mathfrak{e}_{7}$ & $\underset{[\mathfrak{so}_{32}]}{\overset{\mathfrak{sp}_{Q}}{1}} \overset{\mathfrak{so}_{4Q - 16}}{4} \overset{\mathfrak{sp}_{3Q - 24}}{1}\overset{\displaystyle\overset{\mathfrak{sp}_{2Q - 20}}{1}}{\overset{\mathfrak{so}_{8Q - 64}}{4}}\overset{\mathfrak{sp}_{3Q - 28}}{1} \overset{\mathfrak{so}_{4Q - 32}}{4} \overset{\mathfrak{sp}_{Q - 12}}{1}$ & $48Q - 382$ & $18Q - 133$ \\
$\mathfrak{e}_{8}$ & $\underset{[\mathfrak{so}_{32}]}{\overset{\mathfrak{sp}_{Q}}{1}} \overset{\mathfrak{so}_{4Q - 16}}{4} \overset{\mathfrak{sp}_{3Q - 24}}{1} \overset{\mathfrak{so}_{8Q - 64}}{4} \overset{\mathfrak{sp}_{5Q - 48}}{1}\overset{\displaystyle\overset{\mathfrak{sp}_{3Q - 32}}{1}}{\overset{\mathfrak{so}_{12Q - 112}}{4}}\overset{\mathfrak{sp}_{4Q - 40}}{1} \overset{\mathfrak{so}_{4Q - 32}}{4}$ & $120Q - 1078$ & $30Q - 248$ \\

           \bottomrule
        \end{tabular}
    \end{threeparttable}
	\caption{Heterotic Little String Theories of type $Spin(32)/\mathbb{Z}_2$ with a $\mathfrak{so}_{32}$ flavor symmetry, corresponding to a trivial choice of finite coweight of the associated algebra $\mathfrak{g}$.}
    \label{tab:het-so32-unbroken}
\end{table}

\begin{table}
    \centering
    \begin{threeparttable}
        \begin{tabular}[t]{cccc}
				\toprule
				$\mathfrak{g}$ & LST & $\kappa_R$ & $\text{dim}(\text{CB})$ \\\midrule
				$\mathfrak{su}_K$ & $[\mathfrak{e}_8]\,1\,2\overset{\mathfrak{su}_2}{2}\overset{\mathfrak{su}_3}{2}\cdots\overset{\mathfrak{su}_{K-1}}{2}\underbrace{\overset{\mathfrak{su}_{K}}{2}\dots}_{M}\overset{\mathfrak{su}_{K}}{2}\overset{\mathfrak{su}_{K-1}}{2}\cdots\overset{\mathfrak{su}_3}{2}\overset{\mathfrak{su}_2}{2}2\,1\,[\mathfrak{e}_8]$ &
				$K M+(K^2-K+2)$
				& $K M+(K^2+1) $\\
				$\mathfrak{so}_{2K}$ & {\scriptsize$[\mathfrak{e}_8]\,1 \, 2 \overset{\mathfrak{su}_2}{2} \overset{\mathfrak{g}_2}{3} 1 \overset{\mathfrak{so}_9}{4} \overset{\mathfrak{sp}_1}{1} \overset{\mathfrak{so}_{11}}{4} \cdots \overset{\mathfrak{sp}_{K-4}}{1} \underbrace{\overset{\mathfrak{so}_{2K}}{4}\overset{\mathfrak{sp}_{K-4}}{1}\cdots \overset{\mathfrak{so}_{2K}}{4}}_{M} \overset{\mathfrak{sp}_{K-4}}{1}\cdots  \overset{\mathfrak{so}_{11}}{4}\overset{\mathfrak{sp}_1}{1} \overset{\mathfrak{so}_9}{4} 1 \overset{\mathfrak{g}_2}{3}\overset{\mathfrak{su}_2}{2}2\,1\,[\mathfrak{e}_8] $} &
				{$\Gamma M + 4(K-2)^2+2$} &
			{$\Gamma M + 2 (K + 1)^2 - 9 K$}\\
			$\mathfrak{e}_6$ & 	$[\mathfrak{e}_8]\,1 \, 2 \overset{\mathfrak{su}_2}{2} \overset{\mathfrak{g}_2}{3} 1 \overset{\mathfrak{f}_4}{5} 1 \overset{\mathfrak{su}_3}{3} 1 \underbrace{\overset{\mathfrak{e}_6}{6}1 \overset{\mathfrak{su}_3}{3} 1 \cdots 1 \overset{\mathfrak{su}_3}{3} 1}_{M}\overset{\mathfrak{e}_6}{6} 1 \overset{\mathfrak{su}_3}{3} 1 \overset{\mathfrak{f}_4}{5} 1 \overset{\mathfrak{g}_2}{3} \overset{\mathfrak{su}_2}{2} 2 \, 1\,[\mathfrak{e}_8] $ &
				$24  M + 50$ &
				$12  M + 42$\\
				$\mathfrak{e}_7$ & 	$[\mathfrak{e}_8]\,1 \, 2 \overset{\mathfrak{su}_2}{2} \overset{\mathfrak{g}_2}{3} 1 \overset{\mathfrak{f}_4}{5} 1 \overset{\mathfrak{g}_2}{3} \overset{\mathfrak{su}_2}{2} 1 \underbrace{\overset{\mathfrak{e}_7}{8} 1 \overset{\mathfrak{su}_2}{2} \overset{\mathfrak{so}_7}{3} \overset{\mathfrak{su}_2}{2} 1 \cdots  1 \overset{\mathfrak{su}_2}{2} \overset{\mathfrak{so}_7}{3} \overset{\mathfrak{su}_2}{2} 1 }_{M}\overset{\mathfrak{e}_7}{8} 1 \overset{\mathfrak{su}_2}{2} \overset{\mathfrak{g}_2}{3} 1 \overset{\mathfrak{f}_4}{5} 1 \overset{\mathfrak{g}_2}{3} \overset{\mathfrak{su}_2}{2} 2 \, 1\,[\mathfrak{e}_8] $ &
				$48 M + 50$ & $18M + 47$\\
			$\mathfrak{e}_8$ &
			$
			[\mathfrak{e}_8]\,1 \, 2 \overset{\mathfrak{su}_2}{2} \overset{\mathfrak{g}_2}{3} 1 \overset{\mathfrak{f}_4}{5} 1 \overset{\mathfrak{g}_2}{3} \overset{\mathfrak{su}_2}{2} 2 1 \underbrace{\overset{\displaystyle 1}{\overset{\mathfrak{e}_8}{(12)}} \dots
			\overset{\displaystyle 1}{\overset{\mathfrak{e}_8}{(12)}}}_{M-2} 1 \, 2 \overset{\mathfrak{su}_2}{2} \overset{\mathfrak{g}_2}{3} 1 \overset{\mathfrak{f}_4}{5} 1 \overset{\mathfrak{g}_2}{3} \overset{\mathfrak{su}_2}{2} 2\, 1\,[\mathfrak{e}_8]
			$
			& $120M+29$ & $30M-8$\\
           \bottomrule
        \end{tabular}
    \end{threeparttable}
	\caption{Heterotic $E_8\times E_8$ Little String Theories. $M$ refers to insertions of conformal matter so that the curve $\overset{\mathfrak{g}}{n}_{\mathfrak{g}}$ occurs $M$ times. In the M-theory picture this corresponds to the number of $M5$-branes.}
    \label{tab:het-e8-e8-unbroken}
\end{table}

\begin{table}
    \centering
        \begin{tabular}[h]{ccc|cc}
				\toprule
			$\mathfrak{g}$ & $\widetilde{\mathfrak{b}}$ & $Spin(32)/\mathbb{Z}_2$ LST & $\widetilde{\mathfrak{b}}$ & $(SU(16)\times U(1))/\mathbb{Z}_2$ LST \\\midrule
			$\mathfrak{su}_{K=2N\geq4}$ &
			$C^{(1)}_{N}$ &
			$\underset{[\mathfrak{so}_{32}]}{\overset{\mathfrak{sp}_{Q}}{1}}\underbrace{\overset{\mathfrak{su}_{2Q-8}}{2}\cdots\overset{\mathfrak{su}_{2Q-8(N-1)}}{2}}_{N-1}\overset{\mathfrak{sp}_{Q-4N}}{1} ~\times~\mathfrak{u}_1$  & 
			--- & 
			$\lbrack \mathfrak{su}_{16} \rbrack \underset{\left[\mathfrak{su}_{2}\right]}{\overset{\mathfrak{su}_{K^{2}+(Q-1)K}}{0}} ~\times~\mathfrak{u}_1$
			\\\midrule
			$\mathfrak{so}_{8}$ &
			$D_4^{(1)}$ & 
			$\overset{\displaystyle \overset{[\mathfrak{so}_{32}]}{\overset{\mathfrak{sp}_{Q}}{1}}}{\underset{\displaystyle \overset{\mathfrak{sp}_{Q - 8}}{1}}{\vphantom{\underset{[\mathfrak{so}_1]}{\overset{\mathfrak{so}_1}{4}}}}}\overset{\mathfrak{so}_{4Q - 16}}{1}\overset{\displaystyle \overset{\mathfrak{sp}_{Q - 8}}{1}}{\underset{\displaystyle \overset{\mathfrak{sp}_{Q - 8}}{1}}{\vphantom{\underset{[\mathfrak{so}_1]}{\overset{\mathfrak{so}_1}{4}}}}}$
			& 

			$D_3^{(2)}$ & 
			$\underset{[\mathfrak{su}_{16}]}{\overset{\mathfrak{su}_{2Q}}{2}}\overset{\mathfrak{sp}_{2Q-8}}{1}\overset{\mathfrak{su}_{2Q-8}}{2} ~\times~\mathfrak{u}_1$
			\\\midrule
			$\mathfrak{so}_{10}$ &
			$D_5^{(1)}$ &
			$\overset{\mathfrak{sp}_{Q - 8}}{1}\,\overset{\displaystyle \overset{[\mathfrak{so}_{32}]}{\overset{\mathfrak{sp}_{Q}}{1}}}{\overset{\mathfrak{so}_{4Q - 16}}{4}}\,\overset{\mathfrak{sp}_{2Q - 16}}{1}\,\overset{\mathfrak{su}_{2Q - 16}}{2}$ & 
			$D_3^{(2)}$ & 
			$\underset{[\mathfrak{su}_{16}]}{\overset{\mathfrak{su}_{2Q}}{2}}{\overset{\mathfrak{su}_{4Q-16}}{1}}{\overset{\mathfrak{su}_{2Q-8}}{2}} ~\times~\mathfrak{u}_1	$ 
			\\\midrule

	$\mathfrak{so}_{K=4N}$ & 
	$D_{2N}^{(1)}$ &
	{\scriptsize$\begin{matrix}

	\overset{\mathfrak{sp}_{Q-1}}{1}
	\overset{\displaystyle \overset{[\mathfrak{so}_{32}]}{\overset{\mathfrak{sp}_{Q}}{1}}}{\overset{\mathfrak{so}_{4Q-16}}{4}}
	\underbrace{
		\overset{\mathfrak{sp}_{2Q-16}}{1}\,
		\overset{\mathfrak{so}_{4Q-32}}{4}
	    \cdots
		\overset{\mathfrak{sp}_{2Q-8(N-1)}}{1}
		\overset{\displaystyle\overset{\mathfrak{sp}_{Q-4N}}{1}}{\overset{\mathfrak{so}_{4Q-16(N-1)}}{4}}
	}_{2(N-2)\,\text{curves}}\overset{\mathfrak{sp}_{Q-4N}}{1}\end{matrix}$} & 
	$A_{2N-1}^{(2)}$ & 
	$ \begin{matrix}
		\overset{\mathfrak{su}_{2Q-8}}{2}
		\overset{\displaystyle\overset{\overset{\left[\mathfrak{su}_{16} \right]}{\mathfrak{su}_{2Q}}}{2}}
		{\overset{\mathfrak{su}_{4Q-16}}{2}}
		\underbrace{{\overset{\mathfrak{su}_{4Q-24}}{2}}
		\cdots
		{\overset{\mathfrak{su}_{4Q-8i}}{2}}
		\cdots}_{N-3}
		{\overset{\mathfrak{sp}_{2Q-4N}}{1}} 
		~\times~\mathfrak{u}_1\end{matrix}
		$\\\midrule

	$\mathfrak{so}_{K=4N+2}$ &
	$B_k^{(1)}$ & 
	$\begin{matrix}\underset{[\mathfrak{so}_{32}]}{\overset{\mathfrak{sp}_{Q}}{1}}\underbrace{\overset{\mathfrak{su}_{2Q-12}}{2}\cdots\overset{\mathfrak{su}_{2Q+4-8N}}{2}}_{N-1}\overset{\mathfrak{su}_{Q-2-4N}}{1}\end{matrix}$ & 
	$A_{2N-1}^{(2)}$ & 
	$\begin{matrix}\overset{\mathfrak{su}_{2Q-8}}{2}
	\overset{
		\displaystyle\overset{\overset{[\mathfrak{su}_{16}]}{\mathfrak{su}_{2Q}}}{2}
	}{\overset{\mathfrak{su}_{4Q-16}}{2}}
	\underbrace{{\overset{\mathfrak{su}_{4Q-24}}{2}}
	\cdots
	{\overset{\mathfrak{su}_{4Q-8i}}{2}}
	\cdots }_{N-3}
	{\overset{\mathfrak{su}_{4Q-8N}}{1}}  ~\times~\mathfrak{u}_1\end{matrix}
	$\\\midrule

	$\mathfrak{e}_{7}$ & 
	$E_7^{(1)}$ & 
	$\underset{[\mathfrak{so}_{32}]}{\overset{\mathfrak{sp}_{Q}}{4}} \overset{\mathfrak{so}_{4Q - 16}}{1} \overset{\mathfrak{sp}_{3Q - 24}}{4}\overset{\displaystyle\overset{\mathfrak{sp}_{2Q - 20}}{1}}{\overset{\mathfrak{so}_{8Q - 64}}{1}}\overset{\mathfrak{sp}_{3Q - 28}}{4} \overset{\mathfrak{so}_{4Q - 32}}{1} \overset{\mathfrak{sp}_{Q - 12}}{4}$ & 
	$E_6^{(2)}$ & 
	$
	\underset{[\mathfrak{su}_{16}]}{\overset{\mathfrak{su}_{2Q}}{2}}
	\overset{\mathfrak{su}_{4Q-16}}{2}
	\overset{\mathfrak{su}_{6Q-32}}{2}
	\overset{\mathfrak{sp}_{4Q-24}}{1}
	\overset{\mathfrak{so}_{4Q-16}}{4}~\times~\mathfrak{u}_1
	$ 
	\\\midrule
        \bottomrule
        \end{tabular}
		\caption{Exotic T-duality between $Spin(32)/\mathbb{Z}_2$ and $(SU(16)\times U(1))/\mathbb{Z}_2$ theories whose defining coweight corresponds to different algebras. $\mathfrak{g}$ indicates the type of singularity in the F-theory realization.}
    \label{tab:het-so32-unbroken-withoutvs}
\end{table}

\end{landscape}

\section{Derivation of the Anomaly Polynomial}\label{app:anomaly-polynomial}

The anomaly polynomial of a $\mathcal{N}=(1,0)$ SQFT can be written in terms of
characteristic classes. We give here some of the definitions needed to recover
our results. To wit, the A-roof genus $\widehat{A}(T)$, the Hirzebruch genus
$L(T)$, and the Chern character $\text{ch}(F)$ are given up to order eight as:
\begin{equation}
	\begin{aligned}
		\widehat{A}(T) &= 1 - \frac{1}{24}p_1(T) + \frac{1}{5760}\big(7p_1(T)^2-4p_2(T)\big) + \,\cdots \,, \\
		L(T) &= 1 + \frac{1}{3}p_1(T) - \frac{1}{45}\big(p_1(T)^2-7p_2(T)\big) + \,\cdots \,,\\
		\text{ch}_{\bm{R}}(F) &= \text{tr}_{\bm{R}}\, e^{iF} = \text{dim}(\bm{R}) - \frac{1}{2}\text{tr}_{\bm{R}}F^2 + \frac{1}{24}\text{tr}_{\bm{R}}F^4 + \cdots \,,\\
		\text{ch}_{\bm{2}}(R) &= 2 - c_2(R) + \frac{1}{12}c_2(R)^2 + \cdots \,,\qquad c_2(R) = \frac{1}{4}\text{Tr}(R^2)\,,
	\end{aligned}
\end{equation}
where $T, R$ are the background fields corresponding to spacetime and
R-symmetry, respectively, and $F$ the field strength of a gauge or flavor
symmetry. The first and second Pontryagin classes are denoted $p_1(T), p_2(T)$,
respectively. The ``one-loop'' contributions of the various multiplets to the
anomaly polynomial are then
\begin{equation}
	\begin{gathered}
		I_8^\text{hyp}(F) = \left.\widehat{A}(T)\text{ch}_{\mathcal{R}}(F)\right|_\text{8-form}\,,\qquad
		I_8^\text{vec}(F) = - \frac{1}{2}\left.\widehat{A}(T)\text{ch}_{\bm{2}}(R)\text{ch}_{\textbf{adj}}(F)\right|_\text{8-form}\,,\\
		I_8^\text{tensor} = \left.\left(\frac{1}{2}\text{ch}_{\bm{2}}(R)\widehat{A}(T) - \frac{1}{8} L(T)\right)\right|_\text{8-form}\,, \qquad
	\end{gathered}
\end{equation}
If we need to differentiate between the field strength associated with flavor
and gauge symmetries, we will denote them as $F_I$ and $\widetilde{F}_I$,
respectively.

Furthermore, in the case of gauge or flavor symmetries, the traces,
$\text{tr}_{\bm{R}}F^n$, over the background field $F$ depend on the
representation $\bm{R}$ under which the associated fields transform. As
different representations appear in the spectrum, it is important to convert
them to so-called one-instanton normalized traces $\text{Tr} F^n$. Given a
field strength $F$ associated with a Lie algebra, we have the conversion
relations:\footnote{We follow the standard conventions for six-dimensional
theories \cite{Ohmori:2014kda, Heckman:2018jxk, Baume:2021qho}, see Appendix A
of reference \cite{Baume:2023onr} for a review of how trace relations are
obtained for general representations.} 
\begin{equation}\label{app:trace-relations}
		\text{tr}_{\bm{R}} F^2 = h_{\bm{R}} \text{Tr} F^2\,,\qquad \text{tr}_{\bm{R}} F^4 = x_{\bm{R}}\text{Tr}F^4 + y_{\bm{R}}(\text{Tr}F^2)^2\,.
\end{equation}
For the fundamental representation $\bm{F}$ of a classical algebra, we have $(x_{\bm{F}},
y_{\bm{F}})=(1,0)$ and $h_{\bm{F}} = \frac{1}{2}$ for $\mathfrak{su}_k,
\mathfrak{sp}_k$ or $h_{\bm{F}} = 1$ for $\mathfrak{so}_k$. The second-order
quartic Casimir of the adjoint representation is given by $y_{\bm{adj}}=\frac{3}{4}, 3,
\frac{3}{2}$ for $\mathfrak{sp}_k$, $\mathfrak{so}_k$, $\mathfrak{su}_k$,
respectively, while $x_{\bm{adj}}$ satisfies the relation given in equation
\eqref{app:trace-relations}. To encompass all nodes of the generalized quiver
at the same time, we use the following notation:
\begin{equation}
		\begin{gathered}
				\text{Tr}_{\bm{F}} F^2_I = h^J_I\,\text{Tr}F^2_J\,,\qquad \text{Tr}_{\bm{F}} \widetilde{F}^2_I = \widetilde{h}^J_I\,\text{Tr}F^2_J\,, \qquad \sum_I \text{Tr}_{\bm{adj}} \widetilde{F}^2_I = h^I\text{Tr}\widetilde{F}^2_I\,,\\
				\text{tr}_{\bm{F}} \widetilde{F}^4_I = \text{Tr} \widetilde{F}^4_I\,,\qquad
			\sum_I \text{Tr}_{\bm{adj}} \widetilde{F}^4_I = x^I\text{Tr}\widetilde{F}_I^4 + 16 y^{IJ}c_2(\widetilde{F}_I)c_2(\widetilde{F}_J)\,.
		\end{gathered}
\end{equation}
with $h^I=h^\vee_{\mathfrak{g}_I}$ and $h^I_J, \widetilde{h}^I_J, y^{IJ}$
diagonal matrices whose elements are the corresponding coefficients of the
trace relations in Equation \eqref{app:trace-relations}.

Due to the fact that $\mathfrak{so}_{32}$ LSTs have only classical gauge and
flavor symmetries, we can make use of certain group-theoretical relations to
reduce the number of relevant quantities appearing in the anomaly
polynomial, such as the dimension of the fundamental representation $d_J$ and
the Frobenius--Schur indicator $S^I$, see Equation \eqref{classical-relations}. 

Let us compute the one-loop anomaly polynomial given in equation
\eqref{I8-ABJ}. The matrices $A^{IJ}$ and $D^{IJ}$ encode the adjacency of
gauge-gauge and gauge-flavor bifundamental hypermultiplets, differentiating
between half and full hypermultiplets, see discussion around equation
\eqref{I_8-def}. Expanding the anomaly polynomial to make the gauge
contributions $c_2(\widetilde{F}_I)=\frac{1}{4}\text{Tr}F^2_I$ and
$\text{Tr}\widetilde{F}^4_I$ explicit, one finds:
\begin{equation}\label{I8-1loop-app}
		\begin{aligned}
		I_8^\text{1-loop} =& \left.\left(\frac{1}{2}\widehat{A}(T)d_I D^{IJ}\text{ch}_{\bm{F}}(F_J) +\frac{1}{4}d_IA^{IJ}d_J\widehat{A}(T) -\frac{1}{2}\sum_I \text{dim}(\mathfrak{g}_I)\widehat{A}(T)\text{ch}_{\bm{2}}(R)\right)\right|_{8-form}\\
				& + r_{\mathfrak{b}} I_8^\text{tensor} + c_2(\widetilde{F}_I)\left( \big( (\widetilde{h}A\widetilde{h})^{IJ} - \frac{2}{3}y^{IJ}\big)c_2(\widetilde{F}_J) + \delta^{IJ}c_2(F_J)- h^I c_2(R) \right)\\
				& + \frac{1}{24}p_1(T)c_2(\widetilde{F}_I)\left( (D\widetilde{h})^I_Jf^J + 2 (\widetilde{h}A)^{IJ}d_J - 2h^I\right) + \frac{1}{48}\left(f^I D_I^J + d_IA^{IJ} -2 x^J \right)\text{Tr}\widetilde{F}_J^4
		\end{aligned}
\end{equation}
Using the relations between the quartic Casimir in Equation
\eqref{classical-relations} and the definition of the Cartan matrix in Equation
\eqref{def-symm}, we can see that the last term is given by
\begin{equation}
		\text{Tr}\widetilde{F}^4_J\left(D_I^Jf^I + A^{JI}d_I -2 x^J \right) = \text{Tr}\widetilde{F}^4_J D^J_I(f^I - 16S^I - C^{IK}d_K)\,,
\end{equation}
which leads to the anomaly-cancellation condition used throughout this work. We
have used that since $S^I=0$ for $\mathfrak{su}_K$ algebras, then
$D^I_JS^J=S^I$. The rest of the anomaly polynomial can be further simplified by
noticing that since $\mathfrak{sp}_k$ gauge multiplets are rotated by
$\mathfrak{so}_n$ flavor symmetries---and vice versa---we have 
\begin{equation}
		h^J_I =\frac{1}{2}V^J_J\,,\qquad \widetilde{h}^I_J = (D^{-1}V^{-1})^I_J\,\qquad y^{IJ} = 3(V^{-2}D^{-1})^{IJ}\,,
\end{equation}
It is then straightforward to find that the Dirac pairing can be written as
\begin{equation}
		\frac{1}{2}\eta^{IJ} = \frac{2}{3}y^{IJ} - (\widetilde{h}A\widetilde{h})^{IJ} = (V^{-1}CD^{-1}V^{-1})^{IJ}
\end{equation}
The second expression reproduces the coefficient of the term proportional to
$c_2(\widetilde{F})_I\cdot c_2(\widetilde{F})_J$ in Equation \eqref{I8-1loop-app}.
Using the other relations, we are led to the expression given in Equation \eqref{1-loop-so32}.

\section{Of Roots and Weights}\label{app:roots}

We give here a short review of the properties of root systems for finite
algebras and their (possibly-twisted) affine versions.

Let us consider a simple Lie algebra $\mathfrak{g}$ of rank $r_{\mathfrak{g}}$
with Cartan subalgebra $\mathfrak{h}$. The integer root lattice
$\mathcal{Q}\subset \mathfrak{h}^*$ is generated by simple roots $\alpha^a\,, a
= 1,\dots, r_{\mathfrak{g}}$. Given the usual invariant bilinear form
$(\cdot,\cdot)$ on $\mathfrak{h}^*$, the Cartan matrix is obtained through the
natural pairing between $\mathfrak{h}$ and $\mathfrak{h}^*$:
\begin{equation}
		C^{ab} = \left<\alpha^a, \alpha^{b\,\vee}\right> = \frac{2(\alpha^a,\alpha^b)}{(\alpha^b,\alpha^b)}\,,
\end{equation}
with the coroots defined as $\alpha^{b\,\vee} = 2\alpha^b/(\alpha^b,\alpha^b)$,
and span the coroot lattice $\mathcal{Q}^\vee$. Lowercase Latin indices run over the rank of $\mathfrak{g}$:
$a,b=1,\dots,r_{\mathfrak{g}}$.

The simple roots are taken to be in a normalization proportional to the
diagonal matrix $D^{aa} = \frac{1}{2}(\alpha^a, \alpha^a)$, i.e. the longest
simple root is taken to have length two. This is the convention for $D$ used in
Section~\ref{sec:anomalies}, and is furthermore used to define the
\emph{symmetric Cartan matrix}, $G=CD=DC$. 

The fundamental coweights $\omega_a^\vee$ generating the coweight lattice
$\mathcal{P}^{\vee}$ are then defined as the dual basis of the root lattice $\mathcal{Q}$:
\begin{equation}
		\langle\alpha^a, \omega_b^\vee\rangle = \delta^a_b\,,
\end{equation}
In addition, we will also need the notion of dominant coweights, which are
those with non-negative integer coefficients in the basis of fundamental
coweights:
\begin{equation}
		\mathcal{P}^\vee_+ = \{\mu^\vee = \mu^a \,\omega_a^\vee \in \mathcal{P}^\vee\,|\, \mu^a \in \mathbb{N}~\forall a\}\,.
\end{equation}
The weight lattice $\mathcal{P}$ and the set of dominant weights $\mathcal{P}_+$ are defined in a similar manner.

\paragraph{Affine root systems:} to study LSTs, we need to consider the affine
version $\widetilde{\mathfrak{g}}$ of $\mathfrak{g}$. The root lattice
$\mathcal{Q}$ of $\mathfrak{g}$ is then augmented to its affine version
$\widetilde{\mathcal{Q}}$ by adding the zeroth root
$\alpha^0$, such that the affine Cartan matrix is
given by
\begin{equation}\label{affine-cartan-appendix}
		\langle \alpha^A, \alpha^{B\,\vee}\rangle = C^{AB}\,,
\end{equation}
Uppercase Latin indices include the zeroth component:
$A,B=0,1,\dots,r_{\mathfrak{g}}$. The null root
$\widetilde{\theta}=\theta_A\alpha^A$ is then defined as the affine root whose
coefficients, called the marks, satisfy $\theta_AC^{AB}=0$. Similarly, the
comarks $K_A$ are defined as satisfying $C^{AB}K_B=0$.

The weight and coweight lattices $\widetilde{\mathcal{P}}$,
$\widetilde{\mathcal{P}}^\vee$ are then constructed by supplementing the affine
(co)roots with a weight $\Lambda_0$ and two coweights $\mathcal{D}\,,K$ referred to
as the scaling weight and the central element, respectively:\footnote{
		While we mostly follow Kac's conventions \cite{Kac:1990gs}, to
		avoid confusion with the dimension of the fundamental representations
		$d_I$ and the coefficient $\delta$ of the anomaly polynomial, we denote
		the scaling element and null root by $\mathcal{D}$ and $\widetilde{\theta}$,
respectively.}

\begin{equation}
		\widetilde{\mathcal{P}} = \mathcal{P}^{\phantom{\vee}}\,\oplus \,\mathbb{Z}\,\widetilde{\theta}\,\oplus \,\mathbb{Z}\, \Lambda_0\,, \qquad
		\widetilde{\mathcal{P}}^\vee = \mathcal{P}^\vee \, \oplus \, \mathbb{Z} \, \mathcal{D} \, \oplus \, \mathbb{Z} \, K\,.
\end{equation}
We have already encountered the central element---whose coefficients are the
comarks---above, and the other two generators satisfy the relations:
\begin{equation}
		\langle \alpha^A, \mathcal{D} \rangle = \delta^{A0} = \langle\Lambda_0, \alpha^{A\,\vee}\rangle\,,\qquad
		K = K_A\, \alpha^{A\,\vee}\,.
\end{equation}
To ensure that the coweights are generated by a basis dual to the affine roots,
we define the fundamental coweights as their finite versions, shifted by $\mathcal{D}$:
\begin{equation}
	\begin{gathered}
			\widetilde{\omega}_0^\vee = \mathcal{D}\,,\qquad 
			\widetilde{\omega}_a^\vee = \omega_a^\vee + \theta_a\,\mathcal{D}\,,\qquad \left<\alpha^A, \widetilde{\omega}_B^\vee\right> = \delta^A_B\,.
	\end{gathered}
\end{equation}
An affine coweight can then be expanded in this basis as:\footnote{The only
exception is $A^{(2)}_{2k}$, which has $\theta_0=2$, and the affine coweights
must be rescaled appropriately. All other affine Lie algebras have $\theta_0 =
1$ \cite{Kac:1990gs}.} 
\begin{equation}\label{app:coweight-def}
		\mu^\vee = \mu^A\, \widetilde{\omega}^\vee_A + n\, K \, = \, \mu^a\omega_a^\vee + (\mu^0 + \theta_a \mu^a)\mathcal{D} + n\,K\,,
\end{equation}
and is dominant if $\mu^A,n\in\mathbb{N}$. Equivalently, we can see
that if we choose a dominant coweight of the simple algebra $\mu^\vee\in
\mathcal{P}^\vee_+$, we can lift it to a dominant coweight of $\widetilde{\mathfrak{g}}$,
$\widetilde{\mu}^\vee \in \widetilde{\mathcal{P}}^\vee_+$ by choosing a positive integer
$n$, and a level which uniquely fixes the component $\mu^0$.
Every dominant affine coweight
is therefore given by a triple $(\mu^\vee, k, n)$ where $k$ is the level of the
coweight:
\begin{equation}
		\text{lv}(\widetilde{\mu}^\vee)=\langle\widetilde{\theta}, \mu^\vee \rangle = \theta^A\, \mu_A\,.
\end{equation}
Twisted affine algebras can be obtained from their non-twisted versions by
quotienting by an automorphism of their Dynkin diagram. For the purpose of this
work, we will only need to know the root system through its Cartan matrix
in Equation \eqref{affine-cartan-appendix}. Finally, the affine weights
$\widetilde{\omega}_I$ are defined in a similar fashion as in equation
\eqref{app:coweight-def}:
\begin{equation}
	\widetilde{\omega}_0 = \Lambda_0\,,\qquad \widetilde{\omega}_a = \omega_a + K_a\,\Lambda_0\,.
\end{equation}
From this, we find
\begin{equation}
		\langle \alpha^I, \widetilde{\omega}_J^\vee\rangle = \delta^I_J = \langle \widetilde{\omega}_J, \alpha^{I\,\vee}\rangle\,,\qquad \langle \widetilde{\omega}_I , \widetilde{\omega}_J^\vee\rangle = X_{IJ}\,.
\end{equation}
The matrix $X_{AB}$ defined as $X_{0B} = 0 = X_{A0}\,,X_{ab}=(C^{-1})_{ab}$ is
precisely the matrix appearing when inverting the quartic-anomaly cancellation
condition, see Equation \eqref{sol-dI}.

\bibliography{references}
\bibliographystyle{utphys}

\end{document}

%% file: figures/slice_bk.tex
\begin{tikzpicture}
	\node[color=blue] (A1)  {2};
	\node[color=red] (A0) [below=3mm of A1] {$[2k+1]$};
    \draw (A0.north) -- (A1.south);
\end{tikzpicture}